\documentclass[longauth]{aa}

\usepackage[varg]{txfonts}
\usepackage{graphicx}
\usepackage{amsmath}
\usepackage{amssymb}
\usepackage{color}
\usepackage{natbib}	
\usepackage{url}
\usepackage[color=blue!20, bordercolor=orange, disable]{todonotes}
\usepackage{multirow}
\usepackage{threeparttable}      
\usepackage{booktabs}
\usepackage{soul}
\DeclareUnicodeCharacter{2212}{-}
\usepackage{float}
\usepackage[colorlinks=true]{hyperref}

\bibpunct{(}{)}{;}{a}{}{,}

\begin{document}
\title{The ESPRESSO transmission spectrum of HD\,189733\,b}
\subtitle{Extracting the planetary sodium and lithium signatures amid stellar contamination\thanks{Based on Guaranteed Time Observations (GTO) collected at the European Southern Observatory under ESO programme 1102.C-0744 by the ESPRESSO Consortium.}}
\author{D. Mounzer\inst{1}\thanks{\email{dany.mounzer@unige.ch}}
\and W. Dethier\inst{2}
\and C. Lovis\inst{1}
\and V. Bourrier\inst{1}
\and A. Psaridi\inst{1,4,5}
\and H. Chakraborty\inst{1}
\and M. Lendl\inst{1}
\and R. Allart\inst{6,1}
\and J. V. Seidel\inst{7,8}
\and M. R. Zapatero Osorio\inst{9}
\and P. Molaro\inst{10}
\and M. Steiner\inst{1}
\and D. Ehrenreich\inst{1}
\and Y. Alibert\inst{12,13}
\and I. Carleo\inst{10,14,15}
\and S. Cristiani\inst{10,11}
\and J. I. Gonz\'alez Hern\'andez\inst{14,15} 
\and C. J. A. P. Martins\inst{16,2}
\and E. Palle\inst{14,15}
\and J. Rodrigues\inst{2,3,17}
\and N. Santos\inst{2,3}
\and A. Sozzetti\inst{18}
\and A. Su\'arez Mascare\~no \inst{14,15}
}

\institute{Geneva Observatory, University of Geneva, Chemin Pegasi 51b, CH-1290 Versoix,  Switzerland 
\and Instituto de Astrofísica e Ciências do Espaço, Universidade do Porto, CAUP, Rua das Estrelas, 4150-762 Porto, Portugal
\and Departamento de Fisica e Astronomia, Faculdade de
Ciencias, Universidade do Porto, Rua Campo Alegre, 4169-007 Porto, Portugal
\and Institute of Space Sciences (ICE, CSIC), Carrer de Can Magrans S/N, Campus UAB, Cerdanyola del Valles, E-08193, Spain
\and Institut d’Estudis Espacials de Catalunya (IEEC), 08860 Castelldefels (Barcelona), Spain
\and Département de Physique, Institut Trottier de Recherche sur les Exoplanètes, Université de Montréal, Montréal, Québec, H3T 1J4, Canada
\and European Southern Observatory, Alonso de Córdova 3107, Vitacura, Región Metropolitana, Chile
\and Laboratoire Lagrange, Observatoire de la Côte d'Azur, Nice, France
\and Centro de Astrobiología, CSIC-INTA, Camino Bajo del Castillo s/n, 28692 Villanueva de la Cañada, Madrid, Spain
\and INAF – Osservatorio Astrofisico di Trieste, via G. B. Tiepolo 11, 34143, Trieste, Italy
\and IFPU–Institute for Fundamental Physics of the Universe, via Beirut 2, I-34151 Trieste, Italy
\and Center for Space and Habitability, University of Bern, Gesellschaftsstrasse 6, CH-3012 Bern, Switzerland
\and Physics Institute of University of Bern, Gesellschaftsstrasse 6, CH3012 Bern, Switzerland
\and Instituto de Astrof{\'\i}sica de Canarias, E-38205 La Laguna, Tenerife, Spain
\and Universidad de La Laguna, Dept. Astrof{\'\i}sica, E-38206 La Laguna, Tenerife, Spain
\and Centro de Astrof\'{\i}sica da Universidade do Porto, Rua das Estrelas,
4150-762 Porto, Portugal
\and Observatoire François-Xavier Bagnoud -- OFXB, 3961 Saint-Luc, Switzerland
\and INAF – Osservatorio Astrofisico di Torino, Via Osservatorio 20, I10025 Pino Torinese, Italy
}
\date{Received March 27, 2025 / Accepted June 17, 2025}

\abstract{While transmission spectroscopy has allowed us to detect many atomic and molecular species in exoplanet atmospheres, the improvement in resolution and signal-to-noise ratio (S/N) enabled us to become sensitive to planet-occulted line distortions (POLDs) in the spectrum that are induced by center-to-limb variations (CLV) and the Rossiter-McLaughlin effect (RM). POLDs can bias the interpretation of the transmission spectrum, and it is hard to correct for them with stellar models.} 
{We analyzed two ESPRESSO transits ($\mathcal{R}  \sim  140\,000$) of the archetypal hot Jupiter HD\,189733\,b. The transmission spectrum of this aligned system is heavily affected by POLDs, stellar activity, and instrumental effects. It is therefore a challenging study case of how to account for these effects when the planetary signal is retrieved from chemical species through transmission spectroscopy.} 
{We used the \textsc{antaress} workflow to process the datasets to ensure an accurate correction for telluric and instrumental contamination. With improved architectural parameters derived using the RM revolutions technique, we tested several methods of including and correcting the strong POLDs in the transmission spectrum. We then derived the absorption spectrum from sodium through simultaneous forward-modeling of the star and planet using the code called evaporating exoplanets (EvE).} 
{We confirm the previous detections of the sodium doublet signature in the upper atmosphere of HD\,189733\,b. When we accounted for POLDs and isolated the planetary signal from uncorrected stellar residuals, we found a shallower (0.432 $\pm$ 0.027 \%) and more strongly blueshifted (-7.97 $\pm$ 0.28 km s$^{−1}$) signal. We attempted to reinterpret the other high-resolution sodium studies of this system in light of our results. We suggest that the POLDs and stellar activity are insufficiently corrected for in all analyses, including ours. We also detected a planetary lithium signature of 0.102 ± 0.016 \% (6.4$\sigma$) at a blueshift of -2.4 $\pm$ 1.8 km s$^{-1}$.}  
{We have probably reached limitations in the accuracy of theoretical stellar spectra and in our understanding of stellar variability at the timescale of a transit because we are unable to fully correct for the effect of POLDs in HD\,189733\,b transmission spectra. As we shift toward a new generation of ground-based spectrographs on the ELT with an even higher S/N and resolution, addressing these issues will be paramount for a proper characterization of exoplanet atmospheres with transit spectroscopy.}

\titlerunning{}
\maketitle
\hypersetup{citecolor=blue}
\hypersetup{linkcolor=red}

\section{Introduction}\label{section:Intro}

Transmission spectroscopy \citep{SS2000} has greatly contributed to our understanding of the composition and dynamics of exoplanetary atmospheres.
Lower-resolution space-based instruments excel at detecting molecular species through broadband signatures (e.g., \citealt{Rustamkulov23}) and to identify hazes and/or clouds in planet atmospheres (e.g., \citealt{Bell24}) without being hindered by telluric contamination. In contrast, the high-resolution power of a ground-based facility can distinguish the contribution of the atmospheric, planetary, telluric, and stellar lines through the Doppler shift induced by the velocity of each component.

In the optical bandpass, this narrow-band capability can lead to the discovery of chemical species with many absorption lines (e.g.,  \citealt{Bello20}, \citealt{Basil24}) through the cross-correlation technique (CCF; \citealt{Snellen10}), or with fewer absorption lines (e.g., \citealt{Borsa21}). The analysis of the line shape, position, depth, and timing in the absorption spectra can give insights into the dynamical processes of the atmospheres, such as winds, condensation, and evaporation (e.g., \citealt{Seidel23, Radica24, Guill24}, respectively).

When the sensitivity and resolution are high enough, temporal variations during transits that are induced by asymmetries in the atmospheric architecture can also be revealed. For example, on ultra-hot Jupiters, Fe I on WASP-76b \citep{Ehren20} and Na I on WASP-121b \citep{Seidel23, Seidel25} were revealed using ESPRESSO\footnote{Echelle SPectrograph for Rocky Exoplanets and Stable Spectroscopic Observations} \citep{Pepe14,Pepe2021}.

The higher data quality and signal-to-noise ratio (S/N) expose finer effects that are due to planet-occulted line distortions (POLDs; \citealt{Deth23}) arising from the Rossiter-McLaughlin effect (RM; \citealt{Ross1924,McL1924}) and center-to-limb variations (CLV; \citealt{Czesla15, Yan2017}), however. The signature of the RM effect in radial velocity (RV) measurements can uncover stellar and orbital properties such as the projected spin-orbit angle as well as the rotation speed, convective blueshift, differential rotation, and inclination of the star \citep{Cegla16, Bourrier2021, Cristo23}. In transmission spectra, however, POLDs can cause absorption features to be misinterpreted because their combination with a planetary signal (or a lack thereof) can lead to different interpretations of atmospheric (non-)signatures.

This is the case for the benchmark hot Jupiter HD\,209458\,b. As the first exoplanet that was observed to transit \citep{Charb00,Henry00,Mazeh2000}, it also yielded the first detection of an exoplanetary atmosphere through an excess sodium absorption at 589 nm \citep{Charb02} at a spectral resolution of $\mathcal{R}  \sim  5\,500$. Recent studies have revealed that this absorption can exclusively be explained by the effects of POLDs \citep{CB2020, CB2021} at the sodium line cores on various transits with the high-resolution spectrographs HARPS(-N)\footnote{High Accuracy Radial velocity Planet Searcher (for the Northern hemisphere)} \citep{Mayor03,Cosen12}, CARMENES\footnote{Calar Alto high-Resolution search for M dwarfs with Exoearths with Near-infrared and optical Échelle Spectrographs} \citep{Quirr14}, and ESPRESSO. \citet{Cart23} showed, however, that at the resolution of HST/STIS, POLDs cannot explain the signal detected by \citet{Charb02}. This indicates that this signal might originate from the wings of the sodium doublet.

Another extensively studied hot Jupiter, HD\,189733\,b \citep{Bouchy2005}, has a radius of 1.138 $R_\mathrm{J}$, a mass of 1.123 $M_\mathrm{J}$, and an equilibrium temperature of 1712 K. It orbits a bright active K2 star (V = 7.67) every P = 2.22 days \citep{Bonomo17, Krenn23}. It holds the first atmospheric detection from ground-based observations \citep{Redfield08}. The bright host star of these two planets, the high effective temperature, and the low density make them prime candidates for atmospheric characterization, together with some of the highest transmission spectroscopy metric (TSM; \citealt{Kemp18}) at TSM $\sim$ 951 for HD\,209458\,b and TSM $\sim$ 743 for HD\,189733\,b. The two systems are aligned and have a similar stellar rotational velocity \citep{CB2021,Cristo23}, so that we expect their RM effect on the transmission spectrum to be comparable in shape and amplitude.

Since its discovery, a plethora of chemical species and dynamical processes have been detected in the atmosphere of HD\,189733\,b (see the exhaustive overview by \citealt{Blain24} in the visible). We mostly focus on transmission spectroscopy in the Na\,I D doublet ($\lambda_{D1}$ = 5895.924 \AA, $\lambda_{D2}$ = 5889.951 \AA) and on the ways in which the planetary signal can be distinguished from POLDs. The study of the sodium signature in the atmosphere of HD\,189733\,b has a long history from ground-based \citep{Redfield08, Jens11, Wytt15} to space-based facilities \citep{Huit12,Pont13,Sing16}, and the latest high-resolution facilities accounted for RM and CLV \citep{Loud15, Khal2017, Borsa18,Lange22,Sici22,Blain24,Keles24,Sici25}.

This paper is structured as follows: We summarize in Sect. \ref{section:Obs} the ESPRESSO observations of HD\,189733\,b and describe the data reduction with \textsc{antaress}. We derive the orbital architecture parameters using the RM revolutions technique (RMR) in Sect. \ref{section:RM}. We then compute the transmission spectrum in Sect. \ref{section:TS} and propose several ways to account for POLDs in the sodium doublet. Sect. \ref{section:EVE} describes the simultaneous modeling of the planetary signal and POLDs in the absorption spectrum using EvE. We discuss our findings on other atomic species in Sect. \ref{atoms}.
Finally, we discuss our results in Sect. \ref{section:Disc} and compare them with other studies of POLD correction and the sodium signature of HD\,189733\,b before we conclude.

\begin{table*}[ht]
\caption{ESPRESSO observations log of HD189733.}
\label{table:log}
\centering
\begin{tabular}{c c c c c c c c c}
\toprule
Night \# & Date & Telescope & Mode & Spectra & $T_\mathrm{exp} [s]$ & Airmass & Seeing \tablefootmark{a} & S/N at 550nm \tablefootmark{a} \\
\midrule
Night 1 & 2021-08-10 & VLT-UT1 & HR21 & 41 (18 in, 22 out, 1 discarded) & 300 & 2.1-1.5-1.6 & 0.5-0.9 & 129-203 \\
Night 2 & 2021-08-30 & VLT-UT1 & HR21 & 43 (18 in, 25 out) & 300 & 1.7-1.5-2.0 & 0.7-1.7 & 107-194 \\ \toprule
\end{tabular}
\tablefoot{
\tablefoottext{a}{The left value corresponds to the minimum, and the right value corresponds to the maximum.}
}
\end{table*}

\begin{table}[ht]
\caption{Physical and orbital parameters of the system HD\,189733}
\label{table:params}
\small
\centering
\begin{tabular}{l l c}
\toprule \toprule
Parameter & Symbol  [Unit] & Value \\
\toprule
\multicolumn{3}{c}{-- \textit{Stellar Parameters} --} \\

Stellar mass & $M_\mathrm{\star}$ [$M_\mathrm{\odot}$] & 0.783 $\pm$ 0.0041 \\
Stellar radius & $R_\mathrm{\star}$ [$R_\mathrm{\odot}$] & 0.784 $\pm$ 0.007 \\ 
Effective temperature & $T_\mathrm{eff}$ [K] & 4969 $\pm$ 43 \\
Metallicity & [Fe/H] [dex] & -0.07 $\pm$ 0.02  \\
Surface gravity & $log\,g_\mathrm{\star}$ [cgs] &  4.51 $\pm$ 0.03 \\
Limb darkening coefficient & $u_\mathrm{1}$ & 0.6617 \\
Limb darkening coefficient & $u_\mathrm{2}$ & -0.3940 \\
Limb darkening coefficient & $u_\mathrm{3}$ & 1.0507 \\
Limb darkening coefficient & $u_\mathrm{4}$ & -0.4533 \\ 
\\
\multicolumn{3}{c}{-- \textit{Planetary Parameters} --} \\
Planet mass & $M_\mathrm{p}$ [$M_\mathrm{J}$] & 1.123 $\pm$ 0.045  \\
Planet radius & $R_\mathrm{p}$ [$R_\mathrm{J}$] & 1.187 $\pm$ 0.011\\
\\
\multicolumn{3}{c}{-- \textit{System Parameters} --} \\
Transit epoch & $T_\mathrm{c}$ [$BJD_\mathrm{TDB}$]& 2459446.498519  \\
& & $^{+0.000012}_{-0.000013}$ \\
Orbital period & P [d] & 2.2185751979   \\
& & $^{+0.0000000698}_{-0.0000000728}$ \\
Transit duration (P$_1$ to P$_4$) & $T_\mathrm{14}$ [h] & 1.806 $\pm$ 0.033  \\
Transit duration (P$_2$ to P$_3$) & $T_\mathrm{23}$ [h] & 0.992 $\pm$ 0.037  \\
Planet-to-star radius ratio & $R_\mathrm{p}$ / $R_\mathrm{\star}$ & 0.15565 $^{+0.00024}_{-0.00021}$ \\
Orbital semi-major axis & a [$R_\mathrm{\star}$] & 8.8843 $^{+0.0173}_{-0.0177}$  \\
Orbit inclination & i [deg] & 85.705 $\pm$ 0.016  \\
Eccentricity & e & 0 (fixed)\\
Stellar RV semi-amplitude & $K_\mathrm{\star}$ [m s$^{-1}$] & 201.3 $\pm$ 1.6 \\
Planet RV semi-amplitude & $K_\mathrm{p}$ [km s$^{-1}$] & 150.038 $\pm$ 2.619 \\
Systemic velocity (Night 1) & $\gamma_1$ [km s$^{-1}$] & -2.2203 $\pm$ 0.0004 \\
Systemic velocity (Night 2) & $\gamma_2$ [km s$^{-1}$] & -2.2290 $\pm$ 0.0005 \\ 
\\
\multicolumn{3}{c}{-- \textit{Rossiter-McLaughlin Parameters} --} \\ 
Projected spin-orbit angle & $\lambda$ [deg] & -0.81 $\pm$ 0.07 \, \\
Star rotational velocity & $v_{eq}$ [km s$^{-1}$]& 3.65$^{+0.25}_{-0.27}$ \, \\
Stellar inclination & $i_\star$ [deg] & 91.2$^{+4.5}_{-6.7}$  \\
Convective blueshift velocity & $v_{CB}$ [m s$^{-1}$] & -174$^{+37}_{-33}$  \\
Differential rotation coefficient & $\alpha_{rot}$ & 0.32$^{+0.14}_{-0.22}$ \\

\bottomrule
\end{tabular}
\tablefoot{All parameters were taken or derived from \citet{Krenn23}, except for $K_\mathrm{\star}$, $M_\mathrm{p}$ (from \citet{Bonomo17}), the limb darkening coefficients (derived from \citet{Pont13}), the RM parameters, and the systemic velocities, which were derived from our analysis described in Sect. \ref{section:RM}.}
\end{table}

\section{Observations and data reduction}\label{section:Obs}

\begin{figure} 
\includegraphics[trim=0cm 0cm 0cm 0cm,clip=true,width=\columnwidth]{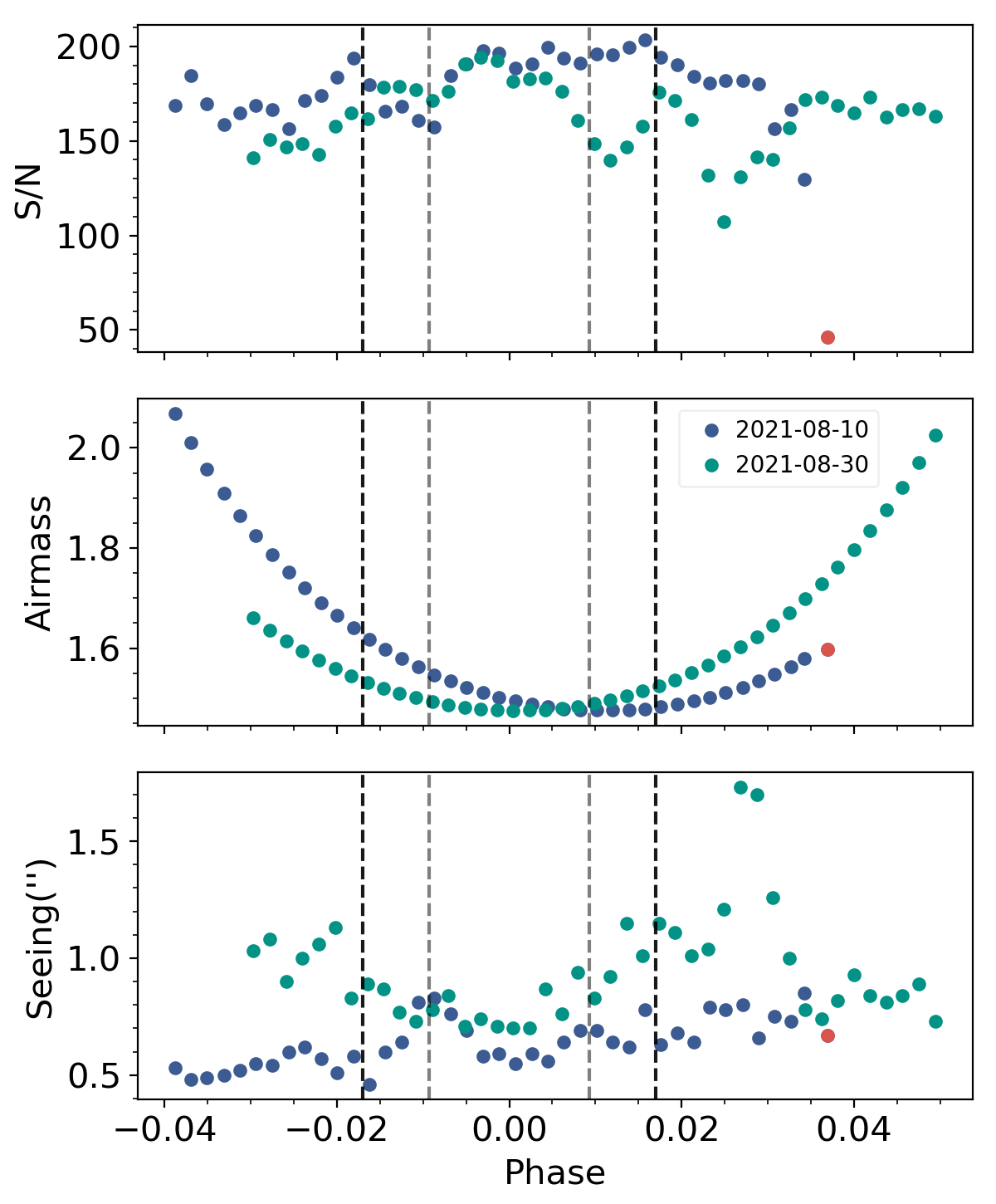}
\centering
\caption[]{Signal-to-noise ratio, airmass, and seeing taken during the two observation nights as a function of the transit phase. Zero is the center of the transit, and the dashed black vertical lines show the transit contact points from $T_1$ to $T_4$. The last spectrum of the first night was discarded because the S/N is far lower.}
\label{fig:ObsLog}
\end{figure}

\subsection{ESPRESSO observations of HD\,189733\,b}

Two transits of HD\,189733\,b were observed using the ESPRESSO spectrograph on one of the four 8.2m Unit Telescopes (UTs) of the ESO Very Large Telescope (VLT) in Paranal, Chile, on the nights of 10 and 30 of August 2021, under program 1104.C-0350 (PI: F. Pepe) as part of the guaranteed time observations (GTO). ESPRESSO is an ultra-stable fiber-fed high-resolution spectrograph spanning the optical bandpass from 3800 to 7880 \AA~at a spectral resolution of $\mathcal{R}  \sim  140\,000$ for the observing mode used (HR21). The data were taken with UT1 and fiber B pointing at the sky.

The observations are summarized in Fig. \ref{fig:ObsLog} and in Table \ref{table:log}, which show the variations in the S/N, airmass, and seeing during the two nights. Forty-one and 43 spectra were taken on the respective nights, and the last spectrum of the first night was discarded because the S/N was lower. Given the brightness of the star, even at a minimum airmass of 1.5, a high S/N -- above 100 on the order of the Na\,I D doublet -- was achieved on all spectra, using an exposure time of 300s.

The starting data products used in this study were the S2D blazed spectra produced by the ESPRESSO data reduction software (DRS) pipeline version 3.0.0, which are now publicly available through the ESO Science Archive\footnote{\url{https://archive.eso.org/cms.html}}. Further corrections to the datasets are described in Sect \ref{section:ANTARESS}.

Most of the parameters used for the data reduction and analysis were taken from \citet{Krenn23}, who refined the transit parameters with two CHEOPS\footnote{Characterizing Exoplanets Satellite} \citep{Benz21} transit observations of HD\,189733 simultaneous to our ESPRESSO transits, combined with 13 CHEOPS occultations and TESS\footnote{Transiting Exoplanet Survey Satellite} \citep{Rick14} transits. The complete list of parameters used is shown in Table \ref{table:params}.

\subsection{Simultaneous EulerCam photometry}
\label{sec:LC}
We observed two full transits of HD\,189733\,b with the EulerCam photometer (\citealt{Lendl2012}), mounted on the 1.2m Euler Swiss telescope at the ESO La Silla Observatory. The observations were conducted simultaneously with ESPRESSO on the nights of August 10 and August 30, 2021, using the Gunn $r'$ filter with exposure times of 30 seconds and 10 seconds, respectively. We processed the EulerCam data using standard calibration methods, including bias subtraction and flat-field correction. Differential aperture photometry was employed to extract the transit light curves, carefully selecting reference stars and apertures of 54 and 29 pixels, respectively, to minimize the RMS of the final data.

We fit an achromatic transit model to the EulerCam light curves, fixing all properties to the values of \cite{Krenn23} and letting free the planet ephemeris. Our results are consistent but less precise than those derived from high-precision CHEOPS photometry in \cite{Krenn23}, and we therefore adopted the latter for our analysis. 

For Visit 2, we used a spot plus transit model from \textsc{PyTranSpot} \citep{Juvan18}. We set the spot parameters including size, temperature, latitude, and longitude as free parameters. In addition, the planet-to-star radius ratio and mid-transit time were set as free parameters for the transit model. The lower Bayesian Information Criterion (BIC) for a spot plus transit model over a pure transit model indicates that the Visit 2 transit is contaminated with a stellar spot crossing ($\Delta\rm{BIC} > 4$). The data are consistent with a spot of angular size 3 degrees (similar to sun spots). A detailed characterization of the occulted spot will be presented in a follow-up paper.

\begin{figure}
\includegraphics[trim=0cm 0cm 0cm 0cm,clip=true,width=\columnwidth]{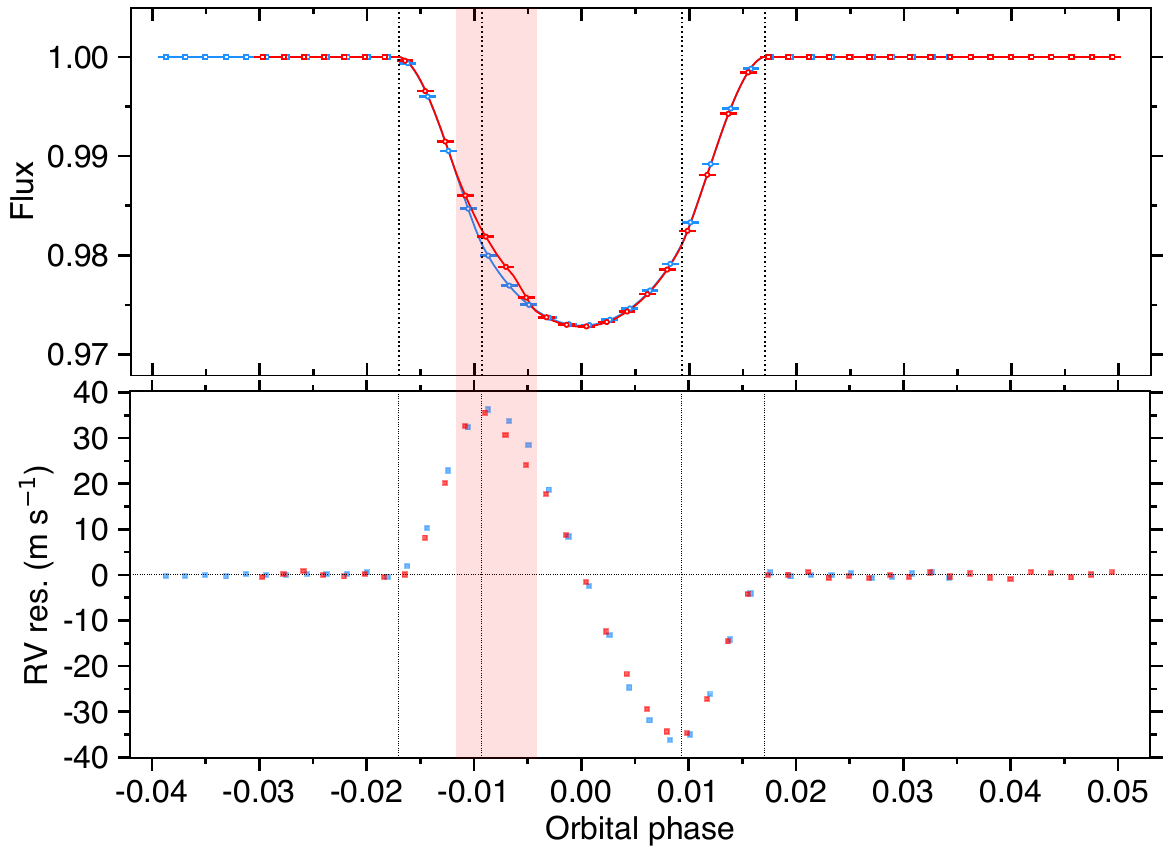}
\centering
\caption[]{Transit comparison between Visit 1 (blue) and Visit 2 (red). ESPRESSO exposures are indicated by the colored symbols. Vertical dashed lines indicate transit contacts. \textit{Top panel:} Best-fit model light curves to the EulerCam photometry obtained simultaneously with the ESPRESSO transits, used to scale disk-integrated CCFs to their correct relative flux level. \textit{Bottom panel:} RV centroids of the disk-integrated CCFs, computed from spectra Doppler-corrected for the stellar Keplerian motion and linear drifts with phase. The spot occultation visible in Visit 2 photometry, occurring within the red-shaded window, induces a clear deviation to the nominal RM anomaly in RV, which has also been found in \cite{Cristo23}.}
\label{fig:LC_RV}
\end{figure}

\subsection{Data reduction with ANTARESS} \label{section:ANTARESS} 
In order to prepare the input data for our analysis, we chose to perform additional correction steps after the initial ESPRESSO data reduction, using the \textsc{antaress} workflow (\citealt{B24}, thereafter B24). \textsc{antaress} is a set of modules that allow for the homogeneous processing of spectral time series, successively cleaned, formatted, and processed to extract the relevant output for a given analysis, with careful propagation of errors. All modules and their connections are listed in Fig. 1 of B24 and are described in detail in the article. 

The following \textsc{antaress} modules were run on the input S2D nonblaze-corrected spectra from the ESPRESSO DRS, in order:
\begin{itemize}
    \item Instrumental response
    \item Telluric correction
    \item Flux balance correction
    \item Cosmics correction
    \item Sine patterns correction (ESPRESSO wiggles)
\end{itemize}
The cross-correlation module was then run in order to convert the cleaned and formatted 2D spectra into CCFs. Those outputs were fit to derive time series of disk-integrated RVs, contrast, and full width at half maximum (FWHM) (Fig. \ref{fig:LC_RV}). Out-of-transit measurements were fit with polynomials to search for deviations from the Keplerian RV model and the average stellar line shape. We identified linear correlations of the RV residuals with time and of the contrast with S/N in both visits, which were then corrected for in the S2D spectra. The out-of-transit RV residuals derived from the corrected spectra then provided the systemic RV of the system in each visit, listed in Table \ref{table:params}. The corrected S2D spectra were then processed as follows:
\begin{itemize}
    \item Alignment in the star rest frame
    \item Broadband flux scaling
    \item Conversion 2D/1D
\end{itemize}
To avoid biases, it is important to scale the spectra using chromatic light curves that account for broadband variations in the planetary absorption and stellar intensity. To this purpose, we generated a set of light curves using the properties from \citet{Krenn23}, and the chromatic spot-corrected transit depths derived by \citet{Pont13} using the limb darkening coefficients from \citet{Hayek12}. We included the spot contribution derived from the EulerCam light curve in the set of light curves for Visit 2. The resulting data products are 1D spectra with the correct relative flux level, aligned in the star rest frame. They were used for the computation and analysis of the transmission spectrum, as described in Sect. \ref{section:TS}. Instead of applying the 2D/1D conversion, the data were also processed with the following modules: 
\begin{itemize}
    \item Extraction of differential spectra
    \item Extraction of intrinsic spectra
    \item Conversion 2D/CCF
    \item RMR analysis
\end{itemize}
This sequence allowed us to extract the planet-occulted spectra along the transit chord that can be used to perform an RMR analysis, as described in Sect. \ref{section:RM}. 
While most of the listed modules' role is self-explanatory, we detail in the following subsections some of the modules that are most pertinent and significant to the field of high-resolution transmission spectroscopy and need further explanation. 

\subsubsection{Telluric correction} 

Telluric correction consists of removing the contamination of Earth's atmosphere in spectral series done with ground-based instruments. Several approaches to do so exist with similar favorable results, for example, atmospheric models-based algorithms such as \textsc{Molecfit} \citep{Smette2015,Kausch2015} or empirical methods such as PCA-based corrections \citep{Arti14,Cret21}. In \textsc{antaress}, the need to have as few input parameters, homogeneous corrections, and simple integration made the automatic telluric correction from \citealt{Allart22} (thereafter ATC) the prime choice of inclusion. ATC is a model-based algorithm that corrects for the main telluric molecules in the optical and near-infrared bandpasses (H$_2$O, O$_2$, CO$_2$ and CH$_4$) by fitting a CCF of a model telluric spectrum for each molecule to the CCF of the exposure in order to get their individual atmospheric properties (e.g., pressure and column density). Then, a telluric spectrum is generated by combining theoretical spectra of all molecules fit at the resolution of the observations, and is divided from the exposure.

For these ESPRESSO datasets, we chose to correct for H$_2$O and O$_2$, the main telluric components in the optical range. The shallow water absorption lines around the sodium doublet and H$\alpha$ (< 10 \%) are well corrected. We chose to mask deep telluric lines, with a threshold of 35\%, as those saturated lines are not well constrained by telluric models when fitting for the CCF. This mostly occurs in the O$_2$ A and B bands in the wavelength range of ESPRESSO, the former being in the same range as two persistent potassium lines $\lambda_{K1}$ = 7698.9645 \AA\, and $\lambda_{K2}$ = 7664.8991 \AA. $\lambda_{K2}$ is completely blended in the O$_2$ telluric absorption lines, while $\lambda_{K1}$ contains several well-corrected shallow lines, but is surrounded by deep lines, which made the study of a potential potassium signature tedious with these observations.

\subsubsection{Correction for ESPRESSO wiggles}
\label{sect:wiggles}
Sinusoidal interference patterns (called wiggles) affect the spectra from ESPRESSO observations in particular. These interferences arise in the Coudé train optics and thus appear on observations with any of the VLT UTs \citep{Allart20, Tab21}. The period and amplitude of the wiggles vary with wavelength and time, and appear to be correlated with the telescope pointing coordinates and guiding star (see Sect. 5.6 in B24). The wiggles are dominated by a beat pattern between two sinusoidal components, which can be well described and corrected for by the analytical model presented in B24. Some datasets, like the ones of HD\,189733, are further contaminated by a third component at lower frequencies, which creates strong and broad S-like features at specific locations in transmission spectra (see Fig. \ref{fig:TS Na wiggle}). This third component is not yet included in \textsc{antaress} wiggle model. Instead, we used \textsc{antaress} filtering correction, which captures the full wiggle pattern in each exposure by fitting a cubic spline to the transmission spectra smoothed with a Savitzky-Golay filter (to prevent capturing high-frequency spectral variations from stellar and planetary lines). Flux spectra were then corrected for the wiggles through division by this model.

\section{Rossiter-McLaughlin analysis}
\label{section:RM}

Intrinsic CCFs derived from the ESPRESSO spectra (Sect.~\ref{section:ANTARESS}) were analyzed with \textsc{antaress} following the RMR procedure detailed in \citealt{Bourrier2021} and \citealt{B24}.

\begin{figure}
\includegraphics[trim=2cm 0cm 3.5cm 4cm,clip=true,width=\columnwidth]{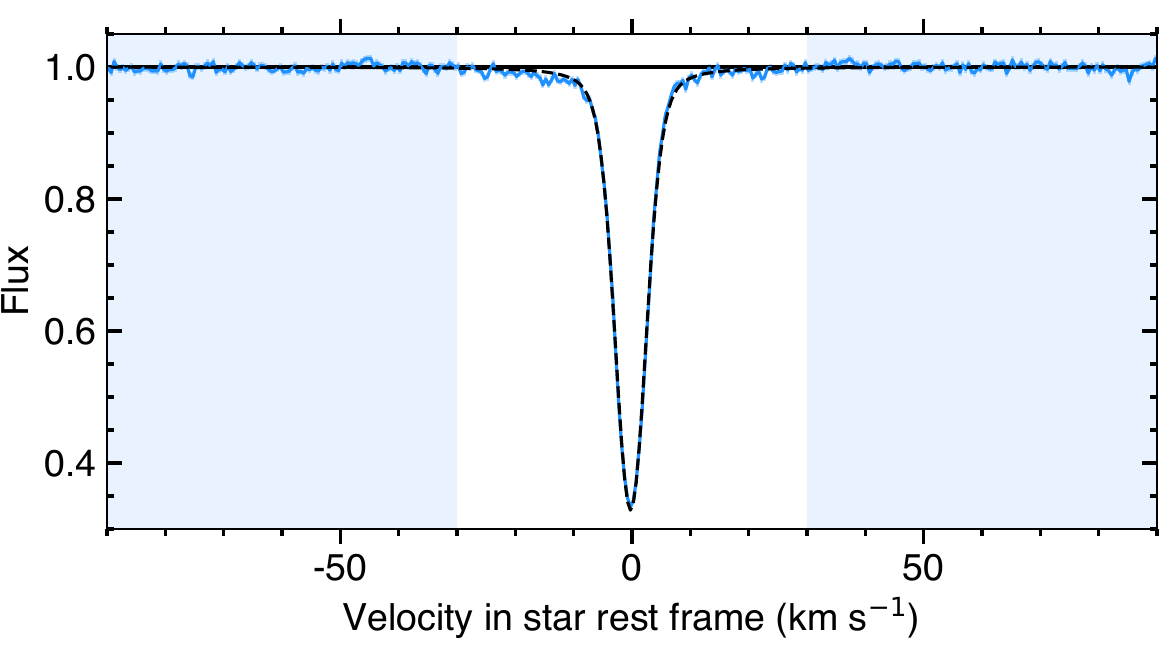}
\centering
\caption[]{Representative intrinsic CCF measured near mid-transit in Visit 1 (blue profile). The dashed black profile is the best-fit Voigt model. The blue range shows the range used to define the continuum.}
\label{fig:Intrinsic_CCF}
\end{figure}

\begin{figure}
\includegraphics[trim=0cm 0cm 0cm 0cm,clip=true,width=\columnwidth]{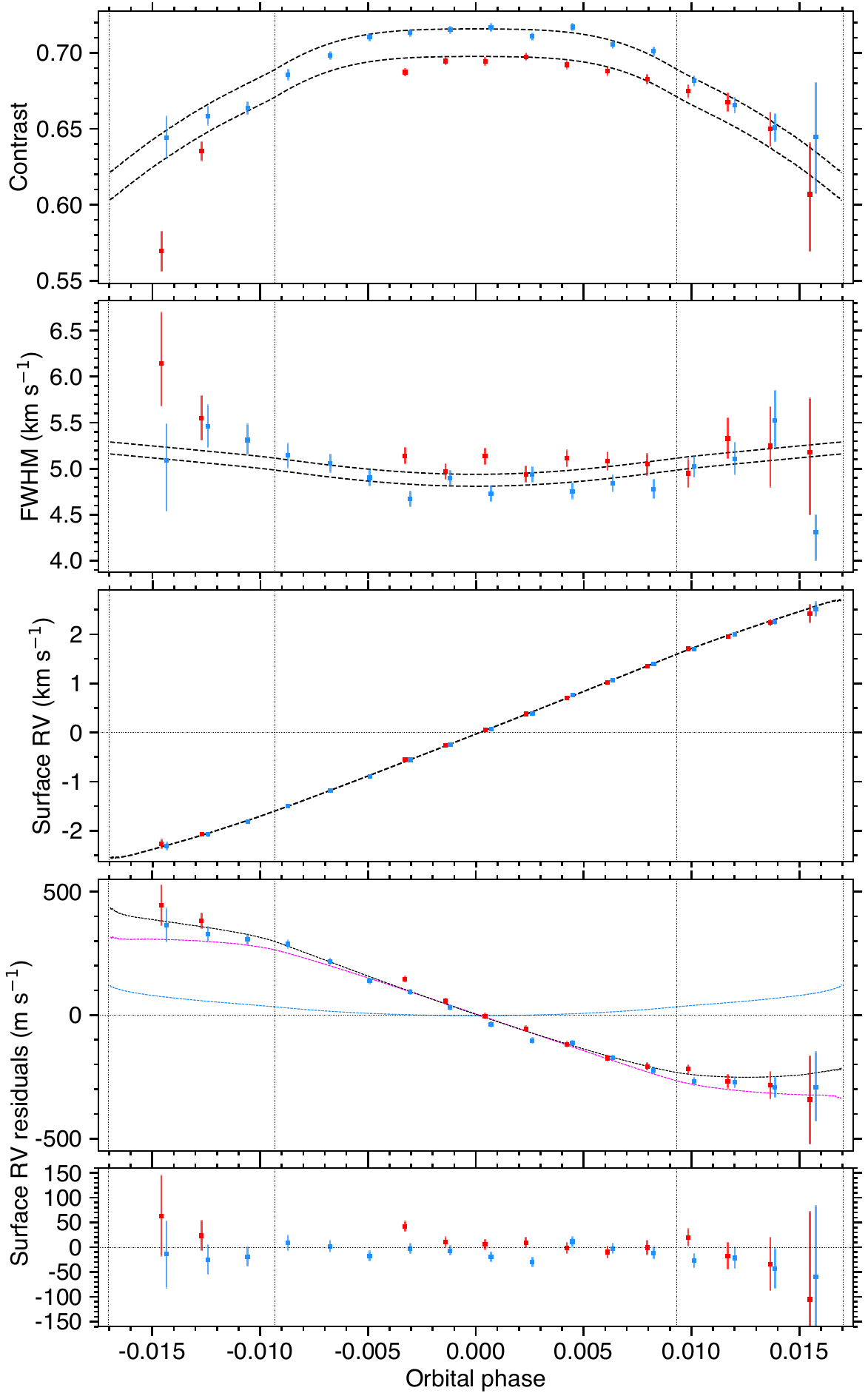}
\centering
\caption[]{Properties of the intrinsic CCFs for HD\,189733. The vertical dashed lines indicate transit contacts. Blue and red points show the contrast, FWHM, and RV centroids derived from fits to individual intrinsic CCFs. Dashed black curves show the models for these properties associated with the best RMR fit to the joint visits. The last two panels show residuals between the RV centroids and the solid-body component (top, calculated with the equatorial velocity) or complete model (bottom) for the stellar surface RVs. In the former panel, the blue and magenta curves show the convective blueshift and differential rotation components, and the black curve shows their combination.}
\label{fig:Intrinsic_properties}
\end{figure}

In a first step, intrinsic CCFs were fit independently to determine which exposures to exclude from the joint time-series fit. They also allowed us to evaluate the best models that describe the photospheric RV field, the local stellar line profile, and its spatial variations across the photosphere. 
The intrinsic line of HD\,189733 was best modeled with a Voigt profile (Fig.~\ref{fig:Intrinsic_CCF}). Individual fits were performed with uniform priors on the line RV centroid ($\mathcal{U}$(-3,3)\,km\,s$^{-1}$, FWHM ($\mathcal{U}$(4, 8)\,km\,s$^{-1}$, contrast ($\mathcal{U}$(0.5, 0.75)), and damping parameter ($\mathcal{U}$(0, 0.5)), based on the variations of each property along the transit chord (Fig.~\ref{fig:Intrinsic_properties}). In both visits, we discarded the first in-transit exposure, during which the planet occulted the edge of the stellar limb where intrinsic CCFs are too faint to be constraining. We also excluded the four spot-contaminated exposures in Visit 2 (Fig.~\ref{fig:LC_RV}), identified using the best-fit transit model to the EulerCam light curve. These exposures display a clear deviation from the nominal properties, visible even in the RM anomaly, and they will be used in a follow-up study to constrain the spot properties. 
The surface RVs of HD\,189733 were measured with exquisite precision with ESPRESSO and allowed us to constrain the differential rotation and a linear convective blueshift. A photospheric RV model that includes both contributions was preferred through BIC comparison over a solid-body rotation model. The intrinsic stellar line also displays clear variations in shape along the transit chord, which were best modeled as polynomials of the sky-projected distance from the star center. Furthermore, the FWHM and contrast of the line in both visits were best described with the same linear and quadratic variations, respectively, with level modulated by a visit-specific coefficient. The damping parameter was best described by the same linear model in both visits.

\begin{figure}
\includegraphics[trim=0cm 0cm 0cm 0cm,clip=true,width=\columnwidth]{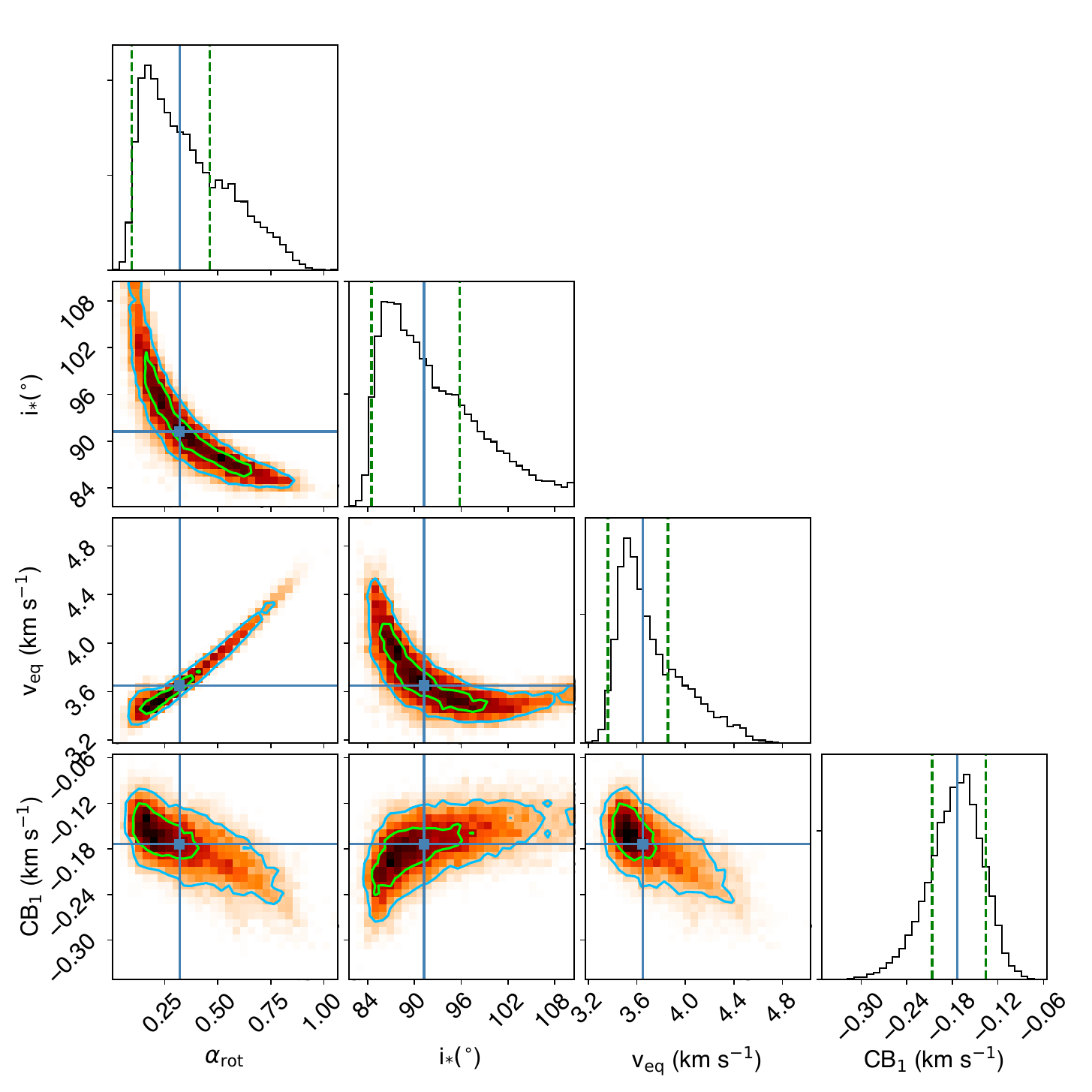}
\centering
\caption[]{Correlation diagrams for the PDFs of the correlated RMR model parameters. The green and blue lines show the 1 and 2$\sigma$ simultaneous 2D confidence regions that contain, respectively, 39.3\% and 86.5\% of the accepted steps. 1D histograms correspond to the distributions projected on the space of each line parameter, with the green dashed lines limiting the 68.3\% HDIs. The blue lines and squares show the median values.}
\label{fig:Corr_diag_ZOOM}
\end{figure}

\begin{figure*}
\begin{minipage}[tbh!]{\textwidth}
\includegraphics[trim=0cm 0cm 0cm 0cm,clip=true,width=\textwidth]{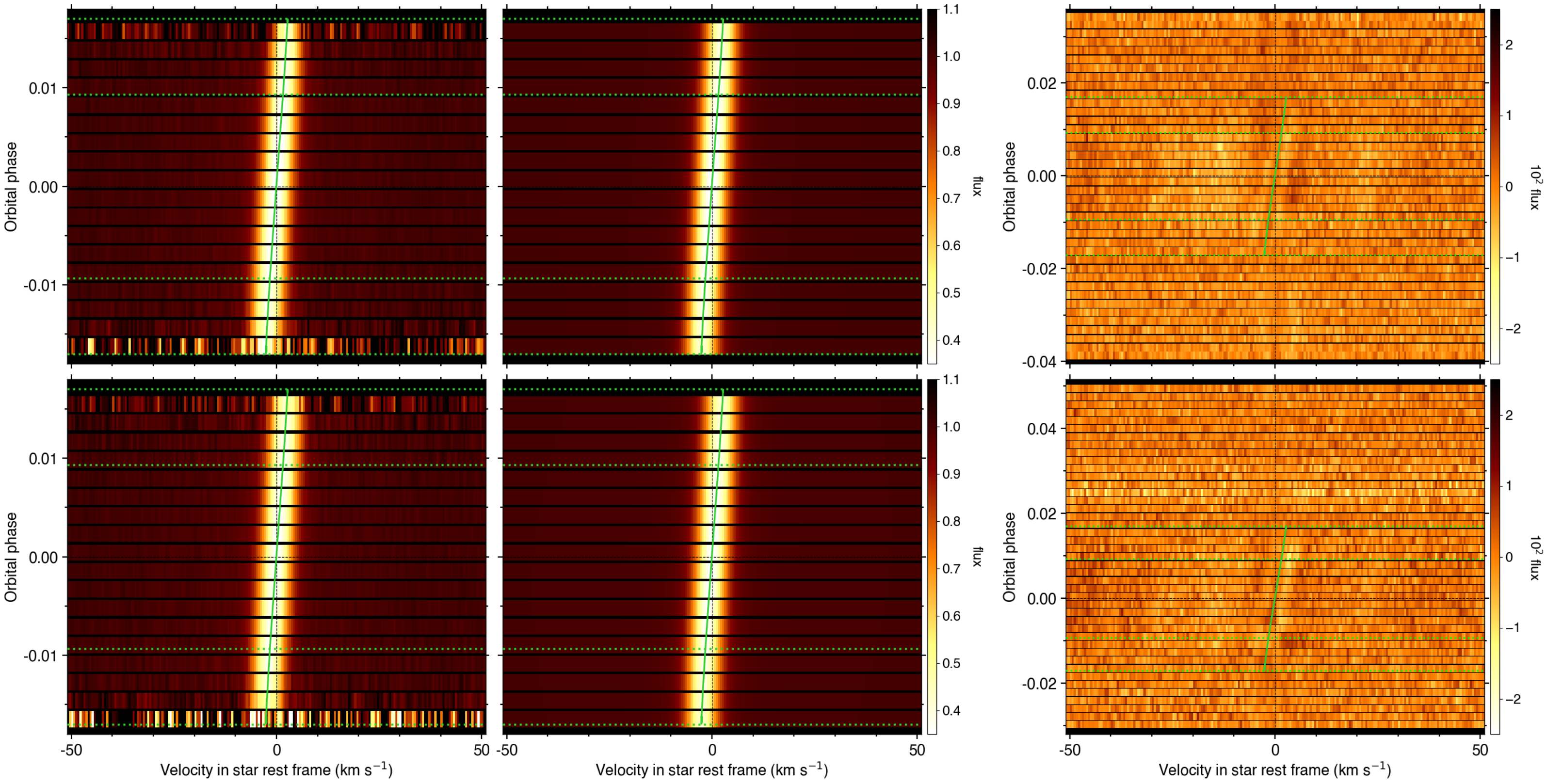}
\centering
\end{minipage}
\caption[]{RMR maps in Visit 1 (top panel) and Visit 2 (bottom panel), plotted as a function of RV in the star rest frame (abscissa) and orbital phase (ordinate). The horizontal dashed green lines show the transit contacts. The solid green line indicates the surface RVs track associated with the best RMR fit. \textit{Left panels}: Measured intrinsic CCFs. \textit{Middle panels}: Best-fit RMR model to the joined intrinsic CCF series. \textit{Right panels}: Residuals between the out-of-transit CCFs and the master-out CCF (outside of the transit window) and between the intrinsic CCFs and their best-fit model (scaled to the level of the out-of-transit residuals).}
\label{fig:Maps_RM}
\end{figure*}

In a second step, we fit a joint model of the stellar line, informed by the first step, to all intrinsic CCFs. We note that in the two steps, the model line profile was convolved with the ESPRESSO line spread function (LSF) before it was fit to the measurements, so that the derived parameters trace the true intrinsic photospheric properties. The constraint on differential rotation breaks the degeneracy on the sky-projected stellar rotational velocity. The joint RMR fit was thus performed using the stellar inclination (as its cosine $\cos i_\star$) and rotational velocity $v_\mathrm{eq}$ as jump parameters, together with the relative differential rotation rate $\alpha$, linear convective blueshift coefficient $CB_\mathrm{1}$, sky-projected spin-orbit angle $\lambda$, and the intrinsic line shape parameters determined in the first step. The joint fit to both visits was constrained well enough that only broad, noninformative uniform priors needed to be set on the model properties. The correlation diagram for all fit properties is shown in Fig.~\ref{fig:Corr_diag_FULL}. While most properties show well-defined Gaussian probability density functions (PDFs), $\cos i_\star$ displays two modes correlated with $\alpha$ and $CB_\mathrm{1}$. The high-inclination mode corresponds to $i_\star$ $\sim$ 117$^{\circ}$ (less consistent with the literature; e.g., \citealt{Cegla16}) and contains much fewer samples than the dominant mode, which was thus isolated for the final analysis (Fig.~\ref{fig:Corr_diag_ZOOM}). We note that this did not significantly impact the results, which are reported in Table \ref{table:params}. Residual maps between the intrinsic CCFs measured in each visit and our best-fit model are shown in Fig.~\ref{fig:Maps_RM}.

We measured $v_{\rm eq}$ = 3.65$\stackrel{+0.25}{_{-0.27}}$\,km\,s$^{-1}$ and $\lambda$ = -0.81$\pm$0.07$^{\circ}$, lower than the values derived by \cite{Cegla16} but still consistent within 2$\sigma$. Our derived stellar inclination, $i_\star$ =  91.2$\stackrel{+4.5}{_{-6.7}}$$^{\circ}$, matches their result very well. \cite{Cristo23} analyzed the same ESPRESSO dataset with a different approach and found somewhat different results on some parameters, but these differences do not significantly impact the projected velocity field on the stellar surface for the purpose of computing the transmission spectrum. Our analysis confirms previous literature results that HD\,189733 is seen equator-on with an aligned planetary system. Our constraints on the stellar differential rotation, with $\alpha$ = 0.32$\stackrel{+0.14}{_{-0.22}}$, are consistent with \cite{Fares10}, \cite{Cegla16} and \cite{Cristo23}. Furthermore, we detected convective blueshift, with a linear variation with $\mu$ by $CB_\mathrm{1}$ = 174$\stackrel{+37}{_{-33}}$\,m\,s$^{-1}$ that corresponds to a net redshift toward the stellar limb, as expected from the radiative 3D MHD simulations of \cite{Cegla16}. Our detections of the intrinsic line shape variations with $\mu$ also match very well the predictions by \cite{Cegla16}, with contrast decreasing toward the stellar limb. Based on their simulations, the increase in the line FWHM toward the limb suggests a low magnetic field strength (on the order of a few tens of G), in line with spectropolarimetry measurements by \cite{Fares10}.

\section{Sodium transmission spectrum of HD\,189733\,b}
\label{section:TS}
\subsection{Computing the transmission spectrum}
The transmission spectrum \citep{SS2000} allows the comparison of spectral properties between an exoplanet atmosphere and its host star during the transit of the former, by dividing in-transit spectra $\tilde{F}_\mathrm{in,i}$ at phase $i$ by an out-of-transit master $\tilde{F}_\mathrm{out}$. POLDs can appear when the disk-integrated spectrum is not representative of the occulted spectra in a given photospheric region, either due to the stellar velocity field (RM) or to the CLVs (see Sect. \ref{RMcorr}). The method for computing the transmission spectrum has evolved through the years, from the most commonly used wavelength-dependent excess absorption $-\tilde{\mathcal{R}} = 1-\tilde{F}_\mathrm{in,i} /  \tilde{F}_\mathrm{out}$ by the planet atmosphere from its opaque disk \citep{Redfield08} to Doppler-shifting each in-transit spectrum to the planetary rest frame so that the potential planetary signal is always aligned with itself during the transit \citep{Wytt15}. This method was then adapted in \citet{Wytt20} and \citet{Moun22} to account for transit depth using the broadband transit light curve $1-\delta_\mathrm{i}$ and for stellar limb darkening LD$_\mathrm{i}$ at in-transit phase $i$ divided by the disk-averaged limb darkening LD$_\mathrm{mean}$. This results in the absorption spectrum expressed in a wavelength-dependent planetary radius over star radius squared $(R_\mathrm{p}(\lambda,t)/R_\mathrm{*}(\lambda))^2$, with a continuum at the level of the white-light radius $(R_\mathrm{p}/R_\mathrm{*})^2$,
\begin{equation}
\frac{R_\mathrm{p}^2(\lambda)}{R_\mathrm{*}^2} = \frac{LD_\mathrm{mean}}{LD_\mathrm{i}} \left(1 - \frac{(1-\delta_\mathrm{i}) \tilde{F}_\mathrm{in,i}(\lambda)}{\tilde{F}_\mathrm{out}(\lambda)} \right).
\label{eq:1}
\end{equation}

The complete derivation of this formula is detailed in Sect. 3 of \citet{Moun22}, and we used the same procedure described in it, namely, the out-of-transit master computation, spectrum normalization, transmission spectrum computation, shift to the planetary rest frame of in-transit spectra and merging. We applied Eq. \ref{eq:1} to all HD\,189733 in-transit spectra, dividing them by their respective transit night master-out, resulting in individual transmission spectra, with the parameters given in Table \ref{table:params}. We used the Python package \textsc{batman} \citep{Kreid2015} to compute the transit light curve. Note that we used Eq. \ref{eq:1} for spectra between transit contact points T$_2$ and T$_3$, meaning that unless explicitly mentioned otherwise, ingress and egress spectra were not used when merging in-transit spectra, due to the lower stellar flux at the limb and lower amount of planetary sodium in front of the star, dampening a potential signature. For each night, this distinction cuts the number of in-transit spectra from 18 to 10, cutting the 4 spectra in ingress and 4 in egress per transit (see Fig. \ref{fig:LC_RV}).

\begin{figure*}
\begin{minipage}[tbh!]{\textwidth}
\includegraphics[trim=0cm 0cm 0cm 0cm,clip=true,width=\textwidth]{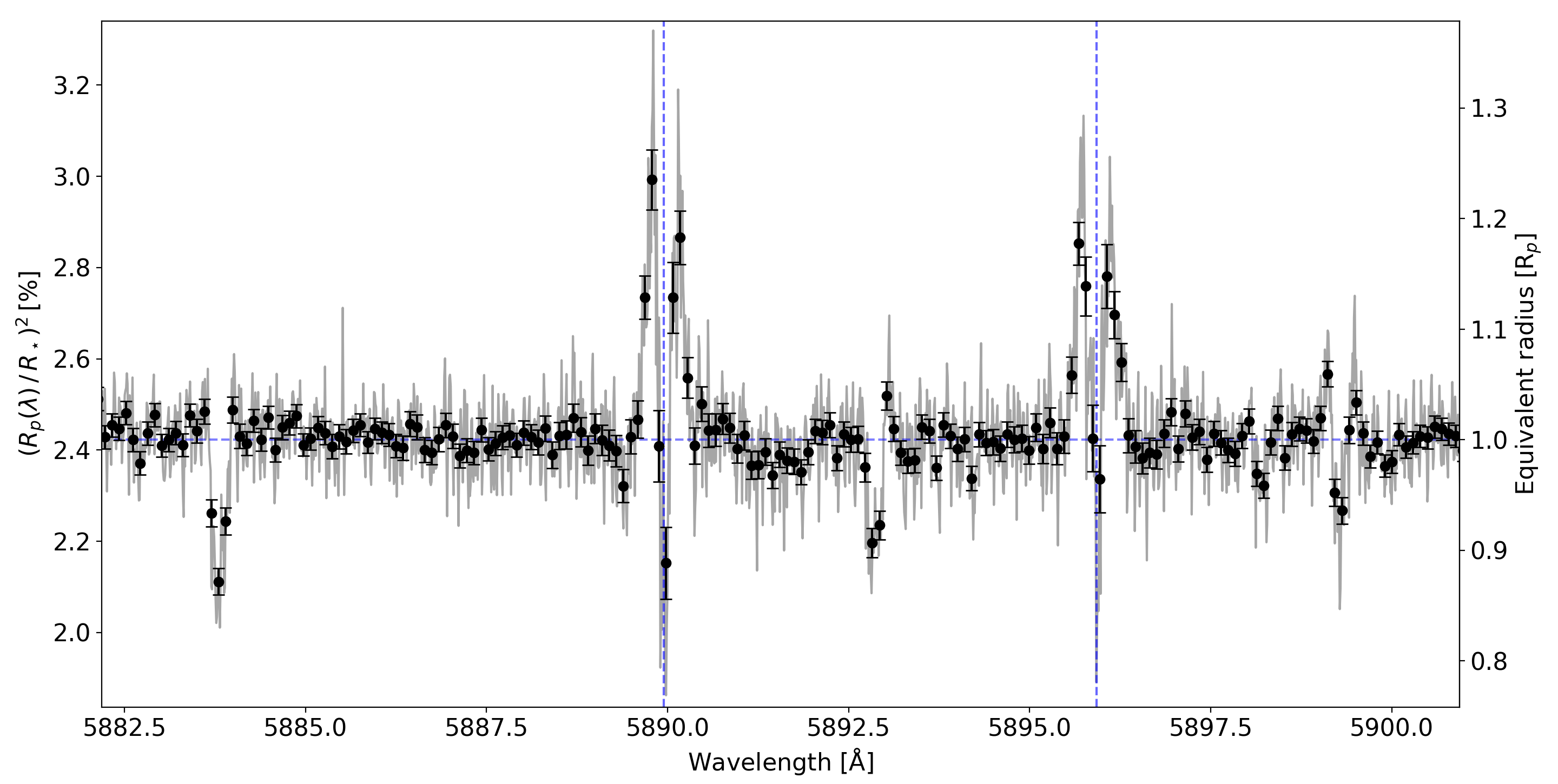}
\centering
\end{minipage}
\caption[]{Binned transmission spectrum of HD\,189733\,b that combines the two transits around the sodium doublet (vertical dashed blue lines). The unbinned transmission spectrum is shown in gray, and the black dots represent the binned spectrum with a step of 0.1 \AA . The horizontal blue line is the white-light radius $(R_\mathrm{p}/R_\mathrm{*})^2$. 
}
\label{fig:TS Na no-corr}
\end{figure*}

\subsection{Sodium signature}
The resulting transmission spectrum combining the two transits around the sodium doublet is shown in Fig. \ref{fig:TS Na no-corr}. The distortion caused by the RM effect is clearly visible on the sodium lines and the atomic lines (Fe, Ni) around it. The depth and shape of the sodium lines' POLDs are in broad agreement with the ones on HD\,209458\,b (\citealt{CB2021}, Fig. 5), as expected because the two host stars have a similar projected rotational velocity and the two systems are aligned. The slight differences in the other atomic lines can be explained by the stars' different spectral type and metallicity. The main distinction between the two comes in the wings of the feature, with a much larger amplitude in absorption for HD\,189733\,b, which can arise, for example, from the different temperatures of host stars, which changes the shape of the Na I lines, or from an added planetary sodium signature. The asymmetry with more absorption in the blue wing corroborates the blueshift detected in all previous HD\,189733\,b sodium detections. We propose and test solutions to disentangle the POLDs and planetary absorption in the next section.

In order to enhance the S/N and better visualize the sodium signature, we can co-add the two sodium lines from the doublet, like a CCF of only two absorption lines, by transforming the wavelength scale to a velocity scale using both lines of the doublet at rest frame as zero velocity, supposing that their gap is always constant. This is done in the same way as \citet{Seidel19}, \citet{Allart20}, and \citet{Moun22}. Co-adding the two lines allows for better S/N, but loses the information on the D2/D1 line ratio, so we only used this as a first-order approach to analyze the shape of the POLD.

First, we interpolated the two sodium lines on a common grid in radial velocity space, ranging from -100 to 100 km s$^{-1}$ (an interval large enough to trace the whole stellar Doppler track during the observations) with a step of 0.5 km s$^{-1}$, corresponding to the ESPRESSO detector's average pixel size\footnote{Around the sodium doublet's rest frame wavelength, this amounts to a $\sim$ 0.01\,\AA\,step.}. We then averaged the excess absorption at those common-velocity pixels.

\subsection{POLDs mitigation} \label{RMcorr}
The star-planet alignment and the relatively fast rotation of the host star induce a high POLD amplitude that partially overlaps with a potential planetary absorption. HD\,189733\,b is more difficult than other cases reported in the literature, where POLDs do not overlap with the atmospheric signal \citep{Prin24}, have negligible amplitude compared to the atmospheric signal \citep{Moun22}, and/or are diluted at lower spectral resolution \citep{Cart23}. In the following subsections, we thus propose and tested several ways of mitigating the POLDs, which range from pre- or post-treatment, using the data alone or numerical models.

\subsubsection{Masking the stellar track}\label{StellTrack}
A solution commonly used to deal with POLDs is to mask the atomic lines' core, following the stellar Doppler track. This technique hides the POLDs originating from the star and counteracts effects due to stellar activity at the core of the line (e.g., \citealt{Keles24}), at the cost of S/N, as masking the track reduces the spectral information. Its efficiency also depends on the overlap between the Doppler and orbital tracks, which depend on the orbital configuration, the rotational velocity and FWHM of the stellar lines, the planetary semi-amplitude, and the velocity of the atmospheric signature. On a sufficiently misaligned system and/or fast-rotating star (e.g., WASP-189b; \citealt{Prin24}) the atmospheric signal is easier to disentangle from the POLDs, and masking the stellar track would hide little or no planetary information. If this optimal configuration is not met, at some point during the transit, the stellar and planetary/atmospheric tracks overlap with each other, and thus, masking the star reduces the S/N of the planetary atmospheric signature. Another issue stems from the sodium doublet lines, which are wider for K-type stars like HD\,189733 and thus generate broader POLDs, which requires masking a larger wavelength range around the lines' core.

In our analysis, we masked 0.4 \AA \,around the sodium doublet cores in the stellar rest frame to completely hide the width of POLDs. This resulted in fully hiding the signal at the core of the lines during the transit, and thus in a noisier transmission spectrum around the lines' center but uncontaminated by the star. The result is shown in the bottom left panel of Fig. \ref{fig:RMCorrs}. On both lines of the sodium doublet, this correction shows a blueshifted absorption signal that should be planetary, as all stellar contributions to the transmission spectrum have been masked. The redshifted absorption component of the noncorrected spectrum has disappeared, suggesting it was mainly induced by POLDs and/or stellar activity.

\subsubsection{Approximation using master-out and local RVs} \label{LocalRV}

\begin{figure*}
\begin{minipage}[tbh!]{\textwidth}
\includegraphics[trim=0cm 0cm 0cm 0cm,clip=true,width=\textwidth]{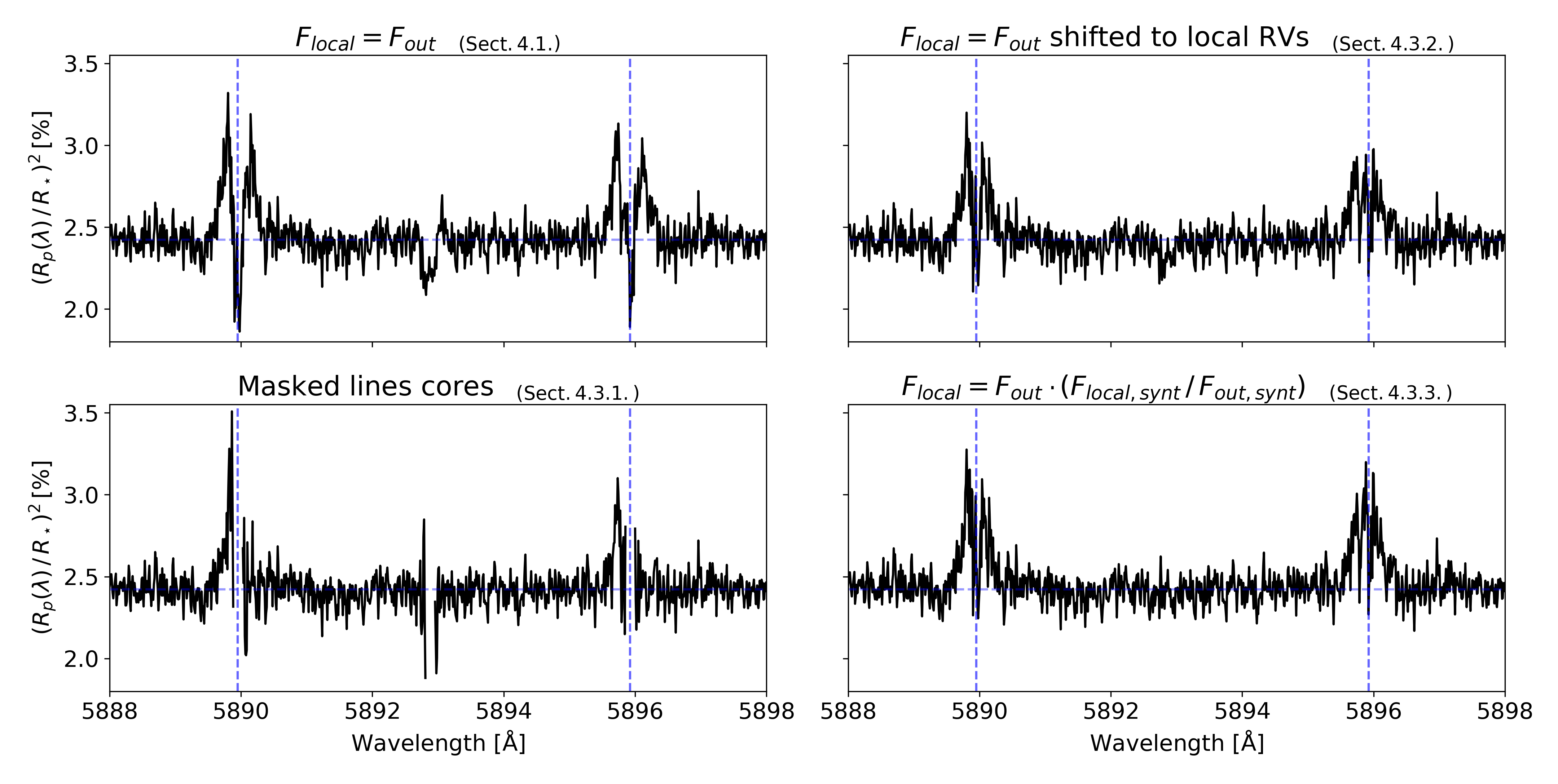}
\centering
\end{minipage}
\caption[]{Comparison of POLDs corrections on the absorption spectrum of HD\,189733\,b. \textit{Top left}: No correction applied, as in Fig. \ref{fig:TS Na no-corr}. \textit{Top right}: Correction using the master-out as a proxy for local spectra, shifted to the surface RVs. \textit{Bottom left}: Masking the stellar Doppler track, with a width of 0.4 \AA. \textit{Bottom right}: Correction using synthetic models.}
\label{fig:RMCorrs}
\end{figure*}

Equation \ref{eq:1} was derived using the assumption that the local planetary-occulted stellar spectra $\tilde{F}_\mathrm{local,i}(\lambda)$ can be approximated by the disk-integrated master-out $\tilde{F}_\mathrm{out}(\lambda)$, which completely neglects the RM and CLV effects in the line profiles. The nonsimplified equation (from \citealt{Moun22}) is
\begin{equation}
\frac{R_\mathrm{p}^2(\lambda)}{R_\mathrm{*}^2} = \frac{LD_\mathrm{mean}}{LD_\mathrm{i}} \frac{\tilde{F}_\mathrm{out}(\lambda) - (1-\delta_\mathrm{i}) \tilde{F}_\mathrm{in,i}(\lambda)}{\tilde{F}_\mathrm{local,i}(\lambda)}.
\label{eq:3}
\end{equation}
A slightly better assumption we can make is to keep the master-out as a good enough approximation of the local line profiles, but Doppler-shift it to the local occulted velocity of the stellar surface. This local velocity was derived from the RM analysis in Sect. \ref{section:RM} (see Fig. \ref{fig:Intrinsic_properties}, middle panel). We computed $\tilde{F}_\mathrm{local,i}(\lambda)$ associated with each $\tilde{F}_\mathrm{in,i}(\lambda)$ and replaced Eq. \ref{eq:1} by Eq. \ref{eq:3} in our pipeline. This resulted in the combined transmission spectrum shown in the top right panel of Fig. \ref{fig:RMCorrs}, which should correct for POLDs if they were the only components that contaminate the signal. This was reported by \citet{CB17, Seidel20}, and the former showed that this correction still leaves a residual signal (their Fig. 6).

This purely data-driven correction seems to effectively dampen the POLDs, as its main shape has mostly disappeared. This is especially the case in the atomic nickel line between the sodium lines, which should only be affected by POLDs, although a small residual remains. For sodium, the correction appears to impact each line in the same way, although a larger dip appears to remain at the core of the D2 line. Overall, Fig. \ref{fig:RMCorrs} (top right panel) still shows significant and broad absorption in both sodium lines, even after the POLDs correction presented here, even if some POLD-induced biases still persist, as shown below.

\subsubsection{Approximation using synthetic stellar spectra}\label{sec:turbospectrum}

\begin{figure}[tbh!]
    \centering
    \includegraphics[trim={3.25cm 0 4.5cm 3cm},clip, width = \linewidth]{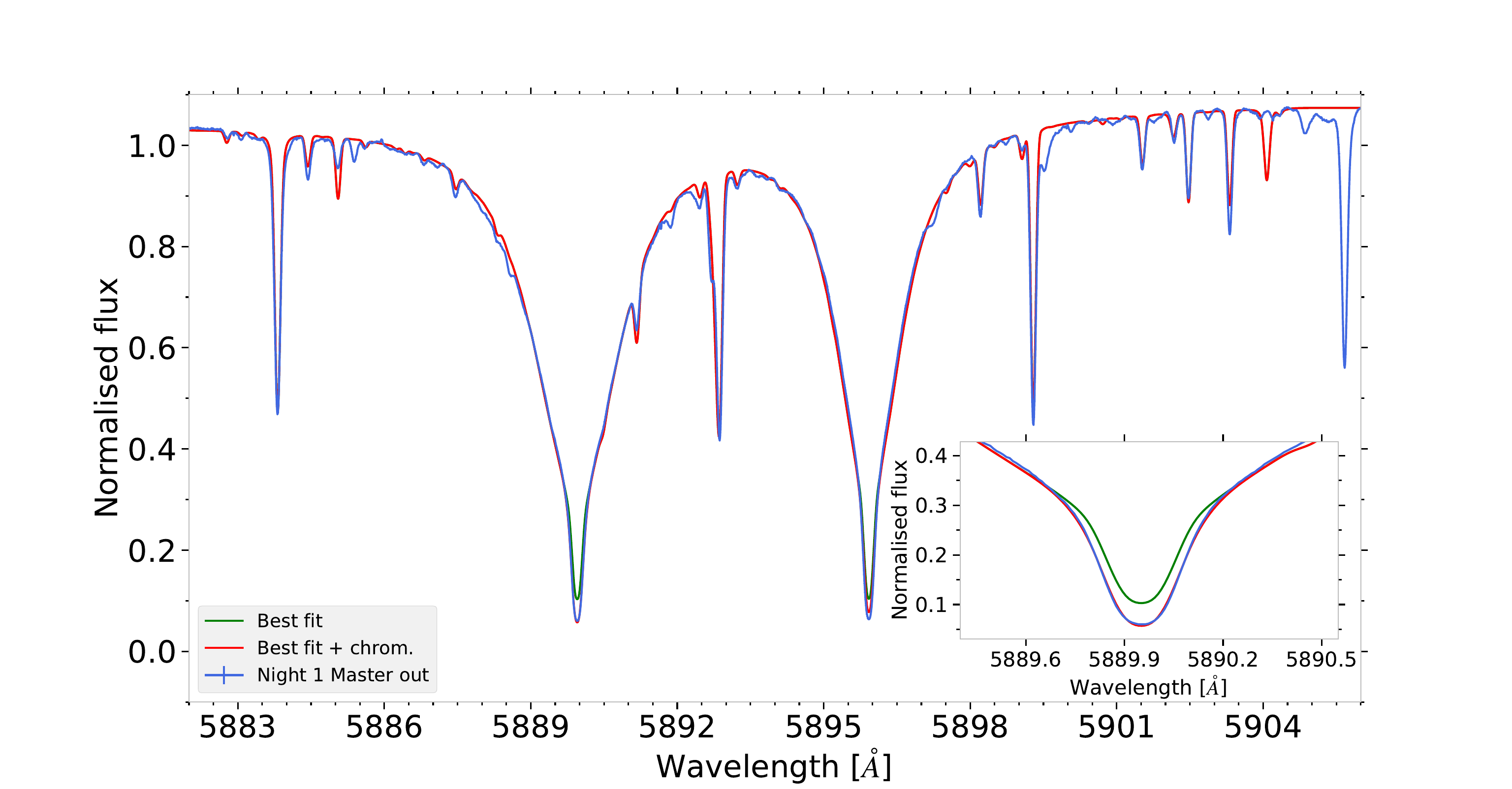}
    \caption{Disk-integrated spectra of HD\,189733 in the sodium region. The blue curve shows the observed master out-of-transit spectrum for the first night, the green curve shows the best fit synthetic \textit{Turbospectrum} model, and the red curve shows the best fit with the additional absorption of sodium. The subplot on the bottom left shows a zoom on the D2 line and its labels are the same as those of the main plot.}
    \label{fig:fout_obs_synt}
\end{figure}

A possible better estimate for the local stellar spectrum comes from synthetic spectra generated from stellar atmosphere models (e.g., \citealt{Yan18, CB2020, CB2021, Jiang23, Keles24}). We computed local synthetic stellar spectra for HD\,189733 using the nonlocal thermodynamic equilibrium (NLTE) version of the \textsc{Turbospectrum} code for spectral synthesis\footnote{\url{https://github.com/bertrandplez/Turbospectrum_NLTE}} \citep{plez1998,plez2012,heiter2021, larsen2022, magg2022} along with MARCS photospheric models \citep{gustafsson2008}\footnote{\url{https://marcs.astro.uu.se}} and spectral line lists from the VALD3 database\footnote{\url{http://vald.astro.uu.se}} \citep{Ryabchikova2015}, following the same method as in \citet{Deth23}. We compared the disc-integrated spectrum built from the grid of local synthetic spectra with the observed master-out spectrum to determine the adequate sodium abundance and photospheric temperature to use in the stellar model (see Fig. \ref{fig:fout_obs_synt}). As HD\,189733 is known to have differential rotation and convective blueshift \citep{Cegla16}, we also included these effects in our synthetic stellar spectrum, using the models and values derived in our RM analysis (Sect.~\ref{section:RM}) and reported in Table \ref{table:params}. Although limb darkening is naturally accounted for in \textit{Turbospectrum}, we scaled the local spectra to the limb darkening values determined from the light curve (see Table \ref{table:params}).\\

We first performed a fit varying the stellar abundance of sodium. We were unable to fully reproduce the depth of the disk-integrated sodium doublet cores (see the close-up view in Fig. \ref{fig:fout_obs_synt}). The reason for this is probably that the MARCS models only simulate the photosphere and that the sodium doublet partly forms in the chromosphere \citep{bruls1992}.

As a first step, we decided to fit the stellar sodium abundance and photosphere temperature only using the wings of the Na\,I D2 and D1 lines, as the core of these lines do not vary significantly with these parameters. The range of wavelength that was used is [5885.000; 5889.737], [5890.215; 5895.700] and [5896.147; 5901.000] \AA. The best-fit values for the stellar sodium abundance and photosphere temperature are A$_*$(Na\,I) $ = 12 + \log_{10} \left( \frac{\text{n(Na\,I)}}{\text{n(H)}}\right)$ = 6.1484 $\pm$ 0.0001, and 5039 $\pm$ 10 K, respectively. 
 
As a second step, in an attempt to better reproduce the whole disc-integrated line profile of sodium, we added to the Turbospectrum profile an analytical absorption signature mimicking the presence of sodium from the stellar chromosphere. The local synthetic spectra from the best fit of the first step were multiplied by an absorption line profile built with the properties of the Na I D2 and D1 transitions. We then fit the signal only in the region of the core of the Na I D2 and D1 lines to determine the best-fit values of the temperature and column density of the ``chromospheric'' sodium. The range of wavelengths used is [5889.737; 5890.215] and [5895.700; 5896.147] \AA. \\
Figure \ref{fig:fout_obs_synt} shows our best-matching synthetic spectrum, including artificial chromospheric absorption. 

\begin{figure*}[tbh!]
\includegraphics[width=\textwidth]{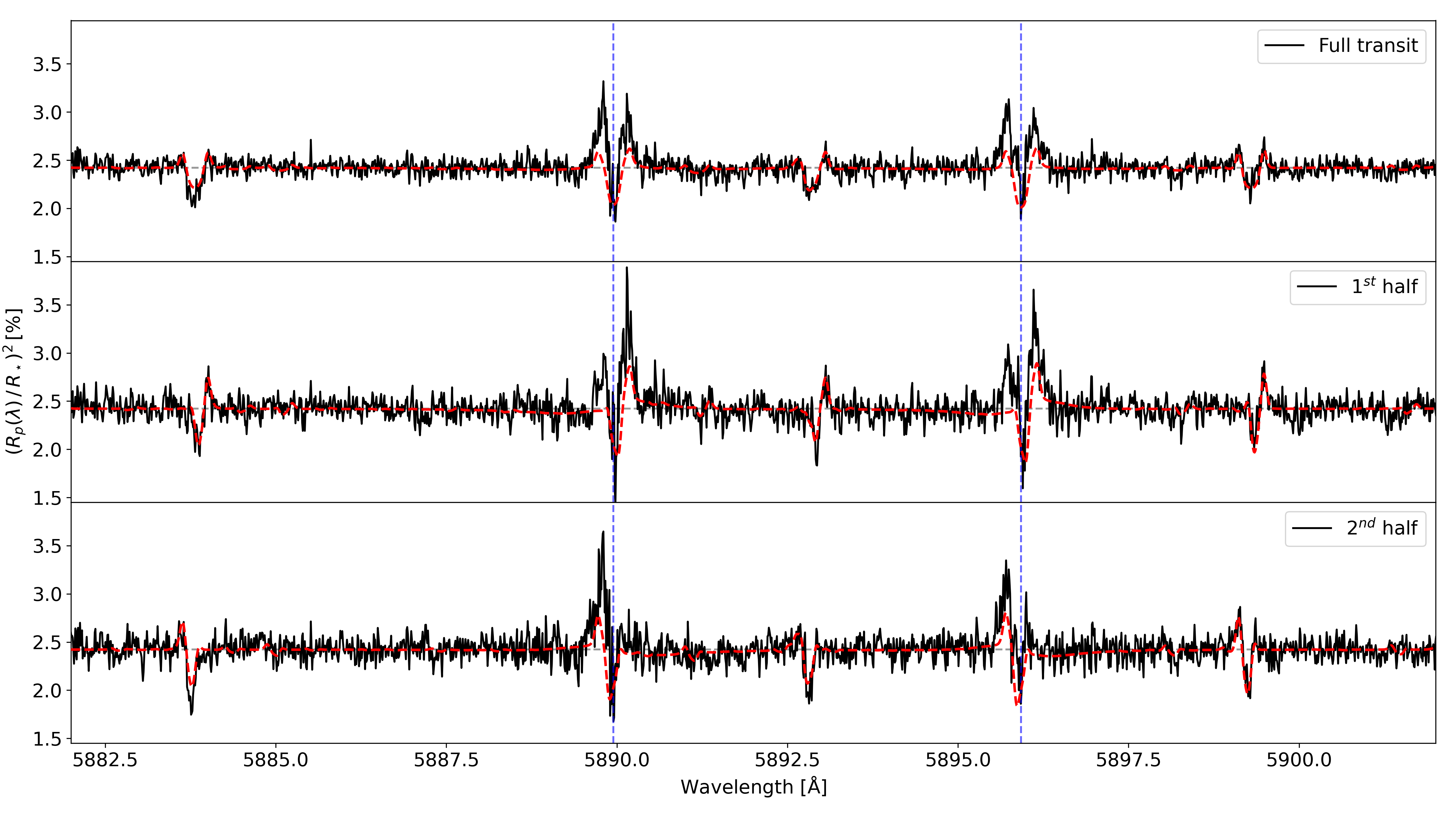}
\centering
\caption[]{Absorption spectrum of HD\,189733\,b (in black) without POLDs correction around the sodium doublet (dashed blue lines) and best-fit synthetic model of the POLDs (dashed red line) for different transit-phase intervals. \textit{Top panel}: Full transit between T$_2$ and T$_3$. \textit{Middle panel}: First half of the transit (from T$_2$ to T$_c$). \textit{Bottom panel}: Second half of the transit (from T$_c$ to T$_3$).}
\label{fig:TS+RM}
\end{figure*}
 
Assuming our synthetic local spectra and master-out are representative of reality, we computed each local spectrum in Eq. \ref{eq:3} to correct for the POLDs using

\begin{equation}
\tilde{F}_\mathrm{local,i}(\lambda) = \frac{\tilde{F}_\mathrm{local,i, synt}(\lambda)}{\tilde{F}_\mathrm{out, synt}(\lambda)} \tilde{F}_\mathrm{out}(\lambda). 
\end{equation}

The bottom right panel in Fig. \ref{fig:RMCorrs} shows the transmission spectrum corrected using those synthetic absorption spectra. We see that the corrected transmission spectrum resembles the one corrected using the local RVs (Sect. \ref{LocalRV}), even though the former used synthetic models and the latter the observed disk-integrated master-out spectrum. The main difference is that the leftover contamination in the nickel line was seemingly better-corrected with the synthetic method and the general amplitude of the absorption in the sodium is slightly higher. In general, most spectral lines seem to be better corrected with the synthetic models, which thus seems to be the most robust technique.

\subsection{Evolution of sodium through the transit} 
In Fig. \ref{fig:TS+RM}, we split the absorption spectra between T$_2$ and T$_3$ in two halves to understand the evolution of the absorption signature through the transit (exposures 17-21/22-26 and 11-15/16-20 for Transit 1 and 2 respectively). Both epochs were combined. The POLDs on the atomic lines around the sodium were well-modeled by the synthetic spectra. For the sodium lines, we see that the temporal behavior of the signal generally matches the predictions by the synthetic model. We note a systematic underestimation of the amplitude by the models, however, which may be due to limitations in the synthetic spectra. Moreover, we can see an excess absorption in the blue wing of both lines that was not reproduced by the models. This excess absorption is present in all parts of the transit, albeit hard to quantify at the end of the transit, as it is merged with the POLDs.

\subsection{Impact of stellar activity}
\begin{figure}
\includegraphics[width=\columnwidth]{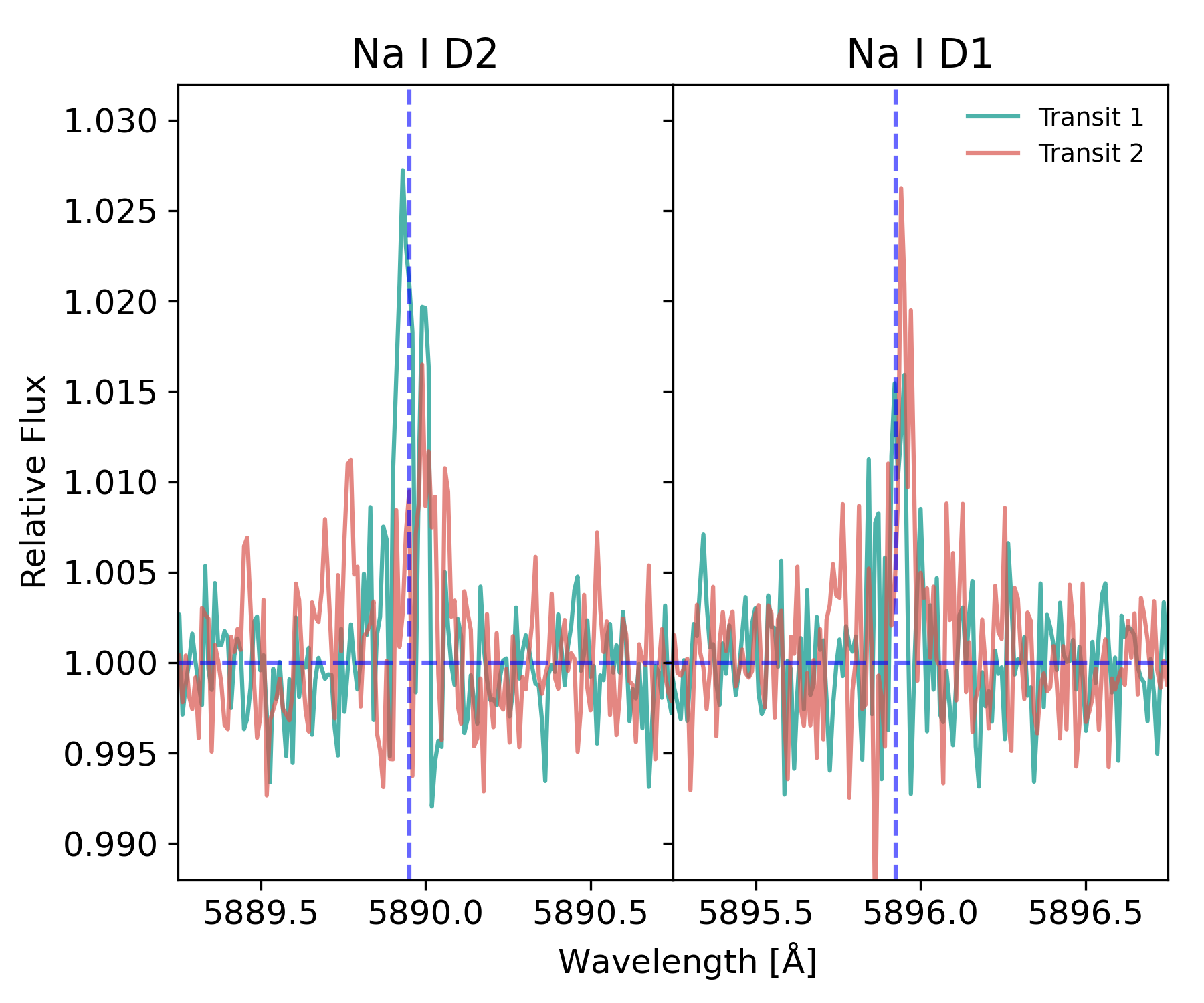}
\includegraphics[width=\columnwidth]{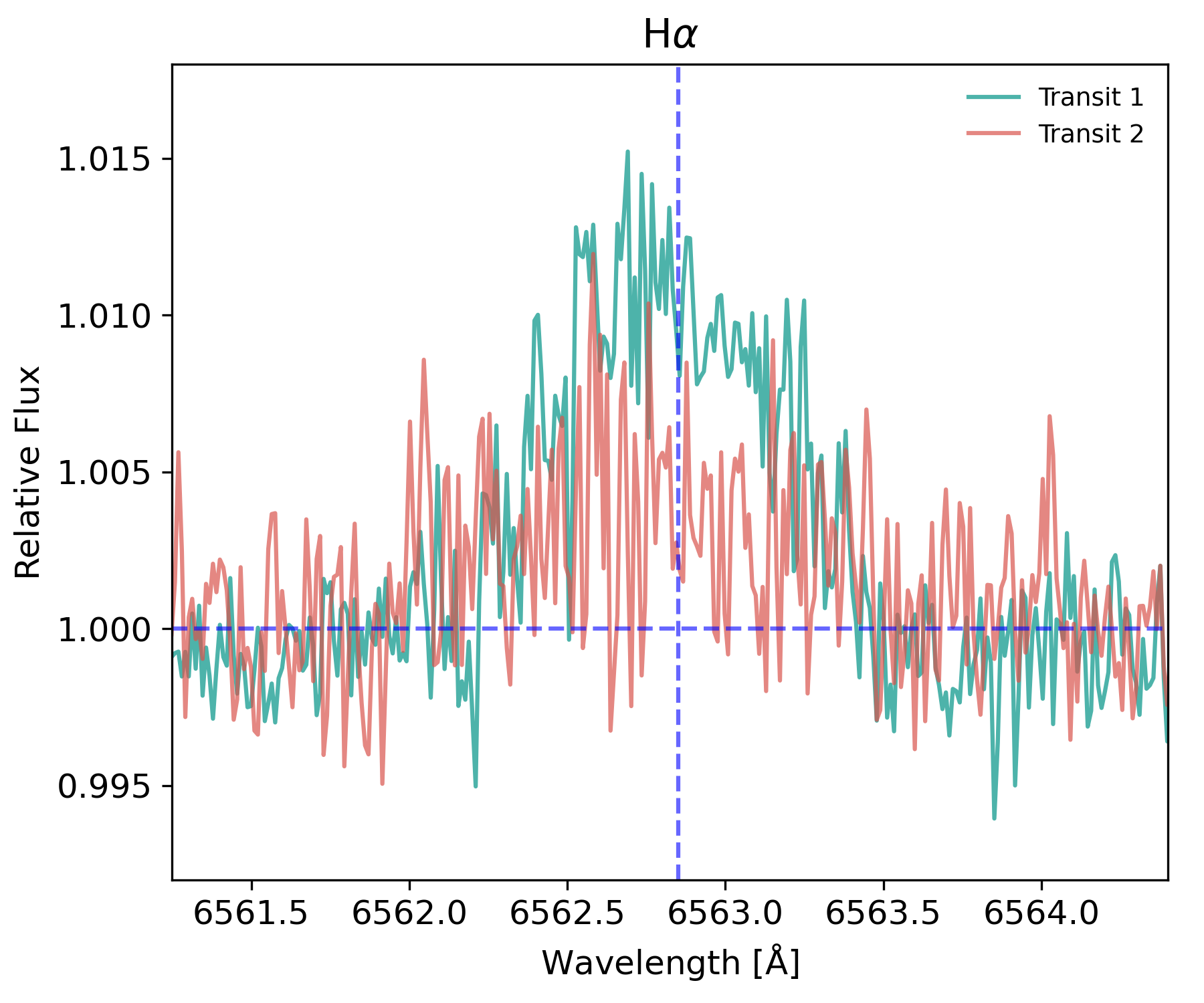}
\centering
\caption[]{Comparison of masters-out before and after transit by dividing the master pre-transit by the master post-transit around the sodium doublet (\textit{top panel}) and H$\alpha$ (\textit{bottom panel}) for each transit.}
\label{fig:Activity}
\end{figure}

HD\,189733 is a known active star \citep{Bouchy2005, Mout07, Mout20} where activity has already proven to influence transmission spectra \citep{Oshagh14, Caul17, Caul18, Guill20}. To investigate a potential impact of stellar activity on our data, we compared the master spectra pre- and post-transit in both nights, focusing on the sodium and H$\alpha$ lines (Fig. \ref{fig:Activity}). We clearly see a temporal variability of the stellar spectrum over the duration of the two nights. The effect is most pronounced in Transit 1, in which the H$\alpha$ and sodium line depths vary by $\sim$ 1 \% and $\sim$ 2 \%, respectively, over about four hours. This is significant compared to the amplitudes of POLDs and potential planetary features. Thus, we have to consider that short-term variability induced by stellar activity will complicate the use of the stellar line cores.

\subsection{Characterization of the planetary sodium signature}
\label{sec:characterization_pl_signature}

The previous sections showed a complex picture regarding the interpretation of the sodium signatures in HD\,189733 data. Our models indicated three components of similar amplitude in the transmission spectrum. First, the POLDs, caused by the RM and CLV effects, are obvious in the data and must be masked or modeled to reveal a potential planetary absorption. Second, stellar activity is an important contributor to line-shape variability across the transits. Third, we also detected a clear blueshifted signal that is a strong candidate for a planetary atmosphere signature. 

\begin{figure*}[tbh!]
\includegraphics[width=\textwidth]{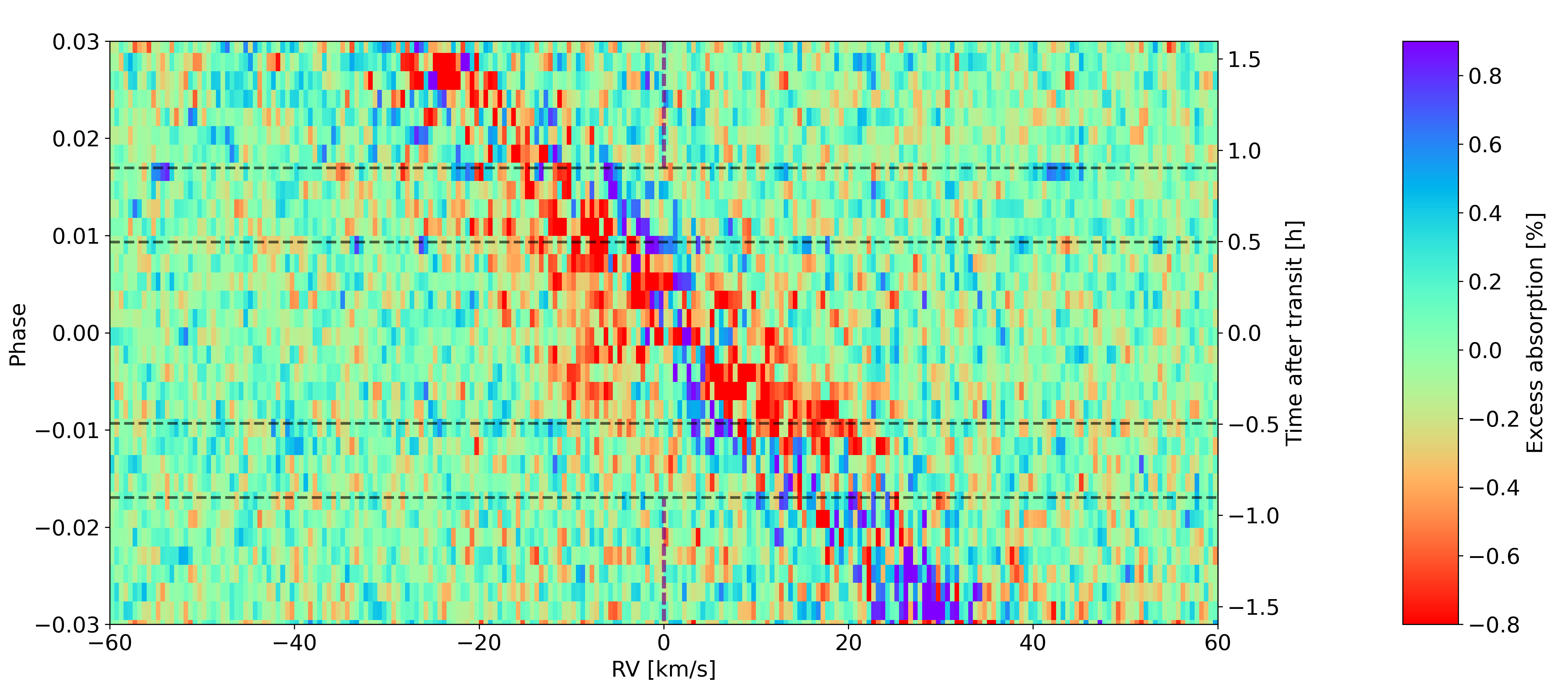}
\caption[]{Transmission spectra map of HD\,189733\,b around the sodium doublet (co-added and both transits combined) after POLDs correction using synthetic models on individual spectra in the planetary rest frame (see Fig. \ref{fig:TS2D} for the uncorrected version). The dashed dark purple line indicates the sodium planetary rest velocity, and the horizontal dashed lines show the transit contacts T$_1$, T$_2$, T$_3$, and T$_4$ from bottom to top.}
\label{fig:TS2DRM}
\end{figure*}

Figure \ref{fig:TS2DRM} synthesizes our findings as a 2D map of excess absorption as a function of time and Doppler shift. Before correction of the POLDs (see Fig. \ref{fig:TS2D}), the raw transmission spectra are dominated by the characteristic patterns of POLDs showing alternating absorption and emission as a function of time. After correction, the transmission spectrum after correction of the POLDs with synthetic spectra. Here, we can clearly see the stellar line core trace evolve from residual emission to residual absorption over time, most likely due to stellar activity. In addition, excess absorption blueward of the stellar trace can be seen in red between T$_2$ and T$_3$ at roughly constant velocity in the planet rest frame. We interpret this as absorption from sodium in the planet's atmosphere since neither the POLDs nor stellar activity in the line cores can explain such a blueshifted signature which seems to be at rest with respect to the planet. 

\begin{figure*}[tbh!]
\includegraphics[width=\textwidth]{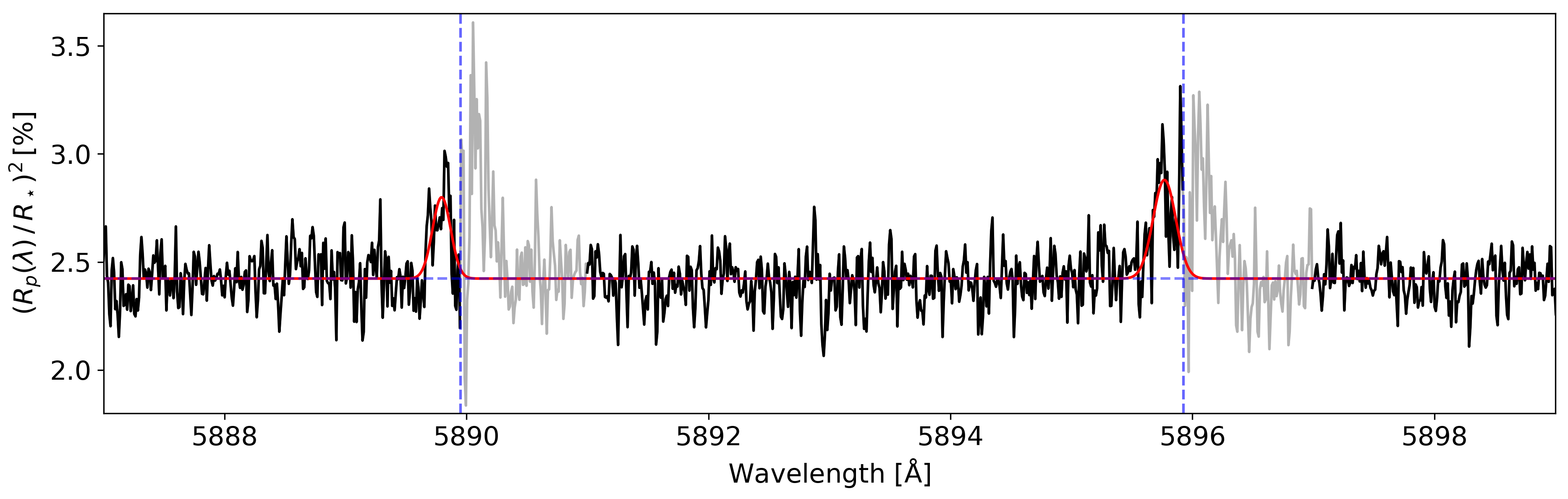}
\caption[]{Absorption spectrum (black line) of HD\,189733\,b around the sodium doublet, combining the first half of the two transits in the planetary rest frame, corrected for the POLDs using synthetic stellar spectra. The gray line shows the masked part of the absorption spectrum, which we excluded from the Gaussian fit (in red). The horizontal dashed blue line indicates the white-light radius ($R_\mathrm{p}$ / $R_\mathrm{\star}$)$^2$, and the vertical lines show the position of the sodium lines in the planet rest frame.}
\label{fig:blue fit}
\end{figure*}

We proceeded to fit this plausible planetary signal by averaging in phase the transmission spectra in the planetary rest frame. We limited ourselves to the first half of the transits to avoid the overlap between planetary and stellar tracks at the end of the transits. We further masked the stellar track on the red side, which was affected by POLD residuals and stellar activity, and fit two Gaussians, separated by the difference in rest wavelengths between the Na\,I D1 and D2 lines, to the putative planetary signal. We obtained a blueshifted radial velocity of -7.92 $\pm$ 0.57 km s$^{-1}$, a Gaussian FWHM of 3.84 $\pm$ 0.77 km s$^{-1}$ and 4.77 $\pm$ 0.72 km s$^{-1}$, and an amplitude of 0.377 $\pm$ 0.058 \% and 0.457 $\pm$ 0.050 \% in excess absorption for the D2 and D1 lines respectively (see Fig. \ref{fig:blue fit}). 

\begin{figure}[tbh!]
\includegraphics[width=0.95\columnwidth]{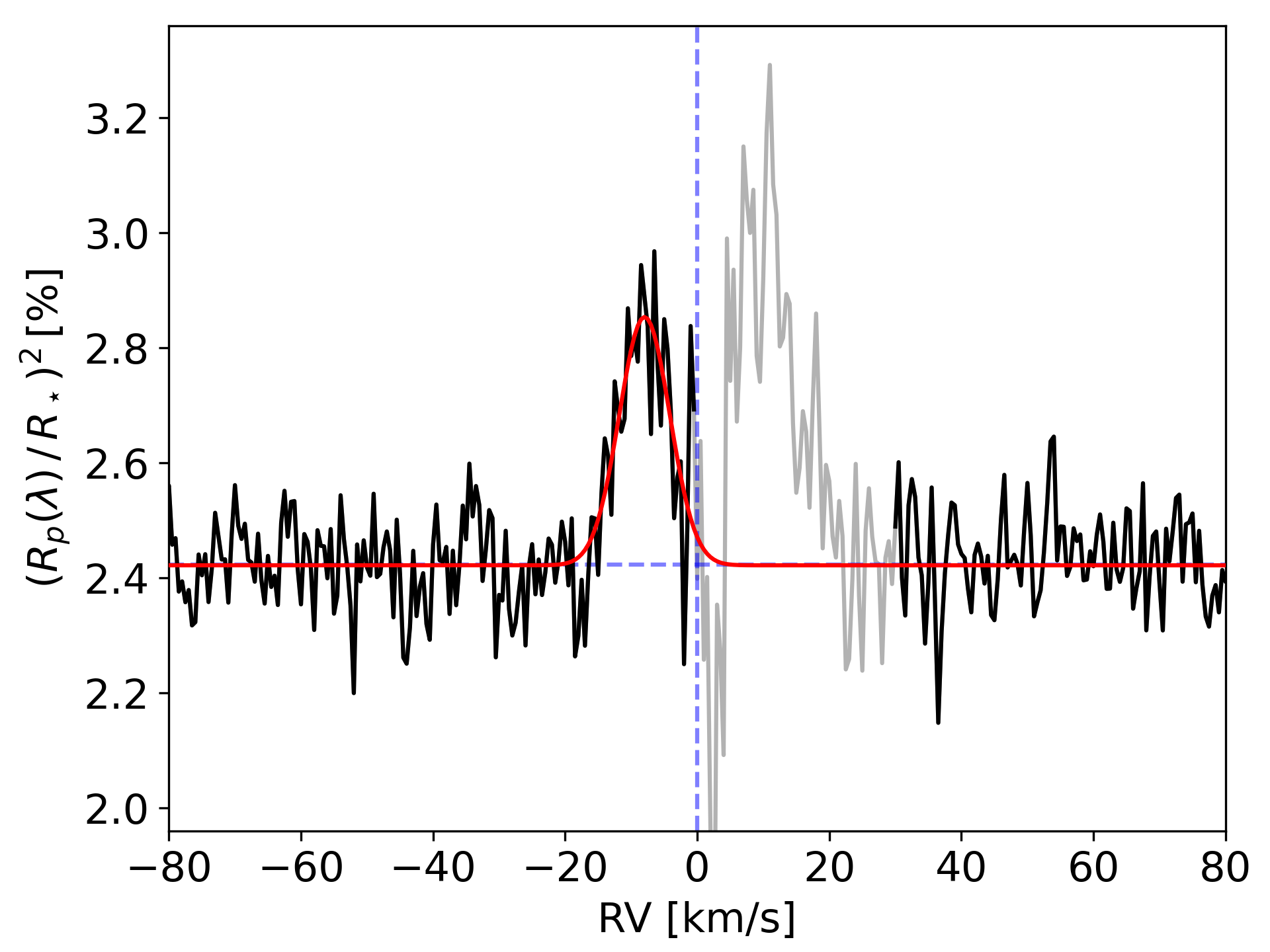}
\centering
\caption[]{Same as Fig. \ref{fig:blue fit} with the two sodium lines co-added.}
\label{fig:blue fit coadded}
\end{figure}

We also performed the same analysis after co-adding the two sodium lines.  We obtained a radial velocity of -7.97 $\pm$ 0.28 km s$^{-1}$, a Gaussian FWHM of 3.82 $\pm$ 0.29 km s$^{-1}$ and an amplitude of 0.432 $\pm$ 0.027 \% in excess absorption (see Fig. \ref{fig:blue fit coadded}). Both fit results are summarized in Table \ref{table:gauss}.

\begin{table}[t]
\centering
\caption{Results of the Gaussian fits for the combined sodium transmission spectrum.}
\label{table:gauss}
\begin{tabular}{l l}
\toprule
Parameter & Fit value\\
\midrule
\multicolumn{2}{c}{-- \textit{Independent sodium lines (Fig. \ref{fig:blue fit})} --} \\
D2 amplitude & 0.377 $\pm$ 0.058 \% \\
D1 amplitude & 0.457 $\pm$ 0.050 \% \\
D2 FWHM & 0.075 $\pm$ 0.015 \AA\ (3.84 $\pm$ 0.77 km s$^{-1}$) \\
D1 FWHM & 0.094 $\pm$ 0.014 \AA\ (4.77 $\pm$ 0.72 km s$^{-1}$) \\
Line center shift & -0.156 $\pm$ 0.011 \AA\ (-7.92 $\pm$ 0.57 km s$^{-1}$) \\
Continuum offset & 0.0003 $\pm$ 0.0021 \%\\
\midrule
\multicolumn{2}{c}{-- \textit{Co-added sodium lines (Fig. \ref{fig:blue fit coadded})} --} \\
Amplitude & 0.432 $\pm$ 0.027 \% \\
FWHM & 3.82 $\pm$ 0.29 km s$^{-1}$ \\
Line center shift & -7.97 $\pm$ 0.28 km s$^{-1}$ \\
Continuum offset & -0.0007 $\pm$ 0.004 \%\\
\bottomrule
\end{tabular}
\end{table}

\section{Combined stellar and planetary modeling} \label{section:EVE}
\begin{figure}
    \centering
    \includegraphics[trim={0.8cm 0 2.35cm 3cm},clip, width = \linewidth]{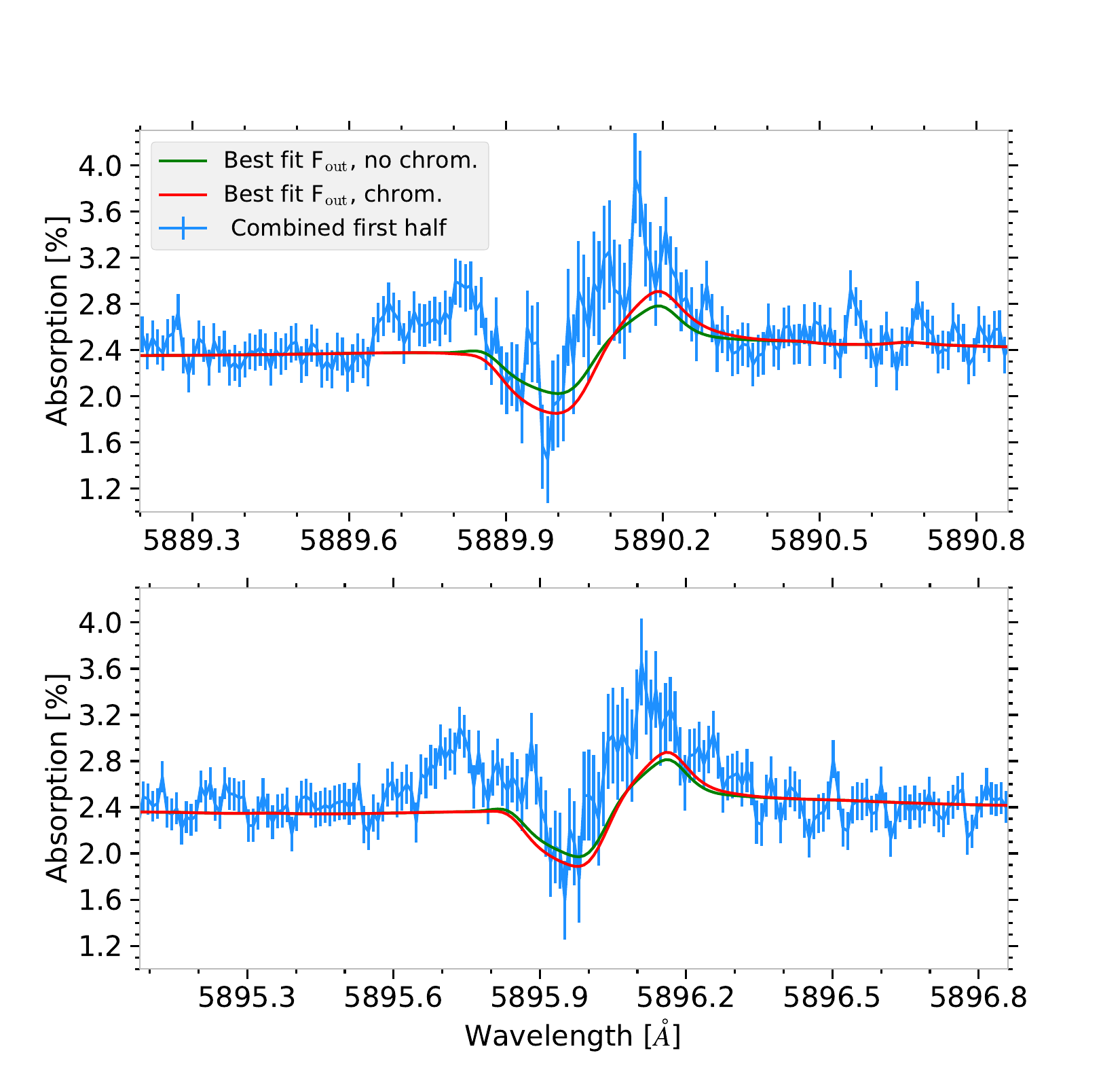}
    \caption{Mean absorption spectra between T$_2$ and T$_0$ of night 1. The green curve was computed using the stellar grid that best matched the observed F$\rm_{out}$ without the additional chromospheric absorption. The red curve was computed using the stellar grid that best matched the observed F$\rm_{out}$ with the additional chromospheric absorption. The top panel focuses on the Na\,I D2 line and the bottom one on the Na\,I D1 line.}
    \label{fig:abs_spec_obs_synt_noatm}
\end{figure}

\begin{figure}
    \centering
    \includegraphics[trim={0.8cm 0 2.35cm 3cm},clip, width = \linewidth]{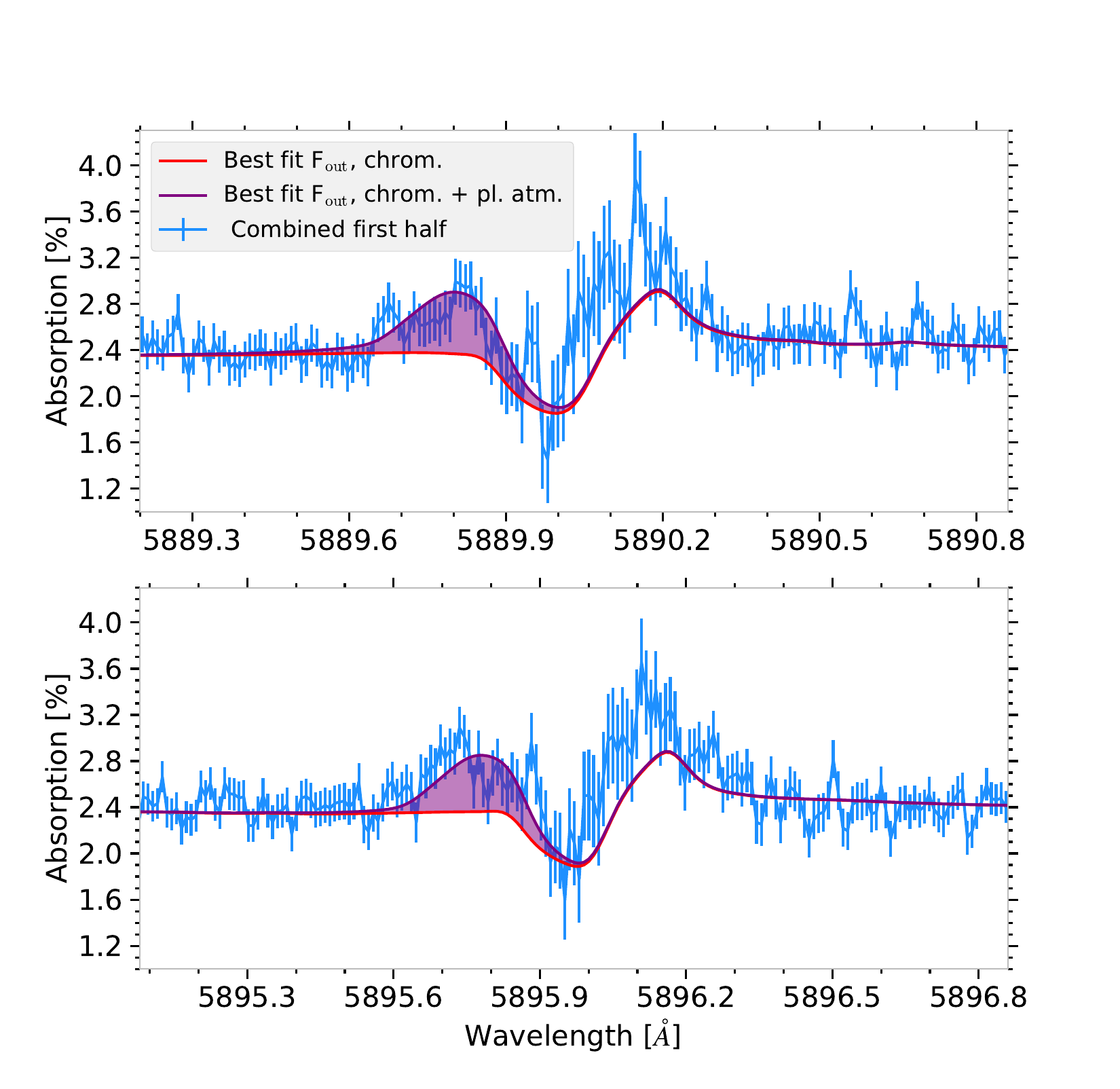}
    \caption{Mean absorption spectra between T$_2$ and T$_0$ of night 1. The red curve was computed using the stellar grid that best matched the observed F$\rm_{out}$ using the additional chromospheric absorption. The purple curve was computed using the stellar grid that best matched the observed F$\rm_{out}$ using the additional chromospheric absorption and contains the planetary atmosphere signature.}
    \label{fig:abs_spec_obs_synt_atm}
\end{figure}

To push our analysis one step further, we used the synthetic stellar grid of spectra from Sect. \ref{sec:turbospectrum} as an input to the EvE code \citep{bourrier2013,bourrier2015,bourrier2016,Deth23}. The system parameters that we used are listed in Table \ref{table:params}. The EvE code allowed us to simulate the absorption of the local stellar spectra during the transit by both the planetary opaque body and an atmosphere containing sodium atoms. The outputs of this procedure are disc-integrated stellar spectra during the transit that contain the absorption signature of the planet and its atmosphere exactly as generated in the observations. With these end products, we can compute absorption spectra that are truly equivalent to those computed from the observations, and make a direct comparison that avoids additional corrections \citep{dethier24}. We thus used the EvE code to fit the observed absorption spectra while varying the properties of the simulated atmosphere. We used a 1D hydrostatic profile that is distributed within regular cubic cells within a 3D spherical grid around the planet, assuming spherical symmetry. The free parameters were the temperature and density of the sodium atoms in the atmosphere. A day-to-nightside wind of 8 km s$^{-1}$ was also added in the planetary atmosphere to help better match the observed signature. This choice was based on the values of the blueshift derived in Sect. \ref{sec:characterization_pl_signature}. To be conservative, we oversampled the number of simulated exposures compared to the actual number of observed exposures. Then, we averaged the simulated exposures that are comprised in the time interval of each of the observed exposures. That way, we could account for the Doppler smearing in the observed spectra due to the planet's orbital motion. The best-fit planetary atmosphere parameters are a temperature of 2750 K and a density of sodium atoms of 1$\times 10^{-21}$ cm$^{-3}$ at the top of the atmosphere (1$\times 10^{-5}$ cm$^{-3}$ at the bottom), set in our simulations to the Roche lobe radius. These best-fit values were obtained by $\chi^2$ minimization (see Fig. \ref{fig:chi2}). The $\chi^2$ was computed on the wavelength range [5889.31, 5889.85] $\AA$ and [5895.45, 5895.85] $\AA$ to focus on and include the planetary signal, and exclude most of the POLDs.\\
Figure \ref{fig:abs_spec_obs_synt_noatm} shows the absorption spectra computed with the EvE code using the two stellar synthetic spectra grids that best matched the observed master out-of-transit spectrum (i.e., with and without the additional chromospheric absorption). Figure \ref{fig:abs_spec_obs_synt_atm} shows the absorption spectra computed with the EvE code using the stellar synthetic spectra grid that best matched the observed master out-of-transit spectrum with the additional chromospheric absorption and the planetary atmosphere. To avoid being too contaminated by the POLDs, we limited the fit to exposures between T$_2$ and T$_c$. 

\begin{figure}
    \centering
    \includegraphics[width = \linewidth]{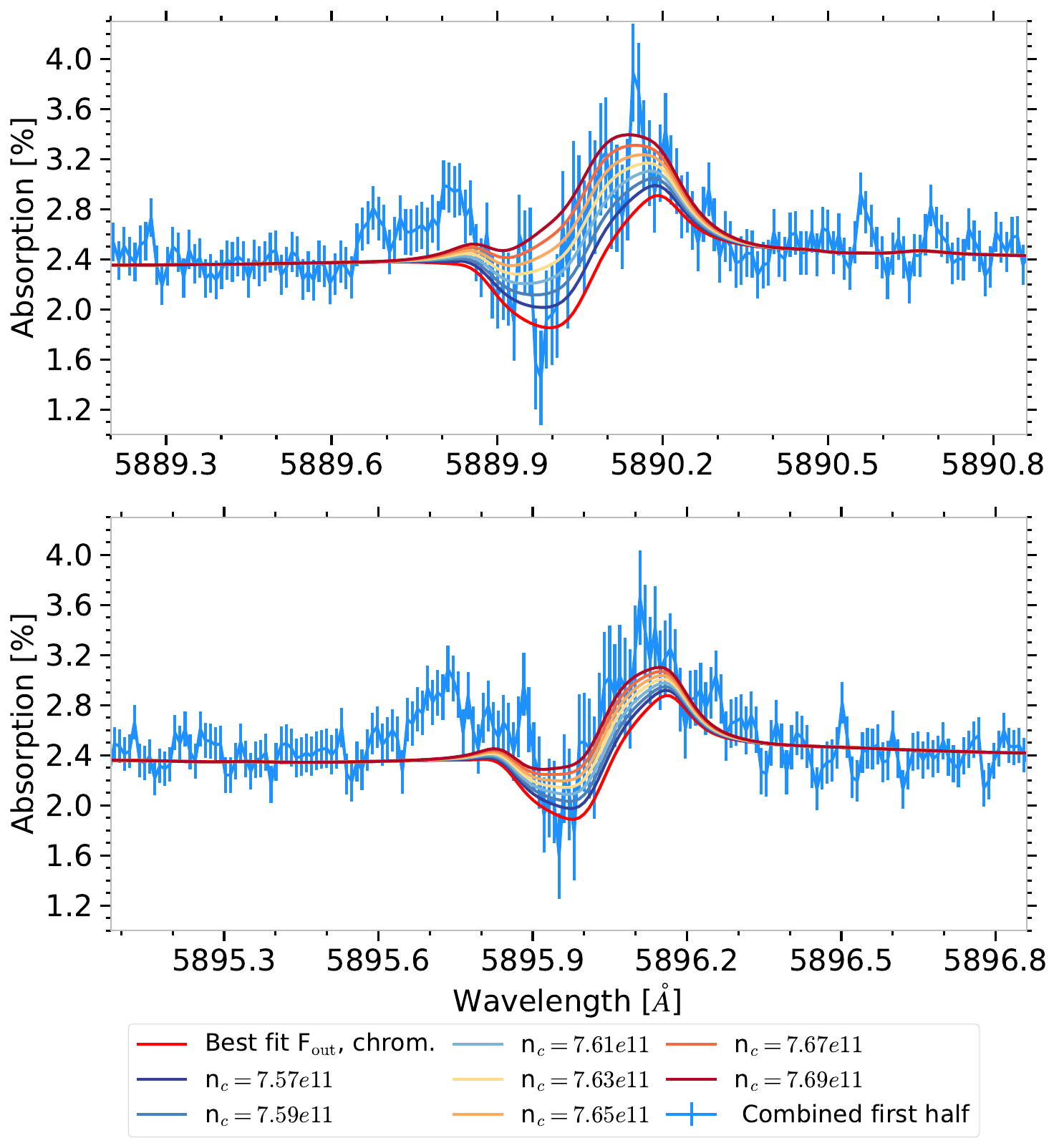} 
    \caption[ ]{Mean absorption spectra between T$_2$ and T$_0$ of night 1. The red curve was computed using the stellar grid that best matched the observed F$\rm_{out}$ using the additional chromospheric absorption. The other curves are computed using the  F$\rm_{out}$ from the best fit and F$\rm_{in}$'s from modified stellar spectral grids. As a reference for the best fit, $n_c = 7.54\times 10^{11}$ atoms cm$^{-3}$. These specific values should only be regarded to get a sense of how the sodium lines were modified, and not as exact values for this specific system.}
    \label{fig:abs_spec_obs_synt_noatm_test}
\end{figure}

From Fig. \ref{fig:abs_spec_obs_synt_noatm}, it is clear that, despite using a stellar spectral grid that matched the observed out-of-transit spectrum, the synthetic POLDs do not correctly reproduce the observed ones. This could be due to the use of 1D stellar atmosphere models which might not be sufficient for this star, and it would require 3D effects to better reproduce the local sodium spectra \citep{canocchi24}. It could also be due to the lack of chromosphere in these models. Although we added extra sodium absorption in the stellar spectrum, it is probably unrealistic and does not reproduce accurately the true local line profiles on the star. These discrepancies suggest that with spectrographs such as ESPRESSO, we have reached such precision that the stellar models start to show their limitations. \\
Thus, we tested another hypothesis to see if we could better reproduce the overall absorption signal by varying the POLDs over time. In this test, we allowed the spectral grid to vary during the planetary transit due, for example, to activity \citep{Cristo23}. To test this possibility, we used two stellar spectral grids. The first was the one that best matched the observed master out-of-transit. For the other, we varied the depth of the additional chromospheric absorption in the stellar grid by increasing the density of sodium atoms in the chromosphere ($n_c$). We used the first grid to compute the out-of-transit spectra in EvE and the second to compute the in-transit spectra in EvE. Figure \ref{fig:abs_spec_obs_synt_noatm_test} shows the absorption spectra resulting from this test.

From this figure, it is clear that a slight change in stellar lines due to inhomogeneities in the transit chord can significantly alter the POLDs in the absorption spectrum. This figure also shows that, by increasing the depth of the sodium lines, we were able to better approach the shape of the observed POLDs. The match is not perfect, however. One possibility for this mismatch could be that we used the same modified stellar grid for all the in-transit exposures, while the observed stellar lines could have changed throughout the transit. We did not investigate further for a better match, as the purpose of this test was to highlight a new degree of complexity in the fitting of absorption spectra, that is, knowing how the stellar activity affects the stellar spectrum even during a transit. \\

\noindent Finally, the simulations performed in this section confirm the presence of sodium in the atmosphere of HD\,189733\,b and hint at the possibility that the stellar line temporal variability affects the planetary sodium characterization. This last statement ought to be compared with simulations that account for 3D stellar atmospheres and more robust chromosphere modeling, however.

\section{Other atomic species}\label{atoms}
\begin{figure}
\includegraphics[width=0.95\columnwidth]{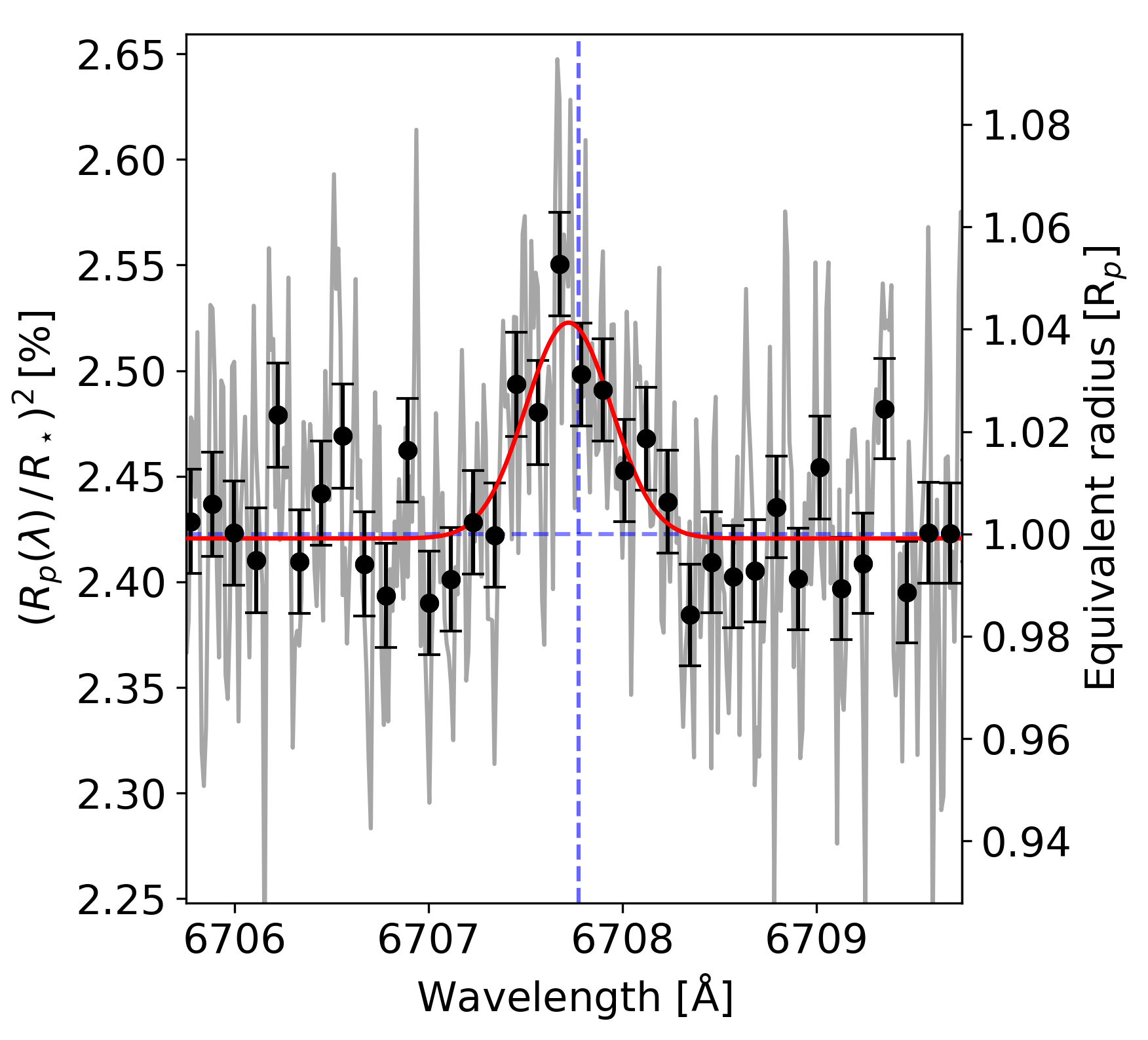}
\centering
\caption[]{Binned absorption spectrum (black dots) of HD\,189733\,b around lithium in the planetary rest frame, corrected using the local RVs (Sect. \ref{LocalRV}). The gray line shows the unbinned absorption spectrum and the red line the Gaussian fit on these data points. The blue horizontal dashed line indicates the white-light radius ($R_\mathrm{p}$ / $R_\mathrm{\star}$)$^2$ and the vertical one the position of the lithium line in the planet rest frame.}
\label{fig:Li}
\end{figure}

Lithium has been found in exoplanet atmospheres in recent years through transmission spectroscopy in hot Jupiters at a low resolution (\citealt{Chen18} for WASP-127b; this was not confirmed with ESPRESSO \citet{Allart20}) and high resolution (\citealt{Jiang23}, WASP-85Ab), and in ultra-hot Jupiters with ESPRESSO (\citealt{Borsa21} on WASP-121b, \citealt{Tab21} on WASP-76b). 
As stars burn their Li content, they show no absorption lines in their spectra, providing as background source a brighter continuum with higher S/N. As such, this line is not affected by POLDs and stellar activity in transmission spectra. We report the detection of Li at 6707.775\,\AA\,in the transmission spectrum of HD\,189733\,b, with an absorption amplitude of 0.102 $\pm$ 0.016 \% (6.4$\sigma$), blueshifted radial velocity of -2.4 $\pm$ 1.8 km s$^{-1}$, and Gaussian FWHM of 9.9 $\pm$ 1.8 km s$^{-1}$, using a Gaussian fit. The Li absorption signature is shown in Fig. \ref{fig:Li}. The lithium transmission spectra corrected using local RVs and uncorrected show very little difference, as expected from the lack of stellar absorption line.

The other examined atomic species did not show an absorption signature in the transmission spectrum. Some showed varying degrees of POLD residuals without significant planetary signal (Mg I, Mn, Ba, K), some lied at the level of the continuum (Ca II H \& K, Mg I), and others were dominated by chromospheric variability (H$\alpha$, H$\beta$). An apparent H$\alpha$ absorption signature appeared during the second transit, but stellar activity (see Fig. \ref{fig:Activity}) and the broadness of the line made it difficult to determine its origin. Both K lines appeared in the telluric O$_2$ A band, the blue completely blended with a telluric line, and the red at the very edge of a masked region and showing some POLD residuals with no significant atmospheric absorption.

\section{Discussion and conclusions} \label{section:Disc}

\subsection{POLDs mitigation}\label{sect:POLD summary}
Accounting for POLDs, caused primarily by the RM effect and CLVs, is paramount for the retrieval of planetary atmospheric signatures. We proposed several approaches, none of which allowed us to fully disentangle the stellar and planetary components during the entire transit. Nevertheless, all of them gave indications about the origin of the different contributions to the signal in the transmission spectrum. 

The main challenge at this stage is the limit in our knowledge of the local stellar spectra. On the modeling side, the sodium line cores of a K2 star are hard to reproduce using 1D models and would require 3D MHD simulations. The chromospheric contribution may also play a role and should be modeled as well. On the data-processing side, variability of stellar activity on short timescales (of the order of the transit duration) prevented the out-of-transit master spectrum from being an ideal proxy of the in-transit stellar spectrum.

\subsection{Planetary sodium signature}

From the transmission spectrum of the first half of the transit, we can confirm the presence of sodium in the upper atmosphere of HD\,189733\,b, as the POLDs and the planetary signature did not coincide in velocity. In the second half of the transit, we suspected that the POLDs and atmospheric signal overlap and therefore could not be accurately disentangled (see Figs. \ref{fig:TS+RM}, \ref{fig:TS2DRM}, \ref{fig:TS2D}). The integrated absorption signature may appear in the second half of the transit as an excess in the transmission light curve (Fig. \ref{fig:Na TLC}), however. We found no sodium absorption at ingress and egress, neither in the transmission spectrum nor in the sodium transmission light curve, similarly to \cite{Borsa18} with HARPS and \cite{Keles24} with PEPSI\footnote{Potsdam Echelle Polarimetric and Spectroscopic Instrument} \citep{Strass15}, which we interpret in Sect. \ref{sect:PEPSI}. 

With the two sodium lines co-added, in the first half of the transit, we measured a sodium absorption amplitude of 0.432 $\pm$ 0.027 \%, a Gaussian FWHM of 3.82 $\pm$ 0.29 km s$^{-1}$, and a radial velocity blueshift of -7.97 $\pm$ 0.28 km s$^{-1}$, in agreement with the single line values (see Table \ref{table:gauss}). We discuss further and compare those model-independent results with other similar studies in the following subsections.

We then modeled both the stellar and planetary contributions to the observed absorption spectrum using the EvE code, simulating the transit with a synthetic grid stellar based on the master-out spectrum and a 1D NLTE profile for the stellar sodium. We could reproduce the observed atmospheric sodium signature with a day-to-night side wind of 8 km s$^{-1}$ and a temperature of 2750 K. We consider that the 8 km s$^{-1}$ wind is compatible with the blueshift from the Gaussian fit to the transmission spectrum given the uncertainties and biases of the two methods. The POLDs were more difficult to reproduce for the reasons described in Sect. \ref{sect:POLD summary}, however. By introducing an additional stellar chromospheric contribution (see Figs. \ref{fig:fout_obs_synt}, \ref{fig:abs_spec_obs_synt_noatm}), we managed to better match the observed disc-integrated stellar spectrum and to model the POLDs more accurately, with a best-fit density of n$_c$ = 7.54 $\times$ 10$^{11}$ atoms cm$^{-3}$.

\subsection{Comparison to other sodium HD\,189733\,b studies}

\subsubsection{Original HARPS observations}
\cite{Wytt15} analyzed three HARPS transits of HD\,189733\,b, shifting each in-transit spectrum to the planetary rest frame for the first time, to properly derive the planetary signature. They measured sodium absorption with an amplitude of 0.64 $\pm$ 0.07 \% and 0.40 $\pm$ 0.07 \% for the D2 and D1 lines, respectively, and strong day-to-night side winds of -8 $\pm$ 2 km s$^{-1}$, with a broad signature of 0.52 $\pm$ 0.08 \AA. Although the amplitude and shift seem to match our findings, the comparison of results is difficult as \cite{Wytt15} results are biased by their noncorrection of the POLDs. A visual comparison between the uncorrected transmission spectra (our Fig. \ref{fig:TS Na no-corr} and their Fig. 2) shows a significantly higher S/N and a clear RM pattern in ESPRESSO, while the HARPS data show noisier absorption features. We believe that the large line width obtained by \cite{Wytt15} is due to POLDs. Other factors that may cause differences between the two datasets are the resolution of the spectrographs, older system parameters used, weather conditions, and the use of different orbital phase ranges (from T$_1$ to T$_4$ rather than T$_2$ to T$_3$).

The HARPS transits have been reanalyzed extensively \citep{Loud15, CB17, Lange22, Sici22} with different methods to correct for telluric contamination, compute the transmission spectrum, and account for POLDs. \cite{Loud15} accounted for the RM effect, but not the CLV, of one HARPS transit, obtaining a global -1.9$^{+0.7}_{-0.6}$ km s$^{-1}$ velocity blueshift. The other studies used all three HARPS transits and modeled the RM and the CLV following the method described in \cite{Yan2017}. All analyzes found similar blueshifts of around -2 km s$^{-1}$, Gaussian FWHM ranging from 0.41 to 0.64 \AA, and absorption amplitudes from 0.43 to 0.72 \%. 

Our results with ESPRESSO seem to differ significantly on these parameters: the blueshift is higher, the FWHM is narrower, and the amplitude lies on the shallower end at about 0.4 \%. To interpret these discrepancies, we first note that we found fundamental limitations in our ability to model the POLDs (see Sections \ref{RMcorr} and \ref{section:EVE}). We believe that these limitations affected all previous works, with additional biases due to their approach in accounting (or not) for the POLDs, as can be seen in transmission spectra dominated by POLD residuals and stellar activity (Figs. \ref{fig:Activity} and \ref{fig:TS2DRM}). Fitting a Gaussian model to these biased transmission spectra leads to smaller blueshifts, larger FWHMs, and larger amplitudes (see Fig. \ref{fig:blue fit}). On our side, we obtained different results because we selected orbital phases that avoided the stellar line cores, which allowed us to properly isolate the planetary signature. The higher S/N of the ESPRESSO data helps us to understand the complex situation in the analysis of the Na\,I D lines of this type of aligned system.

\subsubsection{PEPSI observations}\label{sect:PEPSI}

\cite{Keles24} obtained a transmission spectrum from one transit of HD\,189733\,b with PEPSI ($\mathcal{R}  \sim  130\,000$) at the Large Binocular Telescope (LBT), at a S/N similar to ESPRESSO transits. The POLDs were modeled using 1D LTE synthetic spectra from PHOENIX \citep{Husser13}, following a similar method as \cite{Yan2017}. For their stellar parameters, they used a similar effective temperature, but a higher stellar sodium abundance (A$_*$(Na\,I) = 6.5 dex, compared to our almost solar A$_*$(Na\,I) = 6.14 dex).

After correcting their transmission spectrum from the RM and CLV effects, they found a larger sodium absorption amplitude of 0.96 $\pm$ 0.05 \%, a narrow Gaussian FWHM of 0.28 $\pm$ 0.02 \AA, and a small blueshift of -0.72 $\pm$ 0.41 km s$^{-1}$ for the D2 line, and a contrast of 0.88 $\pm$ 0.12 \%, FWHM of 0.13 $\pm$ 0.02 \AA\, and blueshift of -0.76 $\pm$ 0.47 km s$^{-1}$ for the D1 line. They found no ingress and egress absorption, which they believe could indicate that the signal detected in-transit is not of planetary origin.

On our side, we did not detect sodium absorption at ingress either, but could not determine if there was absorption at egress due to the overlap with the stellar track (see Fig. \ref{fig:TS2DRM}). We believe that the lack of absorption seen at ingress (and egress) could be due to a noisier sodium signal, as the star is darker at the limbs and the amount of planetary sodium in front of the star is lower during ingress and egress. It could also be that the sodium distribution in the atmosphere is inhomogeneous, with the leading limb being less abundant in sodium than the trailing limb, which would lead to less absorption at ingress \citep{Ehren20}, while we remain uncertain of an absorption at egress as it overlaps with the stellar track.

By visually examining both transmission spectra (Fig. 9 center-bottom of \citet{Keles24}, our Figs. \ref{fig:RMCorrs} bottom right panel, and \ref{fig:blue fit}), we believe that the situation is similar to the HARPS analyses. If we had used the full transit for the transmission spectrum computation and Gaussian-fit the entire resulting signature including POLD residuals and stellar activity, we would have also found a higher amplitude, wider FWHM, and lower to no blueshift. Our analysis points to the need to mask the stellar line cores and consider only a restricted range in orbital phases to isolate the planetary signal, which explains our different results. We also believe that the higher S/N that we achieved with two ESPRESSO transits allowed us to identify the faint planetary signal more easily.

Finally, \cite{Keles24} argued that the FWHM they fit was too narrow to be physically plausible. On our side, the EvE simulations were able to qualitatively reproduce the shape of the planetary signature, which indicates that our $\sim$ 4-5 km s$^{-1}$ line width is physically possible.

\subsubsection{HARPS-N observations}
Very recently, \citet{Sici25} published a new analysis with more transits, combining the original three HARPS observations combined with eight HARPS-N transits. Their use of eleven stacked transits enhanced the S/N of their transmission spectrum, compared to the HARPS-only analyses. With similar POLD corrections and computation procedure as in \citet{Sici22}, their Gaussian fit resulted in a sodium absorption amplitude of around 0.3 $\pm$ 0.04 \% (reaching 0.4 $\pm$ 0.05 \% when excluding the three worst transits), blueshifted by $\sim$ 6 km s$^{-1}$ and a large FWHM of $\sim$ 35 km s$^{-1}$ (see their Fig. 4 and summary on page 8). We believe that our results are consistent with their findings, as the discrepancies stem from the same differences in procedure as with HARPS: The resulting transmission spectrum was still affected by POLD residuals and stellar activity, and the combination of all in-transit spectra (from T$_1$ to T$_4$) led to the overlap of the planetary signature and stellar contamination. Their high S/N transmission spectrum seems to show a slightly deeper, narrower absorption component on the blue side of each sodium line, which could be the planetary signature, drowned in stellar residuals, in a similar way to our transmission spectrum (See Fig. \ref{fig:RMCorrs}). Their sodium transit light curve is also very consistent with ours, showing that most of the absorption signature comes from the fully in-transit orbital phases (T$_2$ to T$_3$).

\subsection{Lithium}
The presence of the Li i 670.7 nm line in the exoplanetary atmosphere of WASP-127b was first claimed by \citet{Chen18} who derived a very high lithium abundance of A(Li) = 8.8 $\pm$ 1.5. This detection was later not confirmed by means of higher-quality ESPRESSO observations (\citealt{Allart20}), however. \citet{Tab21,Borsa21} presented the detection of lithium in the planetary atmospheres of WASP-76b and WASP-121b. These signatures were also confirmed in WASP-76b by \citet{Kess22} and in WASP-76b and WASP-121b by \citet{Azev22} from an independent reanalysis of the same ESPRESSO data. The 2D tomographic evidence of the coherence
with the planetary rest frame made these two detections quite robust.
Moreover, in both cases the possibility of residual stellar contamination
producing the feature could be safely ruled out. In WASP-76, the
stellar Li feature had an Equivalent Width of EW = 29 mA, while in WASP-121 there was no detectable Li and the EW is of < 2 mA. This latter star has a T$_{eff}$ of
6586 $\pm$ 56 K, a range of temperatures where stars show the Li-dip. No
quantitative abundance has been attempted for this lithium feature in the
discovery papers.

Lithium has been found in only five exoplanets so far: WASP-127b, WASP-76b, WASP-121b, WASP-85Ab \citep{Jiang23} and now HD\,189733\,b. Interestingly, contrary to their host stars, lithium is not destroyed in planets. Thus, planets may be a useful tracer of lithium abundances at the time of system formation. We are not aware of any studies discussing the abundance of lithium in exoplanets and its significance for the broader context of lithium abundance in the universe.

\subsection{Conclusion}
We attempted to advance the current results about sodium in HD\,189733\,b by leveraging high S/N ESPRESSO data. As in previous studies, accounting for POLDs and variability due to stellar activity caused significant challenges. We were able to identify and isolate a genuine planetary signature with a blueshift of $\sim$ 8 km s$^{-1}$, which indicates day-to-night winds at high altitudes at the terminator of the planet. To proceed, 3D MHD stellar models for the sodium line cores are likely necessary. A better understanding of short-term stellar variability in the sodium line cores is also needed.

\begin{acknowledgements}
This work has been carried out within the framework of the National Centre of Competence in Research PlanetS supported by the Swiss National Science Foundation (SNSF) under grants 51NF40\textunderscore182901 and 51NF40\textunderscore205606. The authors acknowledge the financial support of the SNSF. This project has received funding from the European Research Council (ERC) under the European Union's Horizon 2020 research and innovation programme (project \textsc{Four Aces}, grant agreement No. 724427; project \textsc{Spice Dune}, grant agreement No. 947634). AP acknowledges grants from the Spanish program Unidad de Excelencia María de Maeztu CEX2020$-$001058$-$M, 2021$-$SGR$-$1526 (Generalitat de Catalunya), and support from the Generalitat de Catalunya/CERCA. RA acknowledges the SNSF support under the Post-Doc Mobility grant P500PT\_222212 and the support of the Institut Trottier de Recherche sur les Exoplan\`etes (iREx). DE and MS acknowledge financial support from the SNSF for project 200021\_200726. Funded by the European Union (ERC, FIERCE, 101052347). Views and opinions expressed are however those of the author(s) only and do not necessarily reflect those of the European Union or the European Research Council. Neither the European Union nor the granting authority can be held responsible for them. This work was supported by FCT - Fundação para a Ciência e a Tecnologia through national funds by these grants: UIDB/04434/2020 DOI: 10.54499/UIDB/04434/2020, UIDP/04434/2020 DOI: 10.54499/UIDP/04434/2020 and in the framework of the project 2022.04048.PTDC (Phi in the Sky, DOI 10.54499/2022.04048.PTDC). CJM also acknowledges FCT and POCH/FSE (EC) support through Investigador FCT Contract 2021.01214.CEECIND/CP1658/CT0001 (DOI 10.54499/2021.01214.CEECIND/CP1658/CT0001). JIGH and ASM acknowledge financial support from the Spanish Ministry of Science, Innovation and Universities (MICIU) projects PID2020-117493GB-I00 and PID2023-149982NB-I00. We acknowledge financial support from the Agencia Estatal de Investigaci\'on of the Ministerio de Ciencia e Innovaci\'on MCIN/AEI/10.13039/501100011033 and the ERDF “A way of making Europe” through project PID2021-125627OB-C32, and from the Centre of Excellence “Severo Ochoa” award to the Instituto de Astrof{\'\i}sica de Canarias. The INAF authors acknowledge financial support of the Italian Ministry of Education, University, and Research with PRIN 201278X4FL and the "Progetti Premiali" funding scheme.
\end{acknowledgements}

\bibliographystyle{aa}    
\bibliography{biblio}

\begin{thebibliography}{97}
\expandafter\ifx\csname natexlab\endcsname\relax\def\natexlab#1{#1}\fi

\bibitem[{{Allart} {et~al.}(2022){Allart}, {Lovis}, {Faria}, {Dumusque}, {Sosnowska}, {Figueira}, {Silva}, {Mehner}, {Pepe}, {Cristiani}, {Rebolo}, {Santos}, {Adibekyan}, {Cupani}, {Di Marcantonio}, {D'Odorico}, {Gonz{\'a}lez Hern{\'a}ndez}, {Martins}, {Milakovi{\'c}}, {Nunes}, {Sozzetti}, {Su{\'a}rez Mascare{\~n}o}, {Tabernero}, \& {Zapatero Osorio}}]{Allart22}
{Allart}, R., {Lovis}, C., {Faria}, J., {et~al.} 2022, \aap, 666, A196

\bibitem[{{Allart} {et~al.}(2020){Allart}, {Pino}, {Lovis}, {Sousa}, {Casasayas-Barris}, {Zapatero Osorio}, {Cretignier}, {Palle}, {Pepe}, {Cristiani}, {Rebolo}, {Santos}, {Borsa}, {Bourrier}, {Demangeon}, {Ehrenreich}, {Lavie}, {Lendl}, {Lillo-Box}, {Micela}, {Oshagh}, {Sozzetti}, {Tabernero}, {Adibekyan}, {Allende Prieto}, {Alibert}, {Amate}, {Benz}, {Bouchy}, {Cabral}, {Dekker}, {D'Odorico}, {Di Marcantonio}, {Dumusque}, {Figueira}, {Genova Santos}, {Gonz{\'a}lez Hern{\'a}ndez}, {Lo Curto}, {Manescau}, {Martins}, {M{\'e}gevand}, {Mehner}, {Molaro}, {Nunes}, {Poretti}, {Riva}, {Su{\'a}rez Mascare{\~n}o}, {Udry}, \& {Zerbi}}]{Allart20}
{Allart}, R., {Pino}, L., {Lovis}, C., {et~al.} 2020, \aap, 644, A155

\bibitem[{{Alvarez} \& {Plez}(1998)}]{plez1998}
{Alvarez}, R. \& {Plez}, B. 1998, \aap, 330, 1109

\bibitem[{{Artigau} {et~al.}(2014){Artigau}, {Astudillo-Defru}, {Delfosse}, {Bouchy}, {Bonfils}, {Lovis}, {Pepe}, {Moutou}, {Donati}, {Doyon}, \& {Malo}}]{Arti14}
{Artigau}, {\'E}., {Astudillo-Defru}, N., {Delfosse}, X., {et~al.} 2014, in Society of Photo-Optical Instrumentation Engineers (SPIE) Conference Series, Vol. 9149, Observatory Operations: Strategies, Processes, and Systems V, ed. A.~B. {Peck}, C.~R. {Benn}, \& R.~L. {Seaman}, 914905

\bibitem[{{Azevedo Silva} {et~al.}(2022){Azevedo Silva}, {Demangeon}, {Santos}, {Allart}, {Borsa}, {Cristo}, {Esparza-Borges}, {Seidel}, {Palle}, {Sousa}, {Tabernero}, {Zapatero Osorio}, {Cristiani}, {Pepe}, {Rebolo}, {Adibekyan}, {Alibert}, {Barros}, {Bouchy}, {Bourrier}, {Lo Curto}, {Di Marcantonio}, {D'Odorico}, {Ehrenreich}, {Figueira}, {Gonz{\'a}lez Hern{\'a}ndez}, {Lovis}, {Martins}, {Mehner}, {Micela}, {Molaro}, {Mounzer}, {Nunes}, {Sozzetti}, {Su{\'a}rez Mascare{\~n}o}, \& {Udry}}]{Azev22}
{Azevedo Silva}, T., {Demangeon}, O.~D.~S., {Santos}, N.~C., {et~al.} 2022, \aap, 666, L10

\bibitem[{{Basilicata} {et~al.}(2024){Basilicata}, {Giacobbe}, {Bonomo}, {Scandariato}, {Brogi}, {Singh}, {Di Paola}, {Mancini}, {Sozzetti}, {Lanza}, {Cubillos}, {Damasso}, {Desidera}, {Biazzo}, {Bignamini}, {Borsa}, {Cabona}, {Carleo}, {Ghedina}, {Guilluy}, {Maggio}, {Mainella}, {Micela}, {Molinari}, {Molinaro}, {Nardiello}, {Pedani}, {Pino}, {Poretti}, {Southworth}, {Stangret}, \& {Turrini}}]{Basil24}
{Basilicata}, M., {Giacobbe}, P., {Bonomo}, A.~S., {et~al.} 2024, \aap, 686, A127

\bibitem[{{Bell} {et~al.}(2024){Bell}, {Crouzet}, {Cubillos}, {Kreidberg}, {Piette}, {Roman}, {Barstow}, {Blecic}, {Carone}, {Coulombe}, {Ducrot}, {Hammond}, {Mendon{\c{c}}a}, {Moses}, {Parmentier}, {Stevenson}, {Teinturier}, {Zhang}, {Batalha}, {Bean}, {Benneke}, {Charnay}, {Chubb}, {Demory}, {Gao}, {Lee}, {L{\'o}pez-Morales}, {Morello}, {Rauscher}, {Sing}, {Tan}, {Venot}, {Wakeford}, {Aggarwal}, {Ahrer}, {Alam}, {Baeyens}, {Barrado}, {Caceres}, {Carter}, {Casewell}, {Challener}, {Crossfield}, {Decin}, {D{\'e}sert}, {Dobbs-Dixon}, {Dyrek}, {Espinoza}, {Feinstein}, {Gibson}, {Harrington}, {Helling}, {Hu}, {Iro}, {Kempton}, {Kendrew}, {Komacek}, {Krick}, {Lagage}, {Leconte}, {Lendl}, {Lewis}, {Lothringer}, {Malsky}, {Mancini}, {Mansfield}, {Mayne}, {Evans-Soma}, {Molaverdikhani}, {Nikolov}, {Nixon}, {Palle}, {Petit dit de la Roche}, {Piaulet}, {Powell}, {Rackham}, {Schneider}, {Steinrueck}, {Taylor}, {Welbanks}, {Yurchenko}, {Zhang}, \& {Zieba}}]{Bell24}
{Bell}, T.~J., {Crouzet}, N., {Cubillos}, P.~E., {et~al.} 2024, Nature Astronomy, 8, 879

\bibitem[{{Bello-Arufe} {et~al.}(2022){Bello-Arufe}, {Cabot}, {Mendon{\c{c}}a}, {Buchhave}, \& {Rathcke}}]{Bello20}
{Bello-Arufe}, A., {Cabot}, S. H.~C., {Mendon{\c{c}}a}, J.~M., {Buchhave}, L.~A., \& {Rathcke}, A.~D. 2022, \aj, 163, 96

\bibitem[{{Benz} {et~al.}(2021){Benz}, {Broeg}, {Fortier}, {Rando}, {Beck}, {Beck}, {Queloz}, {Ehrenreich}, {Maxted}, {Isaak}, {Billot}, {Alibert}, {Alonso}, {Ant{\'o}nio}, {Asquier}, {Bandy}, {B{\'a}rczy}, {Barrado}, {Barros}, {Baumjohann}, {Bekkelien}, {Bergomi}, {Biondi}, {Bonfils}, {Borsato}, {Brandeker}, {Busch}, {Cabrera}, {Cessa}, {Charnoz}, {Chazelas}, {Collier Cameron}, {Corral Van Damme}, {Cortes}, {Davies}, {Deleuil}, {Deline}, {Delrez}, {Demangeon}, {Demory}, {Erikson}, {Farinato}, {Fossati}, {Fridlund}, {Futyan}, {Gandolfi}, {Garcia Munoz}, {Gillon}, {Guterman}, {Gutierrez}, {Hasiba}, {Heng}, {Hernandez}, {Hoyer}, {Kiss}, {Kovacs}, {Kuntzer}, {Laskar}, {Lecavelier des Etangs}, {Lendl}, {L{\'o}pez}, {Lora}, {Lovis}, {L{\"u}ftinger}, {Magrin}, {Malvasio}, {Marafatto}, {Michaelis}, {de Miguel}, {Modrego}, {Munari}, {Nascimbeni}, {Olofsson}, {Ottacher}, {Ottensamer}, {Pagano}, {Palacios}, {Pall{\'e}}, {Peter}, {Piazza}, {Piotto}, {Pizarro}, {Pollaco}, {Ragazzoni}, {Ratti}, {Rauer}, {Ribas}, {Rieder},
  {Rohlfs}, {Safa}, {Salatti}, {Santos}, {Scandariato}, {S{\'e}gransan}, {Simon}, {Smith}, {Sordet}, {Sousa}, {Steller}, {Szab{\'o}}, {Szoke}, {Thomas}, {Tschentscher}, {Udry}, {Van Grootel}, {Viotto}, {Walter}, {Walton}, {Wildi}, \& {Wolter}}]{Benz21}
{Benz}, W., {Broeg}, C., {Fortier}, A., {et~al.} 2021, Experimental Astronomy, 51, 109

\bibitem[{{Blain} {et~al.}(2024){Blain}, {S{\'a}nchez-L{\'o}pez}, \& {Molli{\`e}re}}]{Blain24}
{Blain}, D., {S{\'a}nchez-L{\'o}pez}, A., \& {Molli{\`e}re}, P. 2024, \aj, 167, 179

\bibitem[{{Bonomo} {et~al.}(2017){Bonomo}, {Desidera}, {Benatti}, {Borsa}, {Crespi}, {Damasso}, {Lanza}, {Sozzetti}, {Lodato}, {Marzari}, {Boccato}, {Claudi}, {Cosentino}, {Covino}, {Gratton}, {Maggio}, {Micela}, {Molinari}, {Pagano}, {Piotto}, {Poretti}, {Smareglia}, {Affer}, {Biazzo}, {Bignamini}, {Esposito}, {Giacobbe}, {H{\'e}brard}, {Malavolta}, {Maldonado}, {Mancini}, {Martinez Fiorenzano}, {Masiero}, {Nascimbeni}, {Pedani}, {Rainer}, \& {Scandariato}}]{Bonomo17}
{Bonomo}, A.~S., {Desidera}, S., {Benatti}, S., {et~al.} 2017, \aap, 602, A107

\bibitem[{{Borsa} {et~al.}(2021){Borsa}, {Allart}, {Casasayas-Barris}, {Tabernero}, {Zapatero Osorio}, {Cristiani}, {Pepe}, {Rebolo}, {Santos}, {Adibekyan}, {Bourrier}, {Demangeon}, {Ehrenreich}, {Pall{\'e}}, {Sousa}, {Lillo-Box}, {Lovis}, {Micela}, {Oshagh}, {Poretti}, {Sozzetti}, {Allende Prieto}, {Alibert}, {Amate}, {Benz}, {Bouchy}, {Cabral}, {Dekker}, {D'Odorico}, {Di Marcantonio}, {Figueira}, {Genova Santos}, {Gonz{\'a}lez Hern{\'a}ndez}, {Lo Curto}, {Manescau}, {Martins}, {M{\'e}gevand}, {Mehner}, {Molaro}, {Nunes}, {Riva}, {Su{\'a}rez Mascare{\~n}o}, {Udry}, \& {Zerbi}}]{Borsa21}
{Borsa}, F., {Allart}, R., {Casasayas-Barris}, N., {et~al.} 2021, \aap, 645, A24

\bibitem[{{Borsa} \& {Zannoni}(2018)}]{Borsa18}
{Borsa}, F. \& {Zannoni}, A. 2018, \aap, 617, A134

\bibitem[{{Bouchy} {et~al.}(2005){Bouchy}, {Udry}, {Mayor}, {Moutou}, {Pont}, {Iribarne}, {da Silva}, {Ilovaisky}, {Queloz}, {Santos}, {S{\'e}gransan}, \& {Zucker}}]{Bouchy2005}
{Bouchy}, F., {Udry}, S., {Mayor}, M., {et~al.} 2005, \aap, 444, L15

\bibitem[{{Bourrier} {et~al.}(2024){Bourrier}, {Delisle}, {Lovis}, {Cegla}, {Cretignier}, {Allart}, {Al Moulla}, {Tavella}, {Attia}, {Mounzer}, {Vaulato}, {Steiner}, {Vrignaud}, {Mercier}, {Dumusque}, {Ehrenreich}, {Seidel}, {Wyttenbach}, {Dethier}, \& {Pepe}}]{B24}
{Bourrier}, V., {Delisle}, J.~B., {Lovis}, C., {et~al.} 2024, \aap, 691, A113

\bibitem[{{Bourrier} {et~al.}(2015){Bourrier}, {Ehrenreich}, \& {Lecavelier des Etangs}}]{bourrier2015}
{Bourrier}, V., {Ehrenreich}, D., \& {Lecavelier des Etangs}, A. 2015, A\&A, 582, A65

\bibitem[{{Bourrier} \& {Lecavelier des Etangs}(2013)}]{bourrier2013}
{Bourrier}, V. \& {Lecavelier des Etangs}, A. 2013, A\&A, 557, A124

\bibitem[{{Bourrier} {et~al.}(2016){Bourrier}, {Lecavelier des Etangs}, {Ehrenreich}, {Tanaka}, \& {Vidotto}}]{bourrier2016}
{Bourrier}, V., {Lecavelier des Etangs}, A., {Ehrenreich}, D., {Tanaka}, Y.~A., \& {Vidotto}, A.~A. 2016, A\&A, 591, A121

\bibitem[{{Bourrier} {et~al.}(2021){Bourrier}, {Lovis}, {Cretignier}, {Allart}, {Dumusque}, {Delisle}, {Deline}, {Sousa}, {Adibekyan}, {Alibert}, {Barros}, {Borsa}, {Cristiani}, {Demangeon}, {Ehrenreich}, {Figueira}, {Gonz{\'a}lez Hern{\'a}ndez}, {Lendl}, {Lillo-Box}, {Lo Curto}, {Di Marcantonio}, {Martins}, {M{\'e}gevand}, {Mehner}, {Micela}, {Molaro}, {Oshagh}, {Palle}, {Pepe}, {Poretti}, {Rebolo}, {Santos}, {Scandariato}, {Seidel}, {Sozzetti}, {Su{\'a}rez Mascare{\~n}o}, \& {Zapatero Osorio}}]{Bourrier2021}
{Bourrier}, V., {Lovis}, C., {Cretignier}, M., {et~al.} 2021, \aap, 654, A152

\bibitem[{{Bruls} \& {Rutten}(1992)}]{bruls1992}
{Bruls}, J.~H.~M.~J. \& {Rutten}, R.~J. 1992, \aap, 265, 257

\bibitem[{{Canocchi} {et~al.}(2024){Canocchi}, {Morello}, {Lind}, {Carleo}, {Stangret}, \& {Pall{\'e}}}]{canocchi24}
{Canocchi}, G., {Morello}, G., {Lind}, K., {et~al.} 2024, \aap, 692, A43

\bibitem[{{Carteret} {et~al.}(2024){Carteret}, {Bourrier}, \& {Dethier}}]{Cart23}
{Carteret}, Y., {Bourrier}, V., \& {Dethier}, W. 2024, \aap, 683, A63

\bibitem[{{Casasayas-Barris} {et~al.}(2017){Casasayas-Barris}, {Palle}, {Nowak}, {Yan}, {Nortmann}, \& {Murgas}}]{CB17}
{Casasayas-Barris}, N., {Palle}, E., {Nowak}, G., {et~al.} 2017, \aap, 608, A135

\bibitem[{{Casasayas-Barris} {et~al.}(2021){Casasayas-Barris}, {Palle}, {Stangret}, {Bourrier}, {Tabernero}, {Yan}, {Borsa}, {Allart}, {Zapatero Osorio}, {Lovis}, {Sousa}, {Chen}, {Oshagh}, {Santos}, {Pepe}, {Rebolo}, {Molaro}, {Cristiani}, {Adibekyan}, {Alibert}, {Allende Prieto}, {Bouchy}, {Demangeon}, {Di Marcantonio}, {D'Odorico}, {Ehrenreich}, {Figueira}, {G{\'e}nova Santos}, {Gonz{\'a}lez Hern{\'a}ndez}, {Lavie}, {Lillo-Box}, {Lo Curto}, {Martins}, {Mehner}, {Micela}, {Nunes}, {Poretti}, {Sozzetti}, {Su{\'a}rez Mascare{\~n}o}, \& {Udry}}]{CB2021}
{Casasayas-Barris}, N., {Palle}, E., {Stangret}, M., {et~al.} 2021, \aap, 647, A26

\bibitem[{{Casasayas-Barris} {et~al.}(2020){Casasayas-Barris}, {Pall{\'e}}, {Yan}, {Chen}, {Luque}, {Stangret}, {Nagel}, {Zechmeister}, {Oshagh}, {Sanz-Forcada}, {Nortmann}, {Alonso-Floriano}, {Amado}, {Caballero}, {Czesla}, {Khalafinejad}, {L{\'o}pez-Puertas}, {L{\'o}pez-Santiago}, {Molaverdikhani}, {Montes}, {Quirrenbach}, {Reiners}, {Ribas}, {S{\'a}nchez-L{\'o}pez}, \& {Zapatero Osorio}}]{CB2020}
{Casasayas-Barris}, N., {Pall{\'e}}, E., {Yan}, F., {et~al.} 2020, \aap, 635, A206

\bibitem[{{Cauley} {et~al.}(2018){Cauley}, {Kuckein}, {Redfield}, {Shkolnik}, {Denker}, {Llama}, \& {Verma}}]{Caul18}
{Cauley}, P.~W., {Kuckein}, C., {Redfield}, S., {et~al.} 2018, \aj, 156, 189

\bibitem[{{Cauley} {et~al.}(2017){Cauley}, {Redfield}, \& {Jensen}}]{Caul17}
{Cauley}, P.~W., {Redfield}, S., \& {Jensen}, A.~G. 2017, \aj, 153, 185

\bibitem[{{Cegla} {et~al.}(2016){Cegla}, {Lovis}, {Bourrier}, {Beeck}, {Watson}, \& {Pepe}}]{Cegla16}
{Cegla}, H.~M., {Lovis}, C., {Bourrier}, V., {et~al.} 2016, \aap, 588, A127

\bibitem[{{Charbonneau} {et~al.}(2000){Charbonneau}, {Brown}, {Latham}, \& {Mayor}}]{Charb00}
{Charbonneau}, D., {Brown}, T.~M., {Latham}, D.~W., \& {Mayor}, M. 2000, \apjl, 529, L45

\bibitem[{{Charbonneau} {et~al.}(2002){Charbonneau}, {Brown}, {Noyes}, \& {Gilliland}}]{Charb02}
{Charbonneau}, D., {Brown}, T.~M., {Noyes}, R.~W., \& {Gilliland}, R.~L. 2002, \apj, 568, 377

\bibitem[{{Chen} {et~al.}(2018){Chen}, {Pall{\'e}}, {Welbanks}, {Prieto-Arranz}, {Madhusudhan}, {Gandhi}, {Casasayas-Barris}, {Murgas}, {Nortmann}, {Crouzet}, {Parviainen}, \& {Gandolfi}}]{Chen18}
{Chen}, G., {Pall{\'e}}, E., {Welbanks}, L., {et~al.} 2018, \aap, 616, A145

\bibitem[{{Cosentino} {et~al.}(2012){Cosentino}, {Lovis}, {Pepe}, {Collier Cameron}, {Latham}, {Molinari}, {Udry}, {Bezawada}, {Black}, {Born}, {Buchschacher}, {Charbonneau}, {Figueira}, {Fleury}, {Galli}, {Gallie}, {Gao}, {Ghedina}, {Gonzalez}, {Gonzalez}, {Guerra}, {Henry}, {Horne}, {Hughes}, {Kelly}, {Lodi}, {Lunney}, {Maire}, {Mayor}, {Micela}, {Ordway}, {Peacock}, {Phillips}, {Piotto}, {Pollacco}, {Queloz}, {Rice}, {Riverol}, {Riverol}, {San Juan}, {Sasselov}, {Segransan}, {Sozzetti}, {Sosnowska}, {Stobie}, {Szentgyorgyi}, {Vick}, \& {Weber}}]{Cosen12}
{Cosentino}, R., {Lovis}, C., {Pepe}, F., {et~al.} 2012, in Society of Photo-Optical Instrumentation Engineers (SPIE) Conference Series, Vol. 8446, Ground-based and Airborne Instrumentation for Astronomy IV, ed. I.~S. {McLean}, S.~K. {Ramsay}, \& H.~{Takami}, 84461V

\bibitem[{{Cretignier} {et~al.}(2021){Cretignier}, {Dumusque}, {Hara}, \& {Pepe}}]{Cret21}
{Cretignier}, M., {Dumusque}, X., {Hara}, N.~C., \& {Pepe}, F. 2021, \aap, 653, A43

\bibitem[{{Cristo} {et~al.}(2024){Cristo}, {Esparza Borges}, {Santos}, {Demangeon}, {Palle}, {Psaridi}, {Bourrier}, {Faria}, {Allart}, {Azevedo Silva}, {Borsa}, {Alibert}, {Figueira}, {Gonz{\'a}lez Hern{\'a}ndez}, {Lendl}, {Lillo-Box}, {Lo Curto}, {Di Marcantonio}, {Martins}, {Nunes}, {Pepe}, {Seidel}, {Sousa}, {Sozzetti}, {Stangret}, {Su{\'a}rez Mascare{\~n}o}, {Tabernero}, \& {Zapatero Osorio}}]{Cristo23}
{Cristo}, E., {Esparza Borges}, E., {Santos}, N.~C., {et~al.} 2024, \aap, 682, A28

\bibitem[{{Czesla} {et~al.}(2015){Czesla}, {Klocov{\'a}}, {Khalafinejad}, {Wolter}, \& {Schmitt}}]{Czesla15}
{Czesla}, S., {Klocov{\'a}}, T., {Khalafinejad}, S., {Wolter}, U., \& {Schmitt}, J.~H.~M.~M. 2015, \aap, 582, A51

\bibitem[{{Dethier} \& {Bourrier}(2023)}]{Deth23}
{Dethier}, W. \& {Bourrier}, V. 2023, \aap, 674, A86

\bibitem[{{Dethier} \& {Tessore}(2024)}]{dethier24}
{Dethier}, W. \& {Tessore}, B. 2024, \aap, 688, L30

\bibitem[{{Ehrenreich} {et~al.}(2020){Ehrenreich}, {Lovis}, {Allart}, {Zapatero Osorio}, {Pepe}, {Cristiani}, {Rebolo}, {Santos}, {Borsa}, {Demangeon}, {Dumusque}, {Gonz{\'a}lez Hern{\'a}ndez}, {Casasayas-Barris}, {S{\'e}gransan}, {Sousa}, {Abreu}, {Adibekyan}, {Affolter}, {Allende Prieto}, {Alibert}, {Aliverti}, {Alves}, {Amate}, {Avila}, {Baldini}, {Bandy}, {Benz}, {Bianco}, {Bolmont}, {Bouchy}, {Bourrier}, {Broeg}, {Cabral}, {Calderone}, {Pall{\'e}}, {Cegla}, {Cirami}, {Coelho}, {Conconi}, {Coretti}, {Cumani}, {Cupani}, {Dekker}, {Delabre}, {Deiries}, {D'Odorico}, {Di Marcantonio}, {Figueira}, {Fragoso}, {Genolet}, {Genoni}, {G{\'e}nova Santos}, {Hara}, {Hughes}, {Iwert}, {Kerber}, {Knudstrup}, {Landoni}, {Lavie}, {Lizon}, {Lendl}, {Lo Curto}, {Maire}, {Manescau}, {Martins}, {M{\'e}gevand}, {Mehner}, {Micela}, {Modigliani}, {Molaro}, {Monteiro}, {Monteiro}, {Moschetti}, {M{\"u}ller}, {Nunes}, {Oggioni}, {Oliveira}, {Pariani}, {Pasquini}, {Poretti}, {Rasilla}, {Redaelli}, {Riva}, {Santana Tschudi},
  {Santin}, {Santos}, {Segovia Milla}, {Seidel}, {Sosnowska}, {Sozzetti}, {Span{\`o}}, {Su{\'a}rez Mascare{\~n}o}, {Tabernero}, {Tenegi}, {Udry}, {Zanutta}, \& {Zerbi}}]{Ehren20}
{Ehrenreich}, D., {Lovis}, C., {Allart}, R., {et~al.} 2020, \nat, 580, 597

\bibitem[{{Fares} {et~al.}(2010){Fares}, {Donati}, {Moutou}, {Jardine}, {Grie{\ss}meier}, {Zarka}, {Shkolnik}, {Bohlender}, {Catala}, \& {Collier Cameron}}]{Fares10}
{Fares}, R., {Donati}, J.~F., {Moutou}, C., {et~al.} 2010, \mnras, 406, 409

\bibitem[{{Guilluy} {et~al.}(2020){Guilluy}, {Andretta}, {Borsa}, {Giacobbe}, {Sozzetti}, {Covino}, {Bourrier}, {Fossati}, {Bonomo}, {Esposito}, {Giampapa}, {Harutyunyan}, {Rainer}, {Brogi}, {Bruno}, {Claudi}, {Frustagli}, {Lanza}, {Mancini}, {Pino}, {Poretti}, {Scandariato}, {Affer}, {Baffa}, {Baruffolo}, {Benatti}, {Biazzo}, {Bignamini}, {Boschin}, {Carleo}, {Cecconi}, {Cosentino}, {Damasso}, {Desidera}, {Falcini}, {Martinez Fiorenzano}, {Ghedina}, {Gonz{\'a}lez-{\'A}lvarez}, {Guerra}, {Hernandez}, {Leto}, {Maggio}, {Malavolta}, {Maldonado}, {Micela}, {Molinari}, {Nascimbeni}, {Pagano}, {Pedani}, {Piotto}, \& {Reiners}}]{Guill20}
{Guilluy}, G., {Andretta}, V., {Borsa}, F., {et~al.} 2020, \aap, 639, A49

\bibitem[{{Guilluy} {et~al.}(2024){Guilluy}, {D'Arpa}, {Bonomo}, {Spinelli}, {Biassoni}, {Fossati}, {Maggio}, {Giacobbe}, {Lanza}, {Sozzetti}, {Borsa}, {Rainer}, {Micela}, {Affer}, {Andreuzzi}, {Bignamini}, {Boschin}, {Carleo}, {Cecconi}, {Desidera}, {Fardella}, {Ghedina}, {Mantovan}, {Mancini}, {Nascimbeni}, {Knapic}, {Pedani}, {Petralia}, {Pino}, {Scandariato}, {Sicilia}, {Stangret}, \& {Zingales}}]{Guill24}
{Guilluy}, G., {D'Arpa}, M.~C., {Bonomo}, A.~S., {et~al.} 2024, \aap, 686, A83

\bibitem[{{Gustafsson} {et~al.}(2008){Gustafsson}, {Edvardsson, B.}, {Eriksson, K.}, {J\o{}rgensen, U. G.}, {Nordlund, \AA{}.}, \& {Plez, B.}}]{gustafsson2008}
{Gustafsson}, B., {Edvardsson, B.}, {Eriksson, K.}, {et~al.} 2008, A\&A, 486, 951

\bibitem[{{Hayek} {et~al.}(2012){Hayek}, {Sing}, {Pont}, \& {Asplund}}]{Hayek12}
{Hayek}, W., {Sing}, D., {Pont}, F., \& {Asplund}, M. 2012, \aap, 539, A102

\bibitem[{{Heiter} {et~al.}(2021){Heiter}, {Lind}, {Bergemann}, {Asplund}, {Mikolaitis}, {Barklem}, {Masseron}, {de Laverny}, {Magrini}, {Edvardsson}, {J{\"o}nsson}, {Pickering}, {Ryde}, {Bayo Ar{\'a}n}, {Bensby}, {Casey}, {Feltzing}, {Jofr{\'e}}, {Korn}, {Pancino}, {Damiani}, {Lanzafame}, {Lardo}, {Monaco}, {Morbidelli}, {Smiljanic}, {Worley}, {Zaggia}, {Randich}, \& {Gilmore}}]{heiter2021}
{Heiter}, U., {Lind}, K., {Bergemann}, M., {et~al.} 2021, \aap, 645, A106

\bibitem[{{Henry} {et~al.}(2000){Henry}, {Marcy}, {Butler}, \& {Vogt}}]{Henry00}
{Henry}, G.~W., {Marcy}, G.~W., {Butler}, R.~P., \& {Vogt}, S.~S. 2000, \apjl, 529, L41

\bibitem[{{Huitson} {et~al.}(2012){Huitson}, {Sing}, {Vidal-Madjar}, {Ballester}, {Lecavelier des Etangs}, {D{\'e}sert}, \& {Pont}}]{Huit12}
{Huitson}, C.~M., {Sing}, D.~K., {Vidal-Madjar}, A., {et~al.} 2012, \mnras, 422, 2477

\bibitem[{{Husser} {et~al.}(2013){Husser}, {Wende-von Berg}, {Dreizler}, {Homeier}, {Reiners}, {Barman}, \& {Hauschildt}}]{Husser13}
{Husser}, T.~O., {Wende-von Berg}, S., {Dreizler}, S., {et~al.} 2013, \aap, 553, A6

\bibitem[{{Jensen} {et~al.}(2011){Jensen}, {Redfield}, {Endl}, {Cochran}, {Koesterke}, \& {Barman}}]{Jens11}
{Jensen}, A.~G., {Redfield}, S., {Endl}, M., {et~al.} 2011, \apj, 743, 203

\bibitem[{{Jiang} {et~al.}(2023){Jiang}, {Wang}, {Chen}, {Yan}, {Cegla}, {Rojo}, {Shi}, {Ouyang}, {Zhai}, {Liu}, {Zhao}, \& {Chen}}]{Jiang23}
{Jiang}, Z., {Wang}, W., {Chen}, G., {et~al.} 2023, \aap, 677, A110

\bibitem[{{Juvan} {et~al.}(2018){Juvan}, {Lendl}, {Cubillos}, {Fossati}, {Tregloan-Reed}, {Lammer}, {Guenther}, \& {Hanslmeier}}]{Juvan18}
{Juvan}, I.~G., {Lendl}, M., {Cubillos}, P.~E., {et~al.} 2018, \aap, 610, A15

\bibitem[{{Kausch} {et~al.}(2015){Kausch}, {Noll}, {Smette}, {Kimeswenger}, {Barden}, {Szyszka}, {Jones}, {Sana}, {Horst}, \& {Kerber}}]{Kausch2015}
{Kausch}, W., {Noll}, S., {Smette}, A., {et~al.} 2015, \aap, 576, A78

\bibitem[{{Keles} {et~al.}(2024){Keles}, {Czesla}, {Poppenhaeger}, {Hauschildt}, {Carroll}, {Ilyin}, {Baratella}, {Steffen}, {Strassmeier}, {Bonomo}, {Gaudi}, {Henning}, {Johnson}, {Molaverdikhani}, {Nascimbeni}, {Patience}, {Reiners}, {Scandariato}, {Schlawin}, {Shkolnik}, {Sicilia}, {Sozzetti}, {Mallonn}, {Veillet}, {Wang}, \& {Yan}}]{Keles24}
{Keles}, E., {Czesla}, S., {Poppenhaeger}, K., {et~al.} 2024, \mnras, 530, 4826

\bibitem[{{Kempton} {et~al.}(2018){Kempton}, {Bean}, {Louie}, {Deming}, {Koll}, {Mansfield}, {Christiansen}, {L{\'o}pez-Morales}, {Swain}, {Zellem}, {Ballard}, {Barclay}, {Barstow}, {Batalha}, {Beatty}, {Berta-Thompson}, {Birkby}, {Buchhave}, {Charbonneau}, {Cowan}, {Crossfield}, {de Val-Borro}, {Doyon}, {Dragomir}, {Gaidos}, {Heng}, {Hu}, {Kane}, {Kreidberg}, {Mallonn}, {Morley}, {Narita}, {Nascimbeni}, {Pall{\'e}}, {Quintana}, {Rauscher}, {Seager}, {Shkolnik}, {Sing}, {Sozzetti}, {Stassun}, {Valenti}, \& {von Essen}}]{Kemp18}
{Kempton}, E. M.~R., {Bean}, J.~L., {Louie}, D.~R., {et~al.} 2018, \pasp, 130, 114401

\bibitem[{{Kesseli} {et~al.}(2022){Kesseli}, {Snellen}, {Casasayas-Barris}, {Molli{\`e}re}, \& {S{\'a}nchez-L{\'o}pez}}]{Kess22}
{Kesseli}, A.~Y., {Snellen}, I.~A.~G., {Casasayas-Barris}, N., {Molli{\`e}re}, P., \& {S{\'a}nchez-L{\'o}pez}, A. 2022, \aj, 163, 107

\bibitem[{{Khalafinejad} {et~al.}(2017){Khalafinejad}, {von Essen}, {Hoeijmakers}, {Zhou}, {Klocov{\'a}}, {Schmitt}, {Dreizler}, {Lopez-Morales}, {Husser}, {Schmidt}, \& {Collet}}]{Khal2017}
{Khalafinejad}, S., {von Essen}, C., {Hoeijmakers}, H.~J., {et~al.} 2017, \aap, 598, A131

\bibitem[{{Kreidberg}(2015)}]{Kreid2015}
{Kreidberg}, L. 2015, \pasp, 127, 1161

\bibitem[{{Krenn} {et~al.}(2023){Krenn}, {Lendl}, {Patel}, {Carone}, {Deleuil}, {Sulis}, {Collier Cameron}, {Deline}, {Guterman}, {Queloz}, {Fossati}, {Brandeker}, {Heng}, {Akinsanmi}, {Adibekyan}, {Bonfanti}, {Demangeon}, {Kitzmann}, {Salmon}, {Sousa}, {Wilson}, {Alibert}, {Alonso}, {Anglada}, {B{\'a}rczy}, {Barrado Navascues}, {Barros}, {Baumjohann}, {Beck}, {Beck}, {Benz}, {Billot}, {Blecha}, {Bonfils}, {Borsato}, {Broeg}, {Cabrera}, {Charnoz}, {Corral van Damme}, {Csizmadia}, {Cubillos}, {Davies}, {Delrez}, {Demory}, {Ehrenreich}, {Erikson}, {Farinato}, {Fortier}, {Fridlund}, {Gandolfi}, {Gillon}, {G{\"u}del}, {Hoyer}, {Isaak}, {Kiss}, {Kopp}, {Laskar}, {Lecavelier des Etangs}, {Lovis}, {Magrin}, {Maxted}, {Mordasini}, {Nascimbeni}, {Olofsson}, {Ottensamer}, {Pagano}, {Pall{\'e}}, {Peter}, {Piotto}, {Pollacco}, {Ragazzoni}, {Rando}, {Rauer}, {Ribas}, {Santos}, {Scandariato}, {S{\'e}gransan}, {Simon}, {Smith}, {Steller}, {Szab{\'o}}, {Thomas}, {Udry}, {Ulmer}, {Van Grootel}, {Venturini}, \&
  {Walton}}]{Krenn23}
{Krenn}, A.~F., {Lendl}, M., {Patel}, J.~A., {et~al.} 2023, \aap, 672, A24

\bibitem[{{Langeveld} {et~al.}(2021){Langeveld}, {Madhusudhan}, {Cabot}, \& {Hodgkin}}]{Lange22}
{Langeveld}, A.~B., {Madhusudhan}, N., {Cabot}, S. H.~C., \& {Hodgkin}, S.~T. 2021, \mnras, 502, 4392

\bibitem[{{Larsen} {et~al.}(2022){Larsen}, {Eitner}, {Magg}, {Bergemann}, {Moltzer}, {Brodie}, {Romanowsky}, \& {Strader}}]{larsen2022}
{Larsen}, S.~S., {Eitner}, P., {Magg}, E., {et~al.} 2022, \aap, 660, A88

\bibitem[{{Lendl} {et~al.}(2012){Lendl}, {Anderson}, {Collier-Cameron}, {Doyle}, {Gillon}, {Hellier}, {Jehin}, {Lister}, {Maxted}, {Pepe}, {Pollacco}, {Queloz}, {Smalley}, {S{\'e}gransan}, {Smith}, {Triaud}, {Udry}, {West}, \& {Wheatley}}]{Lendl2012}
{Lendl}, M., {Anderson}, D.~R., {Collier-Cameron}, A., {et~al.} 2012, \aap, 544, A72

\bibitem[{{Louden} \& {Wheatley}(2015)}]{Loud15}
{Louden}, T. \& {Wheatley}, P.~J. 2015, \apjl, 814, L24

\bibitem[{{Magg} {et~al.}(2022){Magg}, {Bergemann}, {Serenelli}, {Bautista}, {Plez}, {Heiter}, {Gerber}, {Ludwig}, {Basu}, {Ferguson}, {Gallego}, {Gamrath}, {Palmeri}, \& {Quinet}}]{magg2022}
{Magg}, E., {Bergemann}, M., {Serenelli}, A., {et~al.} 2022, \aap, 661, A140

\bibitem[{{Mayor} {et~al.}(2003){Mayor}, {Pepe}, {Queloz}, {Bouchy}, {Rupprecht}, {Lo Curto}, {Avila}, {Benz}, {Bertaux}, {Bonfils}, {Dall}, {Dekker}, {Delabre}, {Eckert}, {Fleury}, {Gilliotte}, {Gojak}, {Guzman}, {Kohler}, {Lizon}, {Longinotti}, {Lovis}, {Megevand}, {Pasquini}, {Reyes}, {Sivan}, {Sosnowska}, {Soto}, {Udry}, {van Kesteren}, {Weber}, \& {Weilenmann}}]{Mayor03}
{Mayor}, M., {Pepe}, F., {Queloz}, D., {et~al.} 2003, The Messenger, 114, 20

\bibitem[{{Mazeh} {et~al.}(2000){Mazeh}, {Naef}, {Torres}, {Latham}, {Mayor}, {Beuzit}, {Brown}, {Buchhave}, {Burnet}, {Carney}, {Charbonneau}, {Drukier}, {Laird}, {Pepe}, {Perrier}, {Queloz}, {Santos}, {Sivan}, {Udry}, \& {Zucker}}]{Mazeh2000}
{Mazeh}, T., {Naef}, D., {Torres}, G., {et~al.} 2000, \apjl, 532, L55

\bibitem[{{McLaughlin}(1924)}]{McL1924}
{McLaughlin}, D.~B. 1924, \apj, 60, 22

\bibitem[{{Mounzer} {et~al.}(2022){Mounzer}, {Lovis}, {Seidel}, {Attia}, {Allart}, {Bourrier}, {Ehrenreich}, {Wyttenbach}, {Astudillo-Defru}, {Beatty}, {Cegla}, {Heng}, {Lavie}, {Lendl}, {Melo}, {Pepe}, {Pepper}, {Rodriguez}, {S{\'e}gransan}, {Udry}, {Linder}, \& {Sousa}}]{Moun22}
{Mounzer}, D., {Lovis}, C., {Seidel}, J.~V., {et~al.} 2022, \aap, 668, A1

\bibitem[{{Moutou} {et~al.}(2020){Moutou}, {Dalal}, {Donati}, {Martioli}, {Folsom}, {Artigau}, {Boisse}, {Bouchy}, {Carmona}, {Cook}, {Delfosse}, {Doyon}, {Fouqu{\'e}}, {Gaisn{\'e}}, {H{\'e}brard}, {Hobson}, {Klein}, {Lecavelier des Etangs}, \& {Morin}}]{Mout20}
{Moutou}, C., {Dalal}, S., {Donati}, J.~F., {et~al.} 2020, \aap, 642, A72

\bibitem[{{Moutou} {et~al.}(2007){Moutou}, {Donati}, {Savalle}, {Hussain}, {Alecian}, {Bouchy}, {Catala}, {Collier Cameron}, {Udry}, \& {Vidal-Madjar}}]{Mout07}
{Moutou}, C., {Donati}, J.~F., {Savalle}, R., {et~al.} 2007, \aap, 473, 651

\bibitem[{{Oshagh} {et~al.}(2014){Oshagh}, {Santos}, {Ehrenreich}, {Haghighipour}, {Figueira}, {Santerne}, \& {Montalto}}]{Oshagh14}
{Oshagh}, M., {Santos}, N.~C., {Ehrenreich}, D., {et~al.} 2014, \aap, 568, A99

\bibitem[{{Pepe} {et~al.}(2021){Pepe}, {Cristiani}, {Rebolo}, {Santos}, {Dekker}, {Cabral}, {Di Marcantonio}, {Figueira}, {Lo Curto}, {Lovis}, {Mayor}, {M{\'e}gevand}, {Molaro}, {Riva}, {Zapatero Osorio}, {Amate}, {Manescau}, {Pasquini}, {Zerbi}, {Adibekyan}, {Abreu}, {Affolter}, {Alibert}, {Aliverti}, {Allart}, {Allende Prieto}, {{\'A}lvarez}, {Alves}, {Avila}, {Baldini}, {Bandy}, {Barros}, {Benz}, {Bianco}, {Borsa}, {Bourrier}, {Bouchy}, {Broeg}, {Calderone}, {Cirami}, {Coelho}, {Conconi}, {Coretti}, {Cumani}, {Cupani}, {D'Odorico}, {Damasso}, {Deiries}, {Delabre}, {Demangeon}, {Dumusque}, {Ehrenreich}, {Faria}, {Fragoso}, {Genolet}, {Genoni}, {G{\'e}nova Santos}, {Gonz{\'a}lez Hern{\'a}ndez}, {Hughes}, {Iwert}, {Kerber}, {Knudstrup}, {Landoni}, {Lavie}, {Lillo-Box}, {Lizon}, {Maire}, {Martins}, {Mehner}, {Micela}, {Modigliani}, {Monteiro}, {Monteiro}, {Moschetti}, {Murphy}, {Nunes}, {Oggioni}, {Oliveira}, {Oshagh}, {Pall{\'e}}, {Pariani}, {Poretti}, {Rasilla}, {Rebord{\~a}o}, {Redaelli}, {Santana Tschudi},
  {Santin}, {Santos}, {S{\'e}gransan}, {Schmidt}, {Segovia}, {Sosnowska}, {Sozzetti}, {Sousa}, {Span{\`o}}, {Su{\'a}rez Mascare{\~n}o}, {Tabernero}, {Tenegi}, {Udry}, \& {Zanutta}}]{Pepe2021}
{Pepe}, F., {Cristiani}, S., {Rebolo}, R., {et~al.} 2021, \aap, 645, A96

\bibitem[{{Pepe} {et~al.}(2014){Pepe}, {Molaro}, {Cristiani}, {Rebolo}, {Santos}, {Dekker}, {M{\'e}gevand}, {Zerbi}, {Cabral}, {Di Marcantonio}, {Abreu}, {Affolter}, {Aliverti}, {Allende Prieto}, {Amate}, {Avila}, {Baldini}, {Bristow}, {Broeg}, {Cirami}, {Coelho}, {Conconi}, {Coretti}, {Cupani}, {D'Odorico}, {De Caprio}, {Delabre}, {Dorn}, {Figueira}, {Fragoso}, {Galeotta}, {Genolet}, {Gomes}, {Gonz{\'a}lez Hern{\'a}ndez}, {Hughes}, {Iwert}, {Kerber}, {Landoni}, {Lizon}, {Lovis}, {Maire}, {Mannetta}, {Martins}, {Monteiro}, {Oliveira}, {Poretti}, {Rasilla}, {Riva}, {Santana Tschudi}, {Santos}, {Sosnowska}, {Sousa}, {Span{\'o}}, {Tenegi}, {Toso}, {Vanzella}, {Viel}, \& {Zapatero Osorio}}]{Pepe14}
{Pepe}, F., {Molaro}, P., {Cristiani}, S., {et~al.} 2014, Astronomische Nachrichten, 335, 8

\bibitem[{{Plez}(2012)}]{plez2012}
{Plez}, B. 2012, {Turbospectrum: Code for spectral synthesis}

\bibitem[{{Pont} {et~al.}(2013){Pont}, {Sing}, {Gibson}, {Aigrain}, {Henry}, \& {Husnoo}}]{Pont13}
{Pont}, F., {Sing}, D.~K., {Gibson}, N.~P., {et~al.} 2013, \mnras, 432, 2917

\bibitem[{{Prinoth} {et~al.}(2024){Prinoth}, {Hoeijmakers}, {Morris}, {Lam}, {Kitzmann}, {Sedaghati}, {Seidel}, {Lee}, {Thorsbro}, {Borsato}, {Damasceno}, {Pelletier}, \& {Seifahrt}}]{Prin24}
{Prinoth}, B., {Hoeijmakers}, H.~J., {Morris}, B.~M., {et~al.} 2024, \aap, 685, A60

\bibitem[{{Quirrenbach} {et~al.}(2014){Quirrenbach}, {Amado}, {Caballero}, {Mundt}, {Reiners}, {Ribas}, {Seifert}, {Abril}, {Aceituno}, {Alonso-Floriano}, {Ammler-von Eiff}, {Antona Jim{\'e}nez}, {Anwand-Heerwart}, {Azzaro}, {Bauer}, {Barrado}, {Becerril}, {B{\'e}jar}, {Ben{\'\i}tez}, {Berdi{\~n}as}, {C{\'a}rdenas}, {Casal}, {Claret}, {Colom{\'e}}, {Cort{\'e}s-Contreras}, {Czesla}, {Doellinger}, {Dreizler}, {Feiz}, {Fern{\'a}ndez}, {Galad{\'\i}}, {G{\'a}lvez-Ortiz}, {Garc{\'\i}a-Piquer}, {Garc{\'\i}a-Vargas}, {Garrido}, {Gesa}, {G{\'o}mez Galera}, {Gonz{\'a}lez {\'A}lvarez}, {Gonz{\'a}lez Hern{\'a}ndez}, {Gr{\"o}zinger}, {Gu{\`a}rdia}, {Guenther}, {de Guindos}, {Guti{\'e}rrez-Soto}, {Hagen}, {Hatzes}, {Hauschildt}, {Helmling}, {Henning}, {Hermann}, {Hern{\'a}ndez Casta{\~n}o}, {Herrero}, {Hidalgo}, {Holgado}, {Huber}, {Huber}, {Jeffers}, {Joergens}, {de Juan}, {Kehr}, {Klein}, {K{\"u}rster}, {Lamert}, {Lalitha}, {Laun}, {Lemke}, {Lenzen}, {L{\'o}pez del Fresno}, {L{\'o}pez Mart{\'\i}}, {L{\'o}pez-Santiago},
  {Mall}, {Mandel}, {Mart{\'\i}n}, {Mart{\'\i}n-Ruiz}, {Mart{\'\i}nez-Rodr{\'\i}guez}, {Marvin}, {Mathar}, {Mirabet}, {Montes}, {Morales Mu{\~n}oz}, {Moya}, {Naranjo}, {Ofir}, {Oreiro}, {Pall{\'e}}, {Panduro}, {Passegger}, {P{\'e}rez-Calpena}, {P{\'e}rez Medialdea}, {Perger}, {Pluto}, {Ram{\'o}n}, {Rebolo}, {Redondo}, {Reffert}, {Reinhardt}, {Rhode}, {Rix}, {Rodler}, {Rodr{\'\i}guez}, {Rodr{\'\i}guez-L{\'o}pez}, {Rodr{\'\i}guez-P{\'e}rez}, {Rohloff}, {Rosich}, {S{\'a}nchez-Blanco}, {S{\'a}nchez Carrasco}, {Sanz-Forcada}, {Sarmiento}, {Sch{\"a}fer}, {Schiller}, {Schmidt}, {Schmitt}, {Solano}, {Stahl}, {Storz}, {St{\"u}rmer}, {Su{\'a}rez}, {Ulbrich}, {Veredas}, {Wagner}, {Winkler}, {Zapatero Osorio}, {Zechmeister}, {Abell{\'a}n de Paco}, {Anglada-Escud{\'e}}, {del Burgo}, {Klutsch}, {Lizon}, {L{\'o}pez-Morales}, {Morales}, {Perryman}, {Tulloch}, \& {Xu}}]{Quirr14}
{Quirrenbach}, A., {Amado}, P.~J., {Caballero}, J.~A., {et~al.} 2014, in Society of Photo-Optical Instrumentation Engineers (SPIE) Conference Series, Vol. 9147, Ground-based and Airborne Instrumentation for Astronomy V, ed. S.~K. {Ramsay}, I.~S. {McLean}, \& H.~{Takami}, 91471F

\bibitem[{{Radica} {et~al.}(2024){Radica}, {Coulombe}, {Taylor}, {Albert}, {Allart}, {Benneke}, {Cowan}, {Dang}, {Lafreni{\`e}re}, {Thorngren}, {Artigau}, {Doyon}, {Flagg}, {Johnstone}, {Pelletier}, \& {Roy}}]{Radica24}
{Radica}, M., {Coulombe}, L.-P., {Taylor}, J., {et~al.} 2024, \apjl, 962, L20

\bibitem[{{Redfield} {et~al.}(2008){Redfield}, {Endl}, {Cochran}, \& {Koesterke}}]{Redfield08}
{Redfield}, S., {Endl}, M., {Cochran}, W.~D., \& {Koesterke}, L. 2008, \apjl, 673, L87

\bibitem[{{Ricker} {et~al.}(2014){Ricker}, {Winn}, {Vanderspek}, {Latham}, {Bakos}, {Bean}, {Berta-Thompson}, {Brown}, {Buchhave}, {Butler}, {Butler}, {Chaplin}, {Charbonneau}, {Christensen-Dalsgaard}, {Clampin}, {Deming}, {Doty}, {De Lee}, {Dressing}, {Dunham}, {Endl}, {Fressin}, {Ge}, {Henning}, {Holman}, {Howard}, {Ida}, {Jenkins}, {Jernigan}, {Johnson}, {Kaltenegger}, {Kawai}, {Kjeldsen}, {Laughlin}, {Levine}, {Lin}, {Lissauer}, {MacQueen}, {Marcy}, {McCullough}, {Morton}, {Narita}, {Paegert}, {Palle}, {Pepe}, {Pepper}, {Quirrenbach}, {Rinehart}, {Sasselov}, {Sato}, {Seager}, {Sozzetti}, {Stassun}, {Sullivan}, {Szentgyorgyi}, {Torres}, {Udry}, \& {Villasenor}}]{Rick14}
{Ricker}, G.~R., {Winn}, J.~N., {Vanderspek}, R., {et~al.} 2014, in Society of Photo-Optical Instrumentation Engineers (SPIE) Conference Series, Vol. 9143, Space Telescopes and Instrumentation 2014: Optical, Infrared, and Millimeter Wave, ed. J.~{Oschmann}, Jacobus~M., M.~{Clampin}, G.~G. {Fazio}, \& H.~A. {MacEwen}, 914320

\bibitem[{{Rossiter}(1924)}]{Ross1924}
{Rossiter}, R.~A. 1924, \apj, 60, 15

\bibitem[{{Rustamkulov} {et~al.}(2023){Rustamkulov}, {Sing}, {Mukherjee}, {May}, {Kirk}, {Schlawin}, {Line}, {Piaulet}, {Carter}, {Batalha}, {Goyal}, {L{\'o}pez-Morales}, {Lothringer}, {MacDonald}, {Moran}, {Stevenson}, {Wakeford}, {Espinoza}, {Bean}, {Batalha}, {Benneke}, {Berta-Thompson}, {Crossfield}, {Gao}, {Kreidberg}, {Powell}, {Cubillos}, {Gibson}, {Leconte}, {Molaverdikhani}, {Nikolov}, {Parmentier}, {Roy}, {Taylor}, {Turner}, {Wheatley}, {Aggarwal}, {Ahrer}, {Alam}, {Alderson}, {Allen}, {Banerjee}, {Barat}, {Barrado}, {Barstow}, {Bell}, {Blecic}, {Brande}, {Casewell}, {Changeat}, {Chubb}, {Crouzet}, {Daylan}, {Decin}, {D{\'e}sert}, {Mikal-Evans}, {Feinstein}, {Flagg}, {Fortney}, {Harrington}, {Heng}, {Hong}, {Hu}, {Iro}, {Kataria}, {Kempton}, {Krick}, {Lendl}, {Lillo-Box}, {Louca}, {Lustig-Yaeger}, {Mancini}, {Mansfield}, {Mayne}, {Miguel}, {Morello}, {Ohno}, {Palle}, {Petit dit de la Roche}, {Rackham}, {Radica}, {Ramos-Rosado}, {Redfield}, {Rogers}, {Shkolnik}, {Southworth}, {Teske}, {Tremblin},
  {Tucker}, {Venot}, {Waalkes}, {Welbanks}, {Zhang}, \& {Zieba}}]{Rustamkulov23}
{Rustamkulov}, Z., {Sing}, D.~K., {Mukherjee}, S., {et~al.} 2023, \nat, 614, 659

\bibitem[{Ryabchikova {et~al.}(2015)Ryabchikova, Piskunov, Kurucz, Stempels, Heiter, Pakhomov, \& Barklem}]{Ryabchikova2015}
Ryabchikova, T., Piskunov, N., Kurucz, R.~L., {et~al.} 2015, Physica Scripta, 90, 054005

\bibitem[{{Seager} \& {Sasselov}(2000)}]{SS2000}
{Seager}, S. \& {Sasselov}, D.~D. 2000, \apj, 537, 916

\bibitem[{{Seidel} {et~al.}(2023){Seidel}, {Borsa}, {Pino}, {Ehrenreich}, {Stangret}, {Zapatero Osorio}, {Palle}, {Alibert}, {Allart}, {Bourrier}, {Di Marcantonio}, {Figueira}, {Gonz{\'a}lez Hern{\'a}ndez}, {Lillo-Box}, {Lovis}, {Martins}, {Mehner}, {Molaro}, {Nunes}, {Pepe}, {Santos}, \& {Sozzetti}}]{Seidel23}
{Seidel}, J.~V., {Borsa}, F., {Pino}, L., {et~al.} 2023, \aap, 673, A125

\bibitem[{{Seidel} {et~al.}(2020){Seidel}, {Ehrenreich}, {Pino}, {Bourrier}, {Lavie}, {Allart}, {Wyttenbach}, \& {Lovis}}]{Seidel20}
{Seidel}, J.~V., {Ehrenreich}, D., {Pino}, L., {et~al.} 2020, \aap, 633, A86

\bibitem[{{Seidel} {et~al.}(2019){Seidel}, {Ehrenreich}, {Wyttenbach}, {Allart}, {Lendl}, {Pino}, {Bourrier}, {Cegla}, {Lovis}, {Barrado}, {Bayliss}, {Astudillo-Defru}, {Deline}, {Fisher}, {Heng}, {Joseph}, {Lavie}, {Melo}, {Pepe}, {S{\'e}gransan}, \& {Udry}}]{Seidel19}
{Seidel}, J.~V., {Ehrenreich}, D., {Wyttenbach}, A., {et~al.} 2019, \aap, 623, A166

\bibitem[{{Seidel} {et~al.}(2025){Seidel}, {Prinoth}, {Pino}, {dos Santos}, {Chakraborty}, {Parmentier}, {Sedaghati}, {Wardenier}, {Farret Jentink}, {Zapatero Osorio}, {Allart}, {Ehrenreich}, {Lendl}, {Roccetti}, {Damasceno}, {Bourrier}, {Lillo-Box}, {Hoeijmakers}, {Pall{\'e}}, {Santos}, {Su{\'a}rez Mascare{\~n}o}, {Sousa}, {Tabernero}, \& {Pepe}}]{Seidel25}
{Seidel}, J.~V., {Prinoth}, B., {Pino}, L., {et~al.} 2025, \nat, 639, 902

\bibitem[{{Sicilia} {et~al.}(2022){Sicilia}, {Malavolta}, {Pino}, {Scandariato}, {Nascimbeni}, {Piotto}, \& {Pagano}}]{Sici22}
{Sicilia}, D., {Malavolta}, L., {Pino}, L., {et~al.} 2022, \aap, 667, A19

\bibitem[{{Sicilia} {et~al.}(2025){Sicilia}, {Malavolta}, {Scandariato}, {Fossati}, {Lanza}, {Bonomo}, {Borsa}, {Guilluy}, {Nascimbeni}, {Pino}, {Biassoni}, {D'Arpa}, {Pagano}, {Sozzetti}, {Stangret}, {Cosentino}, {Giacobbe}, {Lodi}, {Maldonado}, {Nardiello}, \& {Pedani}}]{Sici25}
{Sicilia}, D., {Malavolta}, L., {Scandariato}, G., {et~al.} 2025, \aap, 693, A316

\bibitem[{{Sing} {et~al.}(2016){Sing}, {Fortney}, {Nikolov}, {Wakeford}, {Kataria}, {Evans}, {Aigrain}, {Ballester}, {Burrows}, {Deming}, {D{\'e}sert}, {Gibson}, {Henry}, {Huitson}, {Knutson}, {Lecavelier Des Etangs}, {Pont}, {Showman}, {Vidal-Madjar}, {Williamson}, \& {Wilson}}]{Sing16}
{Sing}, D.~K., {Fortney}, J.~J., {Nikolov}, N., {et~al.} 2016, \nat, 529, 59

\bibitem[{{Smette} {et~al.}(2015){Smette}, {Sana}, {Noll}, {Horst}, {Kausch}, {Kimeswenger}, {Barden}, {Szyszka}, {Jones}, {Gallenne}, {Vinther}, {Ballester}, \& {Taylor}}]{Smette2015}
{Smette}, A., {Sana}, H., {Noll}, S., {et~al.} 2015, \aap, 576, A77

\bibitem[{{Snellen} {et~al.}(2010){Snellen}, {de Kok}, {de Mooij}, \& {Albrecht}}]{Snellen10}
{Snellen}, I. A.~G., {de Kok}, R.~J., {de Mooij}, E. J.~W., \& {Albrecht}, S. 2010, \nat, 465, 1049

\bibitem[{{Strassmeier} {et~al.}(2015){Strassmeier}, {Ilyin}, {J{\"a}rvinen}, {Weber}, {Woche}, {Barnes}, {Bauer}, {Beckert}, {Bittner}, {Bredthauer}, {Carroll}, {Denker}, {Dionies}, {DiVarano}, {D{\"o}scher}, {Fechner}, {Feuerstein}, {Granzer}, {Hahn}, {Harnisch}, {Hofmann}, {Lesser}, {Paschke}, {Pankratow}, {Plank}, {Pl{\"u}schke}, {Popow}, \& {Sablowski}}]{Strass15}
{Strassmeier}, K.~G., {Ilyin}, I., {J{\"a}rvinen}, A., {et~al.} 2015, Astronomische Nachrichten, 336, 324

\bibitem[{{Tabernero} {et~al.}(2021){Tabernero}, {Zapatero Osorio}, {Allart}, {Borsa}, {Casasayas-Barris}, {Demangeon}, {Ehrenreich}, {Lillo-Box}, {Lovis}, {Pall{\'e}}, {Sousa}, {Rebolo}, {Santos}, {Pepe}, {Cristiani}, {Adibekyan}, {Allende Prieto}, {Alibert}, {Barros}, {Bouchy}, {Bourrier}, {D'Odorico}, {Dumusque}, {Faria}, {Figueira}, {G{\'e}nova Santos}, {Gonz{\'a}lez Hern{\'a}ndez}, {Hojjatpanah}, {Lo Curto}, {Lavie}, {Martins}, {Martins}, {Mehner}, {Micela}, {Molaro}, {Nunes}, {Poretti}, {Seidel}, {Sozzetti}, {Su{\'a}rez Mascare{\~n}o}, {Udry}, {Aliverti}, {Affolter}, {Alves}, {Amate}, {Avila}, {Bandy}, {Benz}, {Bianco}, {Broeg}, {Cabral}, {Conconi}, {Coelho}, {Cumani}, {Deiries}, {Dekker}, {Delabre}, {Fragoso}, {Genoni}, {Genolet}, {Hughes}, {Knudstrup}, {Kerber}, {Landoni}, {Lizon}, {Maire}, {Manescau}, {Di Marcantonio}, {M{\'e}gevand}, {Monteiro}, {Monteiro}, {Moschetti}, {Mueller}, {Modigliani}, {Oggioni}, {Oliveira}, {Pariani}, {Pasquini}, {Rasilla}, {Redaelli}, {Riva}, {Santana-Tschudi}, {Santin},
  {Santos}, {Segovia}, {Sosnowska}, {Span{\`o}}, {Tenegi}, {Iwert}, {Zanutta}, \& {Zerbi}}]{Tab21}
{Tabernero}, H.~M., {Zapatero Osorio}, M.~R., {Allart}, R., {et~al.} 2021, \aap, 646, A158

\bibitem[{{Wyttenbach} {et~al.}(2015){Wyttenbach}, {Ehrenreich}, {Lovis}, {Udry}, \& {Pepe}}]{Wytt15}
{Wyttenbach}, A., {Ehrenreich}, D., {Lovis}, C., {Udry}, S., \& {Pepe}, F. 2015, \aap, 577, A62

\bibitem[{{Wyttenbach} {et~al.}(2020){Wyttenbach}, {Molli{\`e}re}, {Ehrenreich}, {Cegla}, {Bourrier}, {Lovis}, {Pino}, {Allart}, {Seidel}, {Hoeijmakers}, {Nielsen}, {Lavie}, {Pepe}, {Bonfils}, \& {Snellen}}]{Wytt20}
{Wyttenbach}, A., {Molli{\`e}re}, P., {Ehrenreich}, D., {et~al.} 2020, \aap, 638, A87

\bibitem[{{Yan} \& {Henning}(2018)}]{Yan18}
{Yan}, F. \& {Henning}, T. 2018, Nature Astronomy, 2, 714

\bibitem[{{Yan} {et~al.}(2017){Yan}, {Pall{\'e}}, {Fosbury}, {Petr-Gotzens}, \& {Henning}}]{Yan2017}
{Yan}, F., {Pall{\'e}}, E., {Fosbury}, R.~A.~E., {Petr-Gotzens}, M.~G., \& {Henning}, T. 2017, \aap, 603, A73

\end{thebibliography}

\begin{appendix}

\onecolumn

\section{ESPRESSO wiggles impact on transmission spectrum}

\begin{figure*}[th!]
\begin{minipage}[tbh!]{\textwidth}
\includegraphics[trim=0cm 0cm 0cm 0cm,clip=true,width=\textwidth]{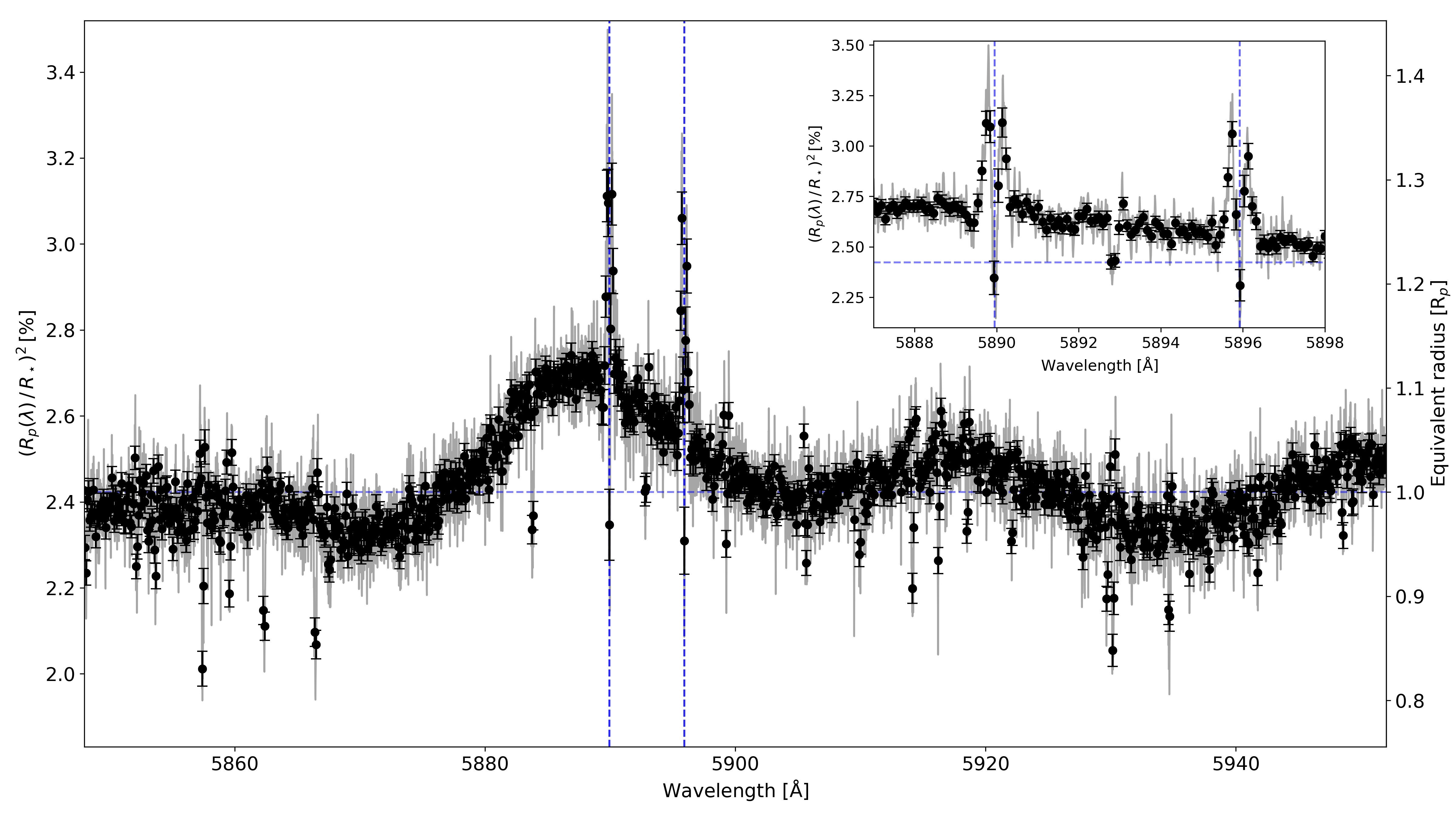}
\centering
\end{minipage}
\caption[]{Binned transmission spectrum of HD\,189733\,b without wiggles correction (see Sect. \ref{sect:wiggles}) combining both transits around the sodium doublet (vertical blue dashed lines). The unbinned transmission spectrum is shown in gray, the black dots representing the binned spectrum with step of 0.1 \AA . The horizontal blue line is the white-light radius $(R_\mathrm{p}/R_\mathrm{*})^2$. The zoomed-in panel shows a close-up view on the sodium doublet with similar wavelength interval to Fig. \ref{fig:TS Na no-corr}. The wiggles occur at a lower frequency than the POLDs and absorption signal. The correction of that pattern does not influence the shape of the signature around the sodium lines.
}
\label{fig:TS Na wiggle}
\end{figure*}

\pagebreak

\section{Rossiter-McLaughlin Revolutions correlation diagrams}

\begin{figure*}[h!]
\begin{minipage}[tbh!]{\textwidth}
\includegraphics[trim=0cm 0cm 0cm 0cm,clip=true,width=\textwidth]{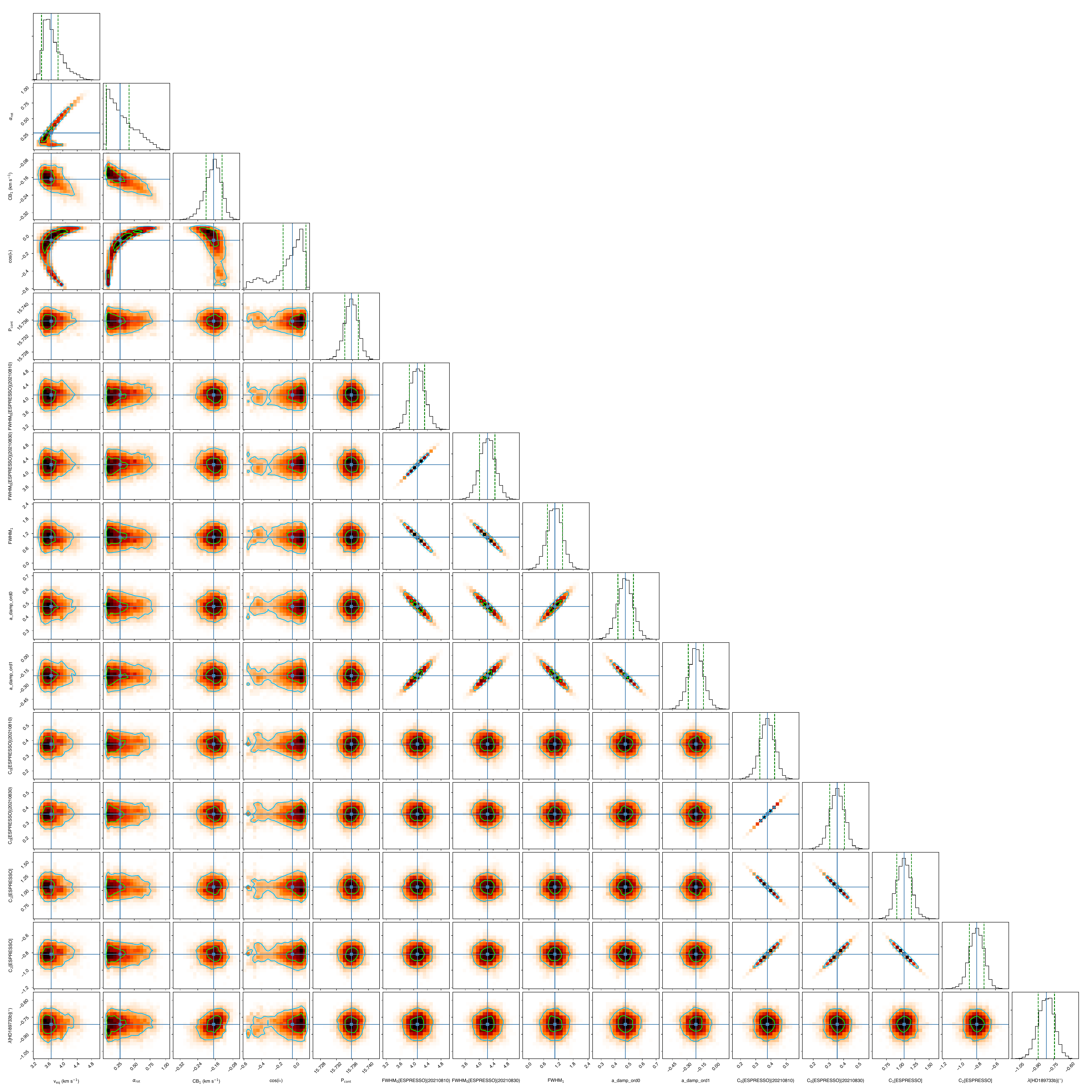}
\centering
\end{minipage}
\caption[]{Correlation diagrams for the PDFs of the raw RMR model parameters. Green and blue lines show the 1 and 2$\sigma$ simultaneous 2D confidence regions that contain, respectively, 39.3\% and 86.5\% of the accepted steps. 1D histograms correspond to the distributions projected on the space of each line parameter, with the green dashed lines limiting the 68.3\% HDIs. The blue lines and squares show the median values.}
\label{fig:Corr_diag_FULL}
\end{figure*}

\pagebreak

\section{Planetary atmosphere parameters fit}

\begin{figure}[tbh!]
\sidecaption
\includegraphics[width=12cm]{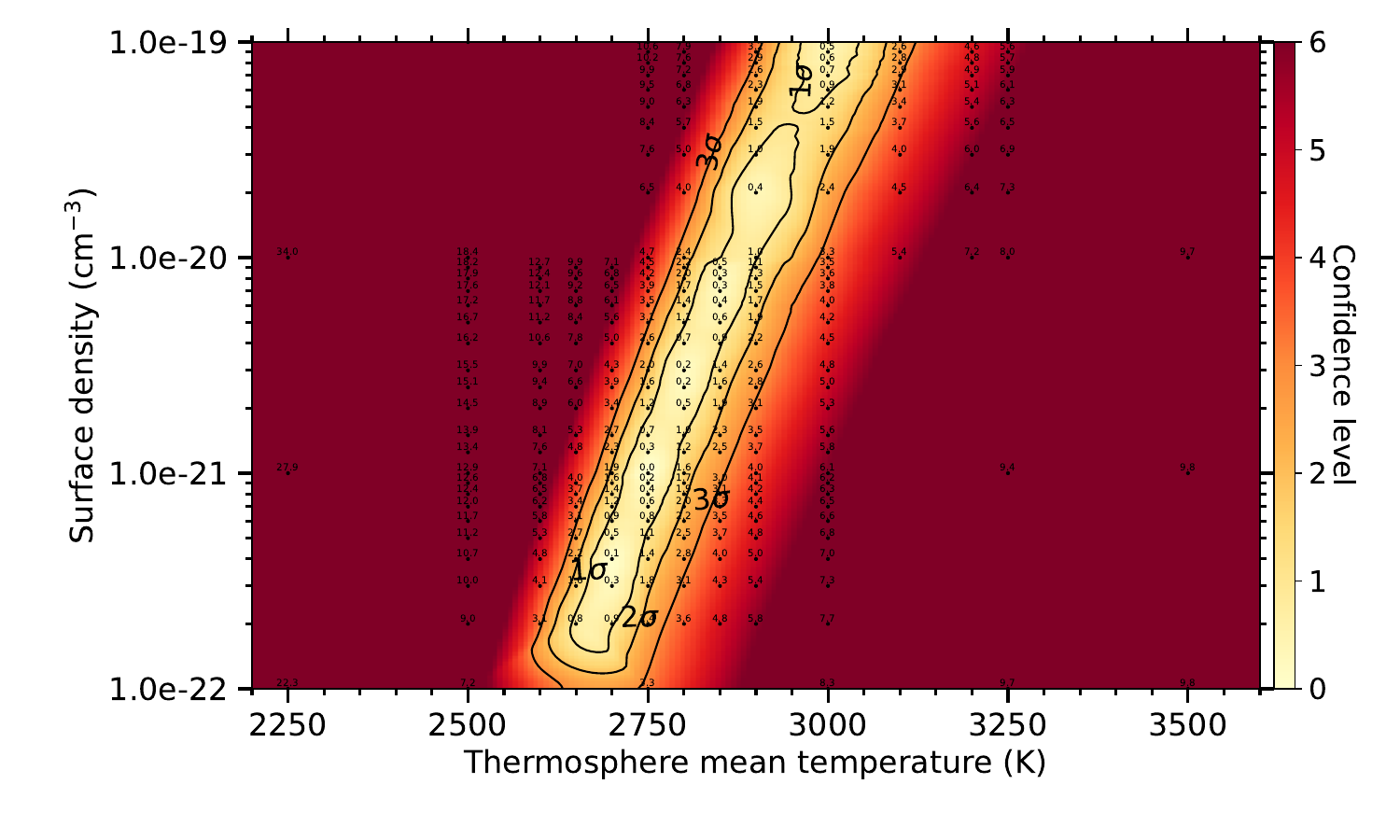}
\caption[]{$\chi^2$ map of temperature and number density of sodium atoms at the top of the atmosphere (at the Roche lobe radius) of HD\,189733\,b (see Sect. \ref{section:EVE} for details).}
\label{fig:chi2}
\end{figure}

\section{Sodium excess absorption maps}

\begin{figure*}[tbh!]
\includegraphics[trim=0cm 0cm 0cm 0cm,clip=true,width=\textwidth]{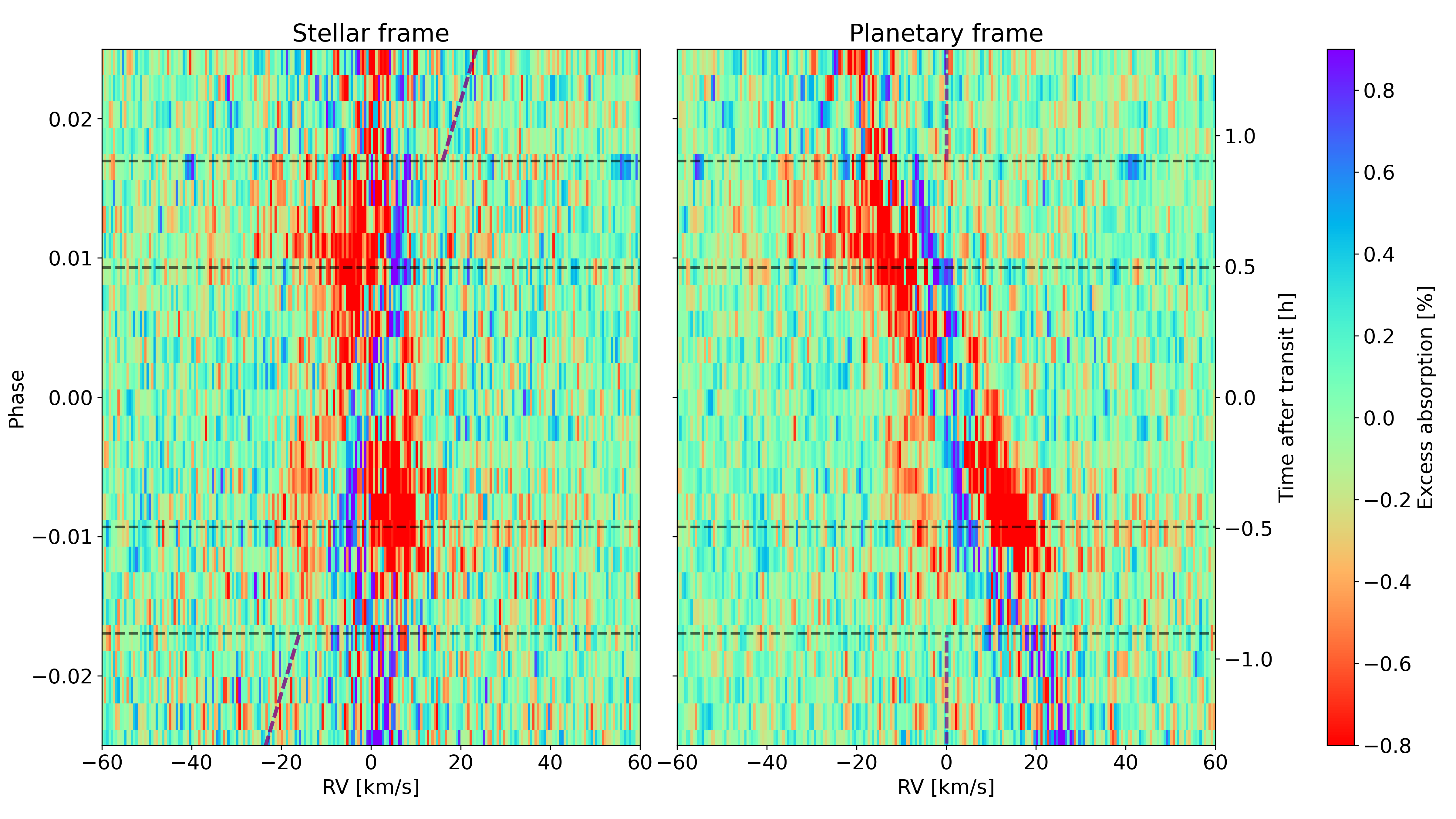}
\centering
\caption[]{Transmission spectra map of HD\,189733\,b around the sodium doublet (co-added and both transits combined) before POLDs correction, in stellar and planetary rest frames respectively (see Fig. \ref{fig:TS2DRM} for the post-correction map). The purple vertical dashed line indicates sodium planetary rest velocity and horizontal dashed lines show transit contacts T$_1$, T$_2$, T$_3$ and T$_4$ from bottom to top.}
\label{fig:TS2D}
\end{figure*}

Absorption maps as a function of phase and wavelength allow us to visually disentangle the planetary and stellar signals, as they follow different Doppler tracks. To have a consistent continuum in all individual in- and out-of-transit transmission spectra, we used the simple form of excess absorption $- \tilde{\mathcal{R}} = 1- \frac{\tilde{F}_\mathrm{i}}{\tilde{F}_\mathrm{out}}$, as normalizing with transit depth and limb darkening shifts the continuum of in-transit spectra to the "white-light" transit depth $(R_\mathrm{p}/R_\mathrm{*})^2$. The spectra are then binned in phase between both transits for better S/N and visualization (see Figs. \ref{fig:TS2DRM} and \ref{fig:TS2D}).

\section{Sodium transmission light curve}
\begin{figure*}[tbh!]
\sidecaption
\begin{minipage}[tbh!]{12cm}
\includegraphics[width=\textwidth]{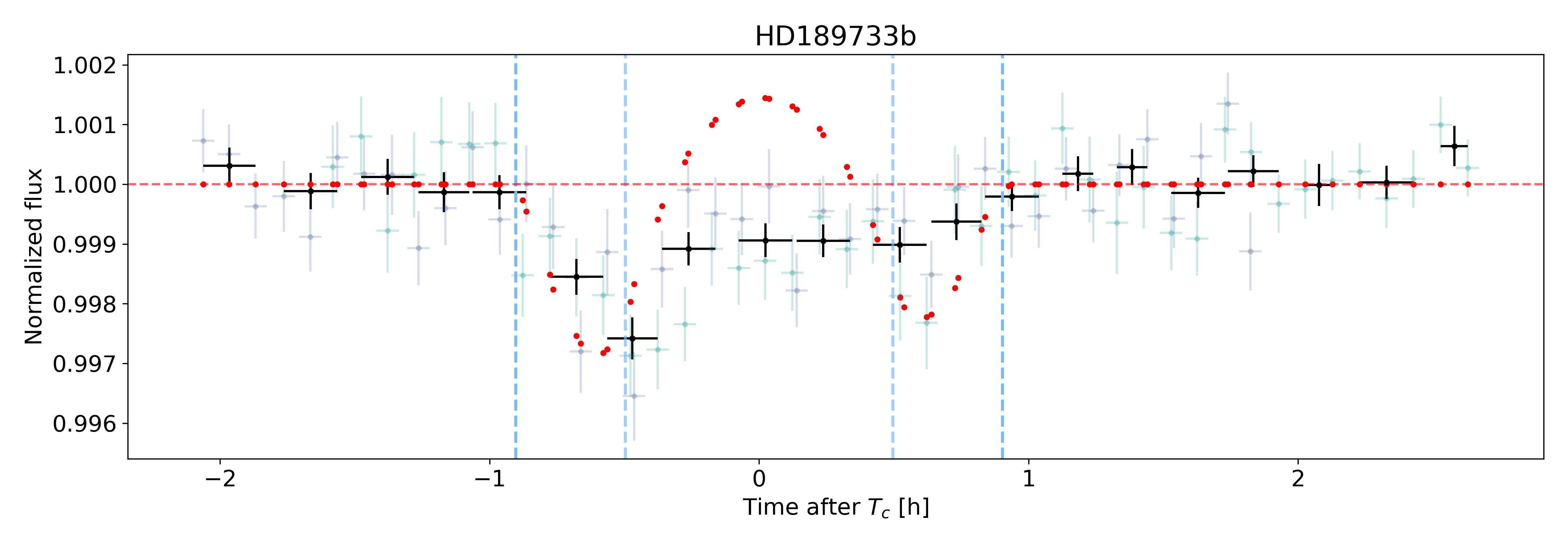}
\includegraphics[width=\textwidth]{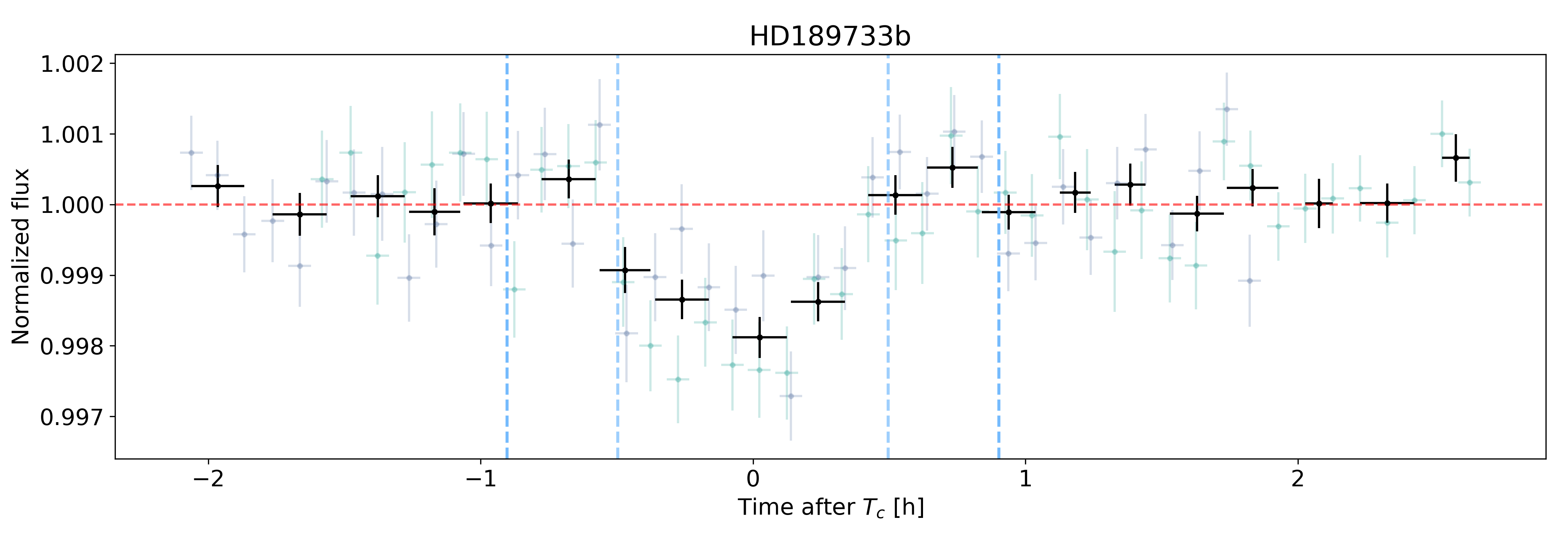}
\end{minipage}
\caption[]{Sodium transmission light curve before (top) and after (bottom) correction of POLDs using synthetic absorption spectra. The red dots in top panel are the transmission light curve points derived from the synthetic absorption spectra. The first transit is shown with blue errorbars, the second in green, and combined in black.}

\label{fig:Na TLC}
\end{figure*}

We compute the sodium transmission light curve by averaging through a passband of $\pm$ 0.4\,\AA\  around the core of each sodium line on individual transmission spectra (e.g., \citealt{Wytt15, Sici22, Keles24}). We do so on the uncorrected, synthetic, and corrected transmission spectra, that we bin together in phase and show in Fig. \ref{fig:Na TLC}. This large passband extends around $\pm$ 20 km s$^{-1}$ and thus also contains the excess absorption from the stellar track that we see in the transmission spectra, which can cause an overestimation of the depth of the absorption. As in \citealt{Keles24}, ingress and egress do not show any atmospheric absorption. We also notice this absence of absorption in the 2D maps (see Fig. \ref{fig:TS2DRM}) and when computing T$_1$-T$_2$ and T$_3$-T$_4$ transmission spectra.

\end{appendix}

\end{document}